\documentclass[11pt, oneside]{article}   	
\usepackage{jheppub}
\usepackage{amsmath}
\usepackage{amssymb}
\usepackage{amsfonts}
\usepackage{amsthm}
\usepackage{amsbsy}
\usepackage{array}
\usepackage{mathtools}
\usepackage[usenames,dvipsnames]{xcolor}
\usepackage{subcaption}
\usepackage{dsdshorthand}
\usepackage{graphicx}
\usepackage[vcentermath]{youngtab}
\usepackage{tikz}
\usepackage{booktabs,multirow}
\usetikzlibrary{decorations.pathreplacing}
\usetikzlibrary{decorations.markings}
\usetikzlibrary{arrows.meta}
\makeatletter
\def\@fpheader{\ }
\makeatother

\title{Angular fractals in thermal QFT}
\author{Nathan Benjamin, Jaeha Lee, Sridip Pal, David Simmons-Duffin, and Yixin Xu}
\affiliation{Walter Burke Institute for Theoretical Physics, Caltech, Pasadena, California 91125, USA}
\emailAdd{nbenjami@caltech.edu, jaeha@caltech.edu, sridip@caltech.edu, dsd@caltech.edu, yixinxu@caltech.edu}

\date{}
\abstract{We show that thermal effective field theory controls the long-distance expansion of the partition function of a $d$-dimensional QFT, with an insertion of any finite-order spatial isometry. Consequently, the thermal partition function on a sphere displays a fractal-like structure as a function of angular twist, reminiscent of the behavior of a modular form near the real line. As an example application, we find that for CFTs, the effective free energy of even-spin minus odd-spin operators at high temperature is smaller than the usual free energy by a factor of $1/2^d$.  Near certain rational angles, the partition function receives subleading contributions from ``Kaluza-Klein vortex defects" in the thermal EFT, which we classify. We illustrate our results with examples in free and holographic theories, and also discuss nonperturbative corrections from worldline instantons.}

\preprint{CALT-TH 2024-021}

\begin{document}

\maketitle
\pagenumbering{roman}
\setcounter{page}{2}
\newpage
\pagenumbering{arabic}
\setcounter{page}{1}

\section{Introduction}
\label{sec:intro}

Many aspects of conformal field theories (CFTs) are universal at high energies. A famous example is Cardy's formula, which states that the entropy of local operators at sufficiently high energies takes a universal form in all unitary, compact 2d CFTs \cite{Cardy:1986ie} (see \cite{Mukhametzhanov:2019pzy, Pal:2019zzr, Mukhametzhanov:2020swe} for a precise formulation).  Equivalently, the partition function of a 2d CFT
\be
\Tr\left[e^{-\beta H+i \theta J}\right]
\label{eq:grandcanonicalpf}
\ee
is universal in the high temperature regime $\beta\to 0$ with $\theta \sim O(\beta)$.

The derivation of Cardy's formula uses  invariance of the torus partition function under the modular transformation $S:\tau\mto -1/\tau$. By instead using the full modular group $\PSL(2,\Z)$, one finds similar universal behavior as $\beta\to 0$, near {\it any rational angle\/} $\theta=\frac{2\pi p}{q}$, see e.g.\ \cite{Benjamin:2019stq}. This leads to universal ``spin-refined" versions of the density of states. For example, in the case $\frac p q = \frac 1 2$, the modular transformation $\tau \mto \frac{-\tau}{2\tau-1}$ gives the universal behavior of
\be
\label{eq:spinrefinedexample}
\mathrm{Tr}\left[e^{-\beta (H - i \Omega J)}(-1)^J\right] = \mathrm{Tr}\left[e^{-\beta H + i \theta J}\right]_{\theta=\pi + \beta \Omega},
\ee 
in the regime $\beta \rightarrow 0$ with $\Omega \sim O(1)$. For any given 2d CFT, the logarithm of (\ref{eq:spinrefinedexample}) is $1/4$ the logarithm of (\ref{eq:grandcanonicalpf}) at high temperature, leading to a universal result for the difference between densities of even- and odd-spin operators in 2d CFTs.\footnote{Modular invariance on higher genus surfaces also leads to universal results for OPE coefficients in 2d CFTs, as derived in \cite{Kraus:2016nwo, Cardy:2017qhl,Das:2017cnv,Brehm:2018ipf,Hikida:2018khg}, and unified in \cite{Collier:2019weq}.}

While modular invariance is not available on $S^{d-1}\x S^1$ in higher dimensions, higher dimensional CFTs still display forms of universality at high energies, both in their density of states \cite{Verlinde:2000wg, Kutasov:2000td, Bhattacharyya:2007vs, Shaghoulian:2015kta, Shaghoulian:2015lcn, Benjamin:2023qsc, Allameh:2024qqp}, and OPE coefficients \cite{Delacretaz:2020nit, Benjamin:2023qsc}. A central insight from \cite{Bhattacharyya:2007vs,Jensen:2012jh,Banerjee:2012iz} is that the high temperature behavior of a CFT can be captured by a ``thermal Effective Field Theory (EFT)" that efficiently encodes the constraints of conformal symmetry and locality. In \cite{Shaghoulian:2015kta, Shaghoulian:2015lcn, Benjamin:2023qsc, Allameh:2024qqp}, thermal EFT plays the role of a surrogate for the modular $S$-transformation (as well as modular transformations on genus-2 surfaces).

In this work, we will be interested in ``spin-refined" information about the CFT density of states in general dimensions. In particular, we will study the partition function (\ref{eq:grandcanonicalpf})
with high temperature ($\beta \ll 1$) and finite angles $\vec{\theta}$. (In higher dimensions we promote $\vec\theta$ and $\vec J$ to vectors with $\lfloor d/2 \rfloor$ components coming from the rank of $SO(d)$.) The regime $\vec{\theta} = \beta \vec \Omega$ with fixed $\vec\Omega$ is captured by thermal EFT as discussed in \cite{Bhattacharyya:2007vs, Jensen:2012jh, Banerjee:2012iz, Shaghoulian:2015lcn, Benjamin:2023qsc}. However, when $\vec{\theta}$ does not scale to zero as $\beta\to 0$, the na\"ive EFT description breaks down.

A simple example of a partition function with finite $\vec\theta$ is (\ref{eq:spinrefinedexample}): the relative density of even-spin and odd-spin operators with respect to some particular Cartan generator $J$ of the rotation group. 
This observable is na\"ively outside the regime of validity of the thermal EFT, since $\theta$ remains finite as $\beta\to 0$. 

\begin{figure}
\centering
\includegraphics[width=0.8\textwidth]{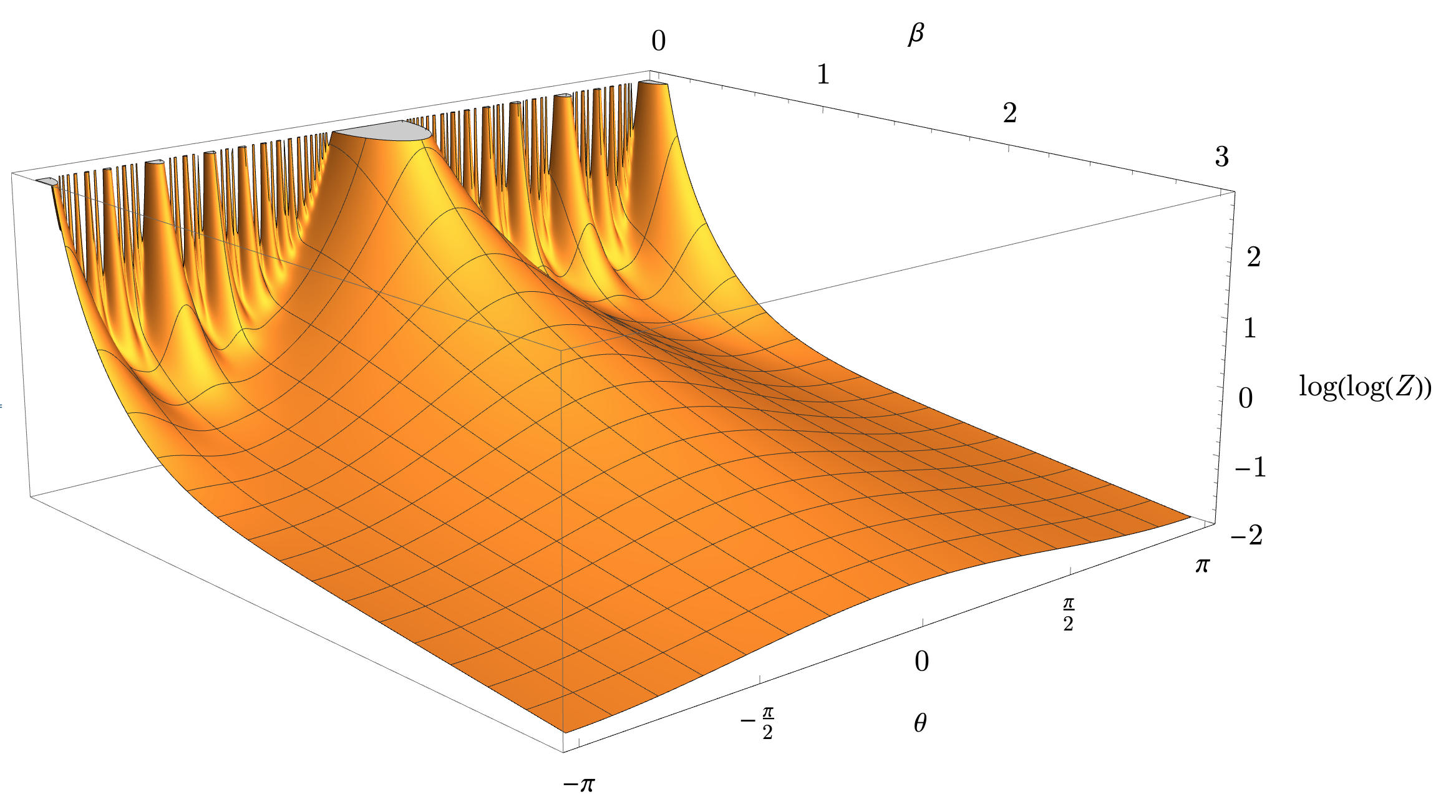}
\caption{\label{fig:isingcartoon}A qualitative picture of $\log(\log(Z))$ in the 3d Ising CFT, where $Z=\Tr(e^{-\beta H+i\theta J})$ is the $S^2 \x S^1$ partition function. To construct this picture, we took the leading terms in the EFT description around each rational angle (up to denominator $15$), and combined them with a root-mean-square. We give more detail in appendix~\ref{sec:pictureappendix}.}
\end{figure}

More generally, we can consider a partition function that includes a rotation by finite rational angles in each of the Cartan directions:\footnote{In parity-invariant theories, we can also include reflections.}
\be
\label{eq:spinrefinedexamplefull}
\mathrm{Tr}\left[e^{-\beta (H - i \vec{\Omega}\cdot \vec{J})}R\right],\quad\textrm{where}\quad R=e^{2\pi i \left(\frac{p_1}{q_1} J_1 + \cdots + \frac{p_n}{q_n} J_n\right)}.
\ee 
Using a trick that was applied in \cite{ArabiArdehali:2021nsx} to study superconformal indices near roots of unity, we will
find a {\it different\/} EFT description for this partition function, in terms of the thermal EFT on a background geometry with inverse temperature $q\beta$ and spatial manifold $S^{d-1}/\Z_q$,  where $ q= \text{lcm}(q_1, \ldots, q_n)$. This determines the small-$\beta$ expansion of (\ref{eq:spinrefinedexamplefull}) in terms of the usual Wilson coefficients of thermal EFT, up to new subleading contributions from ``Kaluza-Klein vortices" that we classify. For example, the effective free energy density of (\ref{eq:spinrefinedexamplefull}), coming from the leading term in the thermal effective action, is smaller than the usual free energy density by a factor of $1/q^d$. In particular, the effective free energy density of even-spin minus odd-spin operators described by (\ref{eq:spinrefinedexample}) is smaller by $1/2^d$. (This generalizes the factor of $1/4$ in 2d.)\footnote{Note that simply taking the density of states computed in \cite{Benjamin:2023qsc} and inserting the phase $R$ into the trace will \emph{not} give the correct answer to the partition function. For a demonstration of this in 2d, see appendix B of \cite{Benjamin:2019stq}.}

The EFT descriptions around each rational angle patch together to create fractal-like behavior in the high-temperature partition function --- see figure~\ref{fig:isingcartoon} for an illustration in the 3d Ising CFT\@.  It is remarkable that effective field theory constrains the asymptotics of the partition function in such an intricate way, even in higher dimensions.

Kaluza-Klein vortices appear whenever the rational rotation $R$ does not act freely on the sphere $S^{d-1}$. Each vortex creates a defect in the thermal EFT, whose action can be written systematically in a derivative expansion in background fields. By contrast, when $R$ generates a group that acts freely, no vortex defects are present, and the complete perturbative expansion of (\ref{eq:spinrefinedexamplefull}) in $\beta$ is determined in terms of thermal EFT Wilson coefficients, with no new undetermined parameters.

While most of our discussion and examples are focused on CFTs, our formalism also applies to general QFTs. In particular, using thermal effective field theory, we derive a relation between the partition function at temperature $T$ with a discrete isometry of order $q$ inserted, to the partition function with no insertion at temperature $T/q$, in the thermodynamic limit.\footnote{We are extremely grateful to Luca Delacretaz for emphasizing the general QFT case to us.} For example, we have
\be
\label{eq:moregeneralresultintro}
-\log \Tr_{\cH(\cM_L)}\left[e^{-\beta H} R\right] &\sim -\frac 1 q \log \Tr_{\cH(\cM_L)}\left[e^{-q\beta H}\right] + \textrm{topological} + \textrm{KK defects} \nn\\
&\hspace{2.57in}\qquad (\textrm{as }L\to \oo).
\ee
Here, $\cM_L$ is a spatial manifold of characteristic size $L$, with associated Hilbert space $\cH(\cM_L)$, $R$ is a discrete isometry of order $q$, and ``$\sim$" denotes agreement to all perturbative orders in $1/L$. The relation (\ref{eq:moregeneralresultintro}) holds whenever the theory is gapped at inverse temperature $q\beta$. We write the most general relation in (\ref{eq:moregeneralresult}), which we check in both massive and massless examples. An interesting consequence of this simple formula is that twists by discrete isometries can be sensitive to lower-temperature phases of the theory. For example, the partition function of QCD at temperature $T>\Lambda_\textrm{QCD}$, twisted by a discrete isometry with order $q$, becomes sensitive to physics below the confinement scale when $T/q<\Lambda_\textrm{QCD}$.

This universality of partition functions with spacetime symmetry insertions is in contrast to the case for global symmetry insertions. The insertion of a global symmetry generator operator is equivalent to turning on a new background field in the thermal EFT\@. The dependence of the effective action on this background field introduces new Wilson coefficients that are not necessarily related in a simple way to the Wilson coefficients without the global symmetry background, see e.g.\ \cite{Pal:2020wwd,Harlow:2021trr,Kang:2022orq}.

The paper is organized as follows. In section~\ref{sec:foldingunfolding}, we present a derivation of our main result: a systematic study of the high temperature expansion of the partition function of any quantum field theory with the insertion of a discrete isometry. In section~\ref{sec:defects}, we look in more detail at the Kaluza-Klein vortices that appear on $S^{d-1}$ when the discrete isometry (which is a rational rotation in this case) has fixed points. In section~\ref{sec:fermions}, we discuss subtleties that appear for fermionic theories. In section~\ref{sec:freetheories}, we give several examples in free theories that illustrate our general results. In section~\ref{sec:2dwithgravitationalanomaly}, we consider thermal effective actions with topological terms. In section~\ref{sec:holography}, we apply our results to holographic CFTs. In section~\ref{sec:journey}, we look at irrational $\theta$. In section~\ref{sec:nonpert}, we discuss non-perturbative corrections in temperature. Finally in section~\ref{sec:discuss}, we conclude and discuss future directions.

\section{Folding and unfolding the partition function}
\label{sec:foldingunfolding}

\subsection{Thermal effective action and finite velocities}
\label{sec:reviewteft}

Equilibrium correlators of generic interacting QFTs at finite temperature are expected to have a finite correlation length. Equivalently, the dimensional reduction of a generic interacting QFT on a Euclidean circle is expected to be gapped. When this is the case, long-distance finite-temperature observables of the QFT can be captured by a local ``thermal effective action" of background fields \cite{Bhattacharyya:2007vs,Jensen:2012jh,Banerjee:2012iz}. For example, consider the partition function of a QFT$_d$ on $\cM_L\x S^1_\beta$, where the spatial $d{-}1$-manifold $\cM_L$ has size $L$. In the thermodynamic limit of large $L$, we have
\be
\label{eq:usethermaleft}
\Tr_{\cH(\cM_{L})}[e^{-\beta H_L}] &= Z_\textrm{QFT}[\cM_{L} \x S^1_\beta]\nn\\
 &= Z_\textrm{gapped}[\cM_{L}]\nn\\
&\sim e^{-S_\textrm{th}[g,A,\f]}+\textrm{nonperturbative in $1/L$} \qquad (L\to \oo),
\ee
where $\cH(\cM_{L})$ is the Hilbert space of states on $\cM_{L}$, and $H_L$ is the Hamiltonian.
Here, the thermal effective action $S_\textrm{th}$ depends on a $d{-}1$-dimensional metric $g_{ij}$, a Kaluza-Klein gauge field $A_i$, and a dilaton $\f$, which can be obtained by placing the $d$-dimensional metric in Kaluza-Klein (KK) form
\be
G_{\mu\nu} dx^\mu dx^\nu &= g_{ij}(\vec x) dx^i dx^j + e^{2\f(\vec x)}(d\tau + A_i(\vec x))^2,
\ee
where $\tau\sim \tau+\beta$ is a periodic coordinate along the thermal circle. The derivative expansion for $S_\textrm{th}$ becomes an expansion in inverse powers of the length $L$.

If the spatial manifold $\cM_{L}$ possesses a continuous isometry $\xi$, then we can additionally twist the partition function by the corresponding charge $Q_\xi$:
\be
\label{eq:twistedobservable}
\Tr_{\cH(\cM_{L})}\left[e^{-\beta( H_L - i \alpha Q_\xi)}\right].
\ee
Geometrically, this twist corresponds to a deformation of the background fields $g,A,\f$ that depends on $\alpha \xi$.  In the thermodynamic limit, we can describe (\ref{eq:twistedobservable}) using the thermal effective action, provided that the background fields $g,A,\f$ remain finite as $L\to \oo$. In particular, the combination $\alpha \xi$ must remain finite as $L\to \oo$. The physical reason is that $i\alpha$ represents the velocity of the system in the direction of $\xi$ in the canonical ensemble. This velocity must remain finite in order to have a good thermodynamic limit.

By contrast, suppose that $\cM_{L}$ possesses a nontrivial discrete isometry $R$ with finite order $R^q=1$. If we twist the partition function by $R$,
\be
\Tr_{\cH(\cM_{L})}\left[e^{-\beta H} R\right],
\ee
then physically this corresponds to a system whose ``velocity" is of order $L$. The background fields $g,A,\f$ na\"ively do not have a good thermodynamic limit, and we cannot apply the thermal effective action in an obvious way.

\subsubsection{Example: CFT partition function}

An important example for us is the partition function of a CFT$_d$ on $S^{d-1} \x S^1_\beta$. Conformal invariance dictates that
\be
\Tr\left[e^{-\beta (H - i \vec \Omega\. \vec J)}\right] &= \Tr_{\cH(S^{d-1}_L)}\left[e^{-L\beta (H_L - i \vec \Omega\. \vec J_L)}\right].
\ee
On the left-hand side, we have the usual partition function of the CFT on a sphere of radius $1$. On the right-hand side, $H_L$ denotes the Hamiltonian on a sphere $S^{d-1}_L$ of radius $L$, and $\vec J_L$ are generators of isometries of the sphere, normalized so that the corresponding Killing vectors are finite in the flat-space limit $L\to \oo$. (For example, for a rotation of the sphere by an angle $\f$, a Killing vector with a finite flat-space limit is $\frac{1}{L} \pdr{}{\f}$.)

When $\beta$ is small, we can set $L=O(1/\beta)$ on the right-hand side and try to apply the thermal effective action (\ref{eq:usethermaleft}). We find that in order to have a good thermodynamic limit as $\beta\to 0$, the angular potentials $\vec\Omega$ must remain finite. Phrased in terms of the rotation angle $\vec \theta=\beta\vec\Omega$, we find that $\vec\theta$ must scale to zero as $\beta\to 0$. Provided this is the case, the $1/L$ expansion of the thermal effective action gives an expansion in small $\beta$ for the CFT partition function.

We can also understand condition $\vec \theta \to 0$ more explicitly from a direct computation using the thermal effective action. In a CFT, the thermal effective action is constrained by $d$-dimensional Weyl invariance. The most general coordinate- and Weyl-invariant action takes the form\footnote{For simplicity, here we assume that the theory is free of gravitational anomalies.}$^,$\footnote{Note that \cite{Benjamin:2023qsc} worked in conventions where $\tau$ has periodicity $1$, and $\beta$ is absorbed into the field $\f$. In this paper, we instead use conventions where $\tau$ has dimensionful periodicity $\beta$ (later we will also have other periodicities) so that explicit powers of $\beta$ appear in the action (\ref{eq:thermalaction}), as required by dimensional analysis. To convert from the conventions of  \cite{Benjamin:2023qsc} to the conventions in this work, one shifts the dilaton by $\f \to \f+\log\beta$.}
\be
\label{eq:thermalaction}
S_\textrm{th} &= \int \frac{d^{d-1} \vec x}{\beta^{d-1}} \sqrt{\hat g} \p{-f + c_1\beta^2  \hat R + c_2 \beta^2  F^2 + \dots} + S_\textrm{anom}.
\ee
Here $\hat g = e^{-2\f}g$, $\hat R$ is the Ricci scalar built from $\hat g$, $F^2$ is a Maxwell term, etc. The term $S_\textrm{anom}$ accounts for Weyl anomalies (which are not important for the present discussion).

On the geometry $S^{d-1}\x S^1_\b$, we can easily determine $g,A,\f$ and evaluate $S_\textrm{th}$ \cite{Benjamin:2023qsc}:
\be
\label{eq:evaluatedaction}
S_\textrm{th} &= \frac{\vol\, S^{d-1}}{\prod_{i=1}^n (1+\Omega_i^2)} \left[-f T^{d-1} + (d-2)\p{(d-1)c_1 + \p{2c_1+\frac 8 d c_2} \sum_{i=1}^n \Omega_i^2}T^{d-3} + \dots \right].
\ee
We see that terms of order $T^{d-1-k}=\b^{k-d+1}$ in the high-temperature expansion of $S_\textrm{th}$ are multiplied by a polynomial in the angular potentials $\Omega_i$ of degree $k$ (see e.g.\ examples in \cite{Bobev:2023ggk}). Consequently, $\theta_i\to 0$ as $\beta\to 0$ is necessary for the high-temperature expansion to be well-behaved.

To summarize, the thermal effective action can describe ``small" angles $\theta\sim \beta\Omega$, where the angular velocity remains finite in the thermodynamic limit. However, results from the thermal effective action like (\ref{eq:evaluatedaction}) break down outside this regime.  How can we access more general angles? 

\subsection{Spin-refined partition functions: warm-up in 2d CFT}
\label{sec:reviewmodular2d}

As a warm-up, in 2d CFT, we can compute partition functions at more general angles using modular invariance. Let us review how this works and derive some example results.
 For convenience, we write the partition function as:
\begin{equation}
Z(\tau,\bar\tau)=\mathrm{Tr} \left[e^{2\pi i\tau \left(L_0-\frac{c}{24}\right)-2\pi i\bar\tau \left(\bar L_0-\frac{c}{24}\right)} \right]\,,
\end{equation}
where $\tau = \frac{i\beta}{2\pi}+\frac{\theta}{2\pi}$ and $\bar\tau = \tau^*$.\footnote{Note that in this section, $\tau$ denotes the modular parameter of the torus, while in other sections $\tau$ denotes Euclidean time. We hope this will not cause confusion.}
The high temperature behavior of $Z(\tau,\bar\tau)$ at small angles can be obtained by performing the modular transformation $\tau\to -1/\tau$ (similarly for $\bar \tau$) and approximating by the contribution of the vacuum state. The result agrees with the thermal effective action:
\be
\label{eq:normalpartitionfunctionintwod}
\Tr\left[e^{-\beta(H-i\Omega J)}\right] &\sim e^{- S_\textrm{th}} = \exp\left[\frac{\vol\, S^1}{ (1+\Omega^2)} \frac f \beta\right] = \exp\left[\frac{4\pi^2}{\beta(1+\Omega^2)} \frac{c}{12}\right]\qquad (\textrm{CFT}_2),
\ee
where $f=\frac{2\pi c}{12}$. Here, we assume $c_L=c_R$ for simplicity. In this case, only the cosmological constant term appears in the thermal effective action in 2d.

Now let us instead assume that $\frac{\theta}{2\pi}$ is close to a nonzero rational angle $\frac{p}{q}$, so that $\tau, \bar\tau$ are very close to $\frac{p}{q}$. Following \cite{Benjamin:2019stq}, we can perform a different modular transformation to map $(\tau, \bar\tau)$ close to $\pm i \infty$ and approximate the partition function by the vacuum state in the new channel. For example, let us study the partition function with an insertion of $(-1)^J$ given in (\ref{eq:spinrefinedexample}). In this case, we have 
\begin{align}
\tau &= \frac12 + \frac{\beta\Omega}{2\pi} + \frac{i\beta}{2\pi} \nonumber \\ \bar\tau &= \frac12 + \frac{\beta\Omega}{2\pi} - \frac{i\beta}{2\pi}, \qquad \beta \ll 1,\ \ \Omega \sim O(1).
\end{align}
Modular invariance is the statement
\begin{equation}
Z(\gamma\circ\tau,\gamma\circ\bar\tau)=Z(\tau,\bar\tau)\,,\quad \gamma\in  \mathrm{PSL}(2,\mathbb{Z}).
\end{equation}
 An appropriate transformation in this case is
\be
\g &= \pm \begin{pmatrix}
-1 & 0 \\
2 & -1
\end{pmatrix}\in  \mathrm{PSL}(2,\mathbb{Z}),
\ee
which leads to
\be
\label{eq:resultfrommodularinvariance}
\mathrm{Tr}\left[e^{-\beta (H - i \Omega J)}(-1)^J\right] &=
\mathrm{Tr} \left[e^{2\pi i \tl \tau \left(L_0-\frac{c}{24}\right)-2\pi i \tl{\bar\tau} \left(\bar L_0-\frac{c}{24}\right)} \right]
\sim \exp\left[\frac 1 4 \frac{4\pi^2}{\b(1+\Omega^2)} \frac c {12}\right], \nn\\
\textrm{where}\quad \tl\tau &= -\frac 1 2 + \frac{\pi i}{2\beta(1-i\Omega)},\quad \tl{\bar \tau} = \tl \tau^*.
\ee
On the right-hand side, we approximated the trace by the contribution of the vacuum state in the $\beta\to 0$ limit.
We find that the partition function weighted by $(-1)^J$ grows exponentially in $1/\beta$, with an exponent that is $1/4$ of the un-weighted case (\ref{eq:normalpartitionfunctionintwod}).

For a general angle $\frac{\theta}{2\pi}$ close to $\frac{p}{q}$, we repeat the same logic above but with a more complicated modular transformation, namely 
\begin{equation}
\gamma = \pm \begin{pmatrix} -(p^{-1})_q & b \\ q & -p \end{pmatrix} \in \text{PSL}(2,\mathbb{Z}),
\label{eq:morecomplicatedmod}
\end{equation}
 where $(p^{-1})_q$ is the inverse of $p$ modulo $q$, and $b$ is chosen so the matrix has determinant $1$. We get
\begin{equation}
\Tr\left[e^{-\beta(H-i\Omega J)} e^{2\pi i \frac pq J}\right] \sim \exp\left[\frac{1}{q^2} \frac{4\pi^2}{\beta(1+\Omega^2)} \frac{c}{12}\right].
\label{eq:mostgeneral2dphase}
\end{equation}
In general, we find that the partition function of a 2d CFT weighted by $e^{2\pi i \frac pq J}$ grows exponentially in $1/\beta$, with an exponent that is $1/q^2$ of the un-weighted case (\ref{eq:normalpartitionfunctionintwod}).

Because modular invariance is not available in higher dimensions, it will be useful to rederive (\ref{eq:mostgeneral2dphase}) in a different way. We now describe two (related) approaches that can generalize to higher dimensions.

\subsection{Folding and unfolding}
\label{sec:foldunfold}

Thermal EFT naively breaks down in spin-refined partition functions like (\ref{eq:spinrefinedexample}) because the large spacetime symmetry $(-1)^J$ moves us outside the thermodynamic limit. One way to recover an EFT description is to perform a change of coordinates that makes $(-1)^J$ look more like a global symmetry. 

For example, consider a spin-refined partition function of a 2d QFT (not necessarily conformal) on $S^1_L \x S^1_\beta$,
\be
\Tr\left[e^{-\beta H}(-1)^J\right],
\ee
where $(-1)^J$ denotes a rotation of the spatial circle $S^1_L$ by $\pi$.
We can reinterpret {\it one\/} copy of the QFT on $S_L^1 \x S^1_\beta$ as {\it two\/} copies of the QFT on $(S_L^1/\Z_2) \x S^1_\beta$, with topological defects that glue the two copies to each other, see the middle of figure~\ref{fig:basicmanipulation}. In this picture, the operator $(-1)^J$ becomes a topological defect that simply permutes the two copies of the QFT as we move along the time direction. 
If we begin in one copy of the QFT and move by $\beta$ in Euclidean time, we pass once through the $(-1)^J$ defect and go to the other copy. Moving by $\beta$ again, we pass through the $(-1)^J$ defect again and end up in the first copy.
Thus, inserting $(-1)^J$ into the partition function creates a new effective thermal circle of length $2\beta$.

\newcommand\singleharrow{
    \draw[thick] (-0.1,0.1) -- (0,0) -- (-0.1,-0.1);
}
\newcommand\singlevarrow{
    \draw[thick] (-0.1,-0.1) -- (0,0) -- (0.1,-0.1);
}
\newcommand\harrow[1]{
  \pgfmathsetmacro{\endValue}{#1}
  \foreach \x in {1,...,\endValue} {
    \begin{scope}[shift={(0.1*\x,0)}]\singleharrow\end{scope}
  }
}
\newcommand\varrow[1]{
  \pgfmathsetmacro{\endValue}{#1}
  \foreach \x in {1,...,\endValue} {
    \begin{scope}[shift={(0,0.1*\x)}]\singlevarrow\end{scope}
  }
}
\newcommand\shift[2]{
  \begin{scope}[shift={#1}] #2 \end{scope}
}
\newcommand\mkgray[1]{
  \begin{scope}[draw=black!40!white] #1 \end{scope}
}
\begin{figure}[!ht]
\centering
\begin{tikzpicture}
    \draw[thick] (2,0) -- (0,0) -- (0,2) -- (2,2);
    \draw[thick,black!40!white] (2,0) -- (4,0) -- (4,2) -- (2,2);
    \shift{(0.95,0)}{\harrow{2}}
    \shift{(0.9,2)}{\harrow{3}}
    \shift{(0,1)}{\varrow{1}}
    \mkgray{
      \shift{(2.9,0)}{\harrow{3}}
      \shift{(2.95,2)}{\harrow{2}}
      \shift{(4,1)}{\varrow{1}}
    }
    \shift{(2,0.8)}{\mkgray{ \varrow{4} }}
    \mkgray{\draw[dashed,thick] (2,0) -- (2,2);}
    \draw[decorate,decoration={brace,amplitude=10pt,mirror}] (0,-0.2) -- (4,-0.2) 
      node[midway,below,yshift=-10pt] {$L$};
    \draw[decorate,decoration={brace,amplitude=10pt}] (-0.2,0) -- (-0.2,2)
      node [pos=0.5,left=10pt,black] {$\beta$}; 
\shift{(6.1,-0.1)}{
  \shift{(0.2,0.2)}{
  \mkgray{
      \draw[thick] (0,0) -- (2,0) -- (2,2) -- (0,2) -- cycle;
      \shift{(0,0.8)}{\varrow{4}}
      \shift{(2,1)}{\varrow{1}}
      \shift{(0.95,0)}{\harrow{3}}
      \shift{(0.95,2)}{\harrow{2}}
    }
  }
    \draw[thick] (0,0) -- (2,0) -- (2,2) -- (0,2) -- cycle;
    \shift{(0,1)}{\varrow{1}}
    \shift{(2,0.8)}{\varrow{4}}
    \shift{(0.95,0)}{\harrow{2}}
    \shift{(0.95,2)}{\harrow{3}}
}
\shift{(11.5,0)}{
  \shift{(0,1)}{
  \mkgray{
    \draw[thick] (0,0) -- (2,0) -- (2,2) -- (0,2) -- cycle;
    \shift{(0,0.8)}{\varrow{4}}
    \shift{(2,1)}{\varrow{1}}
    \shift{(0.95,0)}{\harrow{3}}
    \shift{(0.95,2)}{\harrow{2}}
  }
  }
  \shift{(0,-1)}{
    \draw[thick] (0,0) -- (2,0) -- (2,2) -- (0,2) -- cycle;
    \shift{(0,1)}{\varrow{1}}
    \shift{(2,0.8)}{\varrow{4}}
    \shift{(0.95,0)}{\harrow{2}}
    \shift{(0.95,2)}{\harrow{3}}
   }
   \draw[decorate,decoration={brace,amplitude=10pt}] (-0.2,-1) -- (-0.2,3)
      node [pos=0.5,left=10pt,black] {$2\beta$}; 
   \draw[decorate,decoration={brace,amplitude=10pt,mirror}] (0,-1.2) -- (2,-1.2) 
      node[midway,below,yshift=-10pt] {$L/2$};
}
\node at (5,1) {$\implies$};
\node at (9.5,1) {$\implies$};
\end{tikzpicture}
\caption{\label{fig:basicmanipulation}{\bf Left:}  The torus partition function with spatial cycle of length $L$, inverse temperature $\beta$, and an insertion of $(-1)^J$. The $(-1)^J$ insertion means we must glue the top and bottom of the figure with a half shift around the spatial circle. We split the figure into a left and right half using the trivial defect (vertical dashed line), and for convenience we color the right half grey. {\bf Middle:} Placing the black and grey rectangles on top of each other, we can interpret this same observable as the partition function of two copies of the QFT (black and grey) on an $(L/2)\x \beta$ rectangle, with boundary conditions inherited from the left figure. {\bf Right:} Finally, we can re-stack the two copies of the QFT, resulting in a single copy of the QFT with a new spatial circle of length $L/2$ and an effective thermal circle of length $2\beta$. Note that the effective thermal circle is nontrivially fibered over the new spatial circle. }
\end{figure}
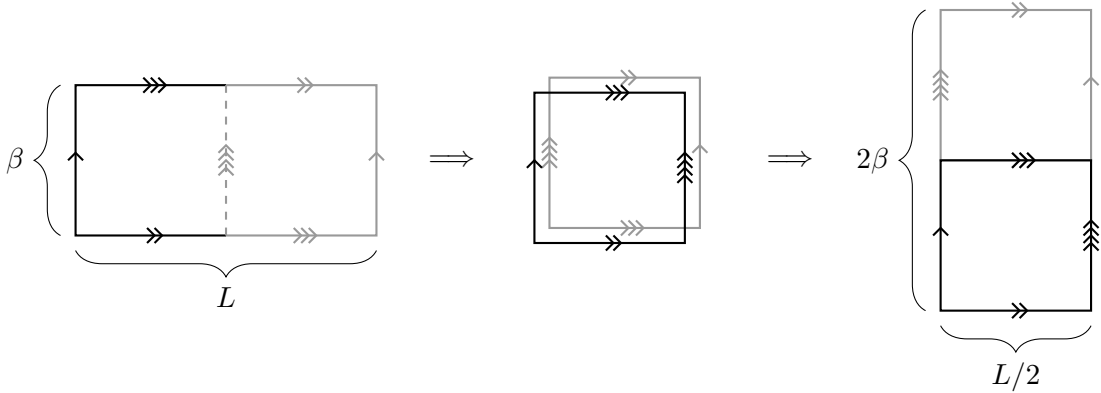

This reinterpretation of the path integral with a $(-1)^J$ insertion is illustrated in figure~\ref{fig:basicmanipulation}. One wrinkle (that is clear in the figure) is that the effective thermal $S^1_{2\beta}$ is nontrivially fibered over the spatial circle $S_L^{1}/\Z_2$: when we go once around the new spatial circle, the $S^1_{2\beta}$ shifts by $\beta$.

So far, we have considered a rotation angle of $\pi$. However, it is straightforward to study nearby rotation angles of the form $\theta=\pi + \beta\Omega$. On the left-hand side of figure~\ref{fig:basicmanipulation}, we simply insert an additional topological operator along the spatial cycle that implements the small rotation $e^{i\beta\Omega J}$. Following the manipulations in the figure, we end up with a product of two such operators on the new spatial cycle $S_L^{1}/\Z_2$, which together implement a rotation of $2\beta\Omega$.

The advantage of this rewriting of the path integral is that we can now smoothly take the thermodynamic limit $L\to \oo$ and use the thermal effective action. The effective inverse temperature is $2\beta$, the rotation angle is $2\beta\Omega$, and the effective spatial cycle is $S^1_L/\Z_2$. 

In fact, the above construction is straightforward to generalize to twists by any rational angle:
\begin{equation}
\label{eq:rationaltwist}
\mathrm{Tr}\left[e^{-\beta(H-i\Omega J)+2\pi i\frac{p}{q} J}\right].
\end{equation}
We interpret (\ref{eq:rationaltwist}) as the partition function of $q$ copies of the QFT on the space $S^1_L/\Z_q$, with appropriate topological defects that glue the copies together. The operator $e^{2\pi i \frac{p}{q} J}$ becomes a topological defect that permutes the copies of the QFT as we move around the Euclidean time circle. This creates an effective thermal circle $S^1_{q\beta}$, which is fibered over $S^1_L/\Z_q$. We can now apply thermal EFT on $S^1_L/\Z_q$.

\subsubsection{Example: 2d CFT}

As an example application, we can recover our previous answer for the spin-refined partition function of a 2d CFT\@. For a twist by $(-1)^J$, we find
\be
\mathrm{Tr}\left[e^{-\beta (H - i \Omega J)}(-1)^J\right] &\sim e^{-S_\textrm{th}[S^1/\Z_2 \x S^1_{2\beta}]} = \exp\left[\frac{2\pi c}{12} \frac{\vol (S^1/\Z_2)}{2\beta(1+\Omega^2)}\right] = \exp\left[\frac{1}{4} \frac{4\pi^2}{\beta(1+\Omega^2)} \frac{c}{12}\right].
\ee
In the action, we obtain one factor of $\frac 1 2$ from the smaller spatial cycle $S^1/\Z_2$, and another factor of $\frac 1 2$ from the larger thermal circle, resulting in an overalll factor of $\frac 1 4$ that agrees with the result from modular invariance (\ref{eq:resultfrommodularinvariance}).\footnote{The fact that the thermal circle is nontrivially fibered plays no role here because the thermal effective action is the integral of a local coordinate-invariant quantity that does not detect global features of the bundle. In a theory with a gravitational anomaly, the thermal effective action would contain an additional $1$-dimensional Chern-Simons for the Kaluza-Klein gauge field, which can detect the nontrivial topological structure of the thermal circle bundle, see section~\ref{sec:2dwithgravitationalanomaly}. The nontrivial topology also enters into nonperturbative corrections, see section~\ref{sec:nonpert}.} 

More generally, for a twist by $\frac{2\pi p}{q}$, the thermal effective action gives
\be
\label{eq:twodresultrationalangle}
\mathrm{Tr}\left[e^{-\beta (H - i \Omega J)}e^{ \frac{2\pi i p}{q} J}\right] &\sim e^{-S_\textrm{th}[S^1/\Z_q\x S^1_{q\beta}]} = \exp\left[\frac{2\pi c}{12}\frac{\vol (S^{1}/\Z_q)}{q\beta(1+\Omega^2)} \right] = \exp\left[\frac{1}{q^2} \frac{4\pi^2}{\beta(1+\Omega^2)}\frac{c}{12}\right].
\ee
We find that the effective free energy at high temperature for the spin-refined partition function (\ref{eq:rationaltwist}) is down by a factor of $q^2$, in agreement with (\ref{eq:mostgeneral2dphase}). 
 Note that the precise permutation of the copies of the CFT implemented by $e^{2\pi i \frac{p}{q} J}$ depends on $p$, but the length of the resulting thermal circle does not. Consequently the partition function is independent of $p$, up to nonperturbative corrections as $\beta\to 0$.\footnote{There is $p$-dependence if the theory is fermionic (see section~\ref{sec:fermions}) or has a gravitational anomaly (see section~\ref{sec:2dwithgravitationalanomaly}).}

\subsubsection{Higher dimensions}

The above construction works for $d>2$ as well, and on more general geometries. Consider a QFT$_d$ on any $(d{-}1)$-dimensional spatial manifold $\cM_L$ with a discrete isometry $R$ of finite order $R^q=1$. Again, we can reinterpret one copy of the QFT on $\cM_L \x S^1_\beta$ as $q$ copies of the QFT on $(\cM_L/\Z_q) \x S^1_\beta$, with topological defects that glue the copies to each other. In this picture, $R$ is represented as a topological defect that simply permutes the $q$ copies of the QFT as we move along the time direction, creating an effective inverse temperature $q\beta$.

\subsection{The EFT bundle}
\label{sec:EFTbundle}

Before exploring further consequences of this idea, it will be helpful to adopt a more abstract, geometrical perspective on this construction. Consider again a $d$-dimensional QFT with spatial manifold $\cM_L$. Given an isometry $U\in \textrm{Iso}(\cM_L)$, the partition function twisted by $U$\footnote{Here, we abuse notation and write $U$ for both the isometry and the operator implementing its action on the Hilbert space $\cH(\cM_L)$.}
\be
\label{eq:thetracehigherd}
\Tr_{\cH(\cM_L)}\left[e^{-\beta H} U\right].
\ee
is computed by the path integral of the CFT on the mapping torus
\be
M_{\beta,U} &\equiv (\cM_L \x \R)/\Z,
\ee
where $\Z=\<h\>$ is generated by
\be
h&:\cM_L \x \R \to \cM_L \x \R,\nn\\
h&:(\vec x,\tau)\ \ \ \ \mapsto (U \vec x, \tau+\beta),
\ee
where $\vec x$ is a coordinate on $\cM_L$.

Now let us specialize to $U=R$, where $R$ has order $q$. In this case, the $q$-th power of $h$ acts very simply: 
it leaves $\cM_L$ invariant, and shifts $\tau$ by $q\beta$:
\be
h^q:(\vec x,\tau)\mto(\vec x,\tau+q \beta).
\ee
Consequently, it is useful to decompose $\Z \cong q \Z \x \Z_q = \<h^q\> \x (\<h\>/\<h^q\>)$, and obtain the mapping torus $M_{\beta,R}$ via two successive quotients. We first quotient by $q\Z \cong \<h^q\>$ (which turns $\R$ into $S^1_{q \beta}$), and then quotient by $\Z_q=\<h\>/\<h^q\>$:
\be
\label{eq:mrasaquotient}
M_{\beta,R} &= ((\cM_L\x\R)/q \Z)/\Z_q = (\cM_L \x S^1_{q\b})/\Z_q.
\ee

The quotient $(\cM_L \x S^1_{q\b})/\Z_q$ on the right-hand side of (\ref{eq:mrasaquotient}) can be viewed as a bundle in two different ways. Firstly, it is a $\cM_L$-bundle over $S^1_{q\beta}/\Z_q \cong S^1_\beta$. This is the usual point of view of the trace as a spatial manifold evolving over Euclidean time $\beta$. However, we can alternatively view $(\cM_L \x S^1_{q\b})/\Z_q$ as an $S^1_{q\beta}$ bundle over $\cM_L/\Z_q$. We call this latter description the ``EFT bundle." In section (\ref{sec:foldunfold}), the EFT bundle was a nontrivial $S^1_{q\beta}$ bundle over $S^1_L/\Z_q$. As we saw, the virtue of the EFT bundle is that the thermodynamic limit $L\to \oo$ is straightforward: we can dimensionally reduce along the effective thermal circle $S^1_{q\beta}$ without leaving the thermodynamic limit. The theory is then described by thermal EFT with effective inverse temperature $q\beta$ and spatial cycle $\cM_L/\Z_q$.

Suppose for the moment that the action of $R$ on $\cM_L$ is free, so that $\cM_L/\Z_q$ is smooth. (This is the case, for example, for a rational rotation of the spatial circle in 2d.) For any term in the thermal effective action that is the integral of a local density, the effect of the quotient by $\Z_q$ is simply to multiply its contribution by $1/q$. Thus, we conclude
\be
\label{eq:onlyrequires}
-\log \Tr\left[e^{-\beta H} R\right] &\sim -\frac 1 q \log \Tr\left[e^{- q\beta H}\right] + \textrm{topological} \qquad (\textrm{if the $R$ action is free}).
\ee
Here, ``$\sim$" denotes agreement to all perturbative orders in the $1/L$ expansion. The term ``topological" indicates potential contributions from a finite number of terms capable of detecting the topology of the EFT bundle, which cannot be written as the integral of a local gauge/coordinate-invariant density.  We discuss such terms in section~\ref{sec:2dwithgravitationalanomaly}.

Let us pause to note that the result (\ref{eq:onlyrequires}) really only requires that the theory be gapped at inverse temperature $q\beta$ (not necessarily at inverse temperature $\beta$), since we only use locality of the thermal effective action on the right-hand side.

\subsubsection{Adding ``small" isometries}

Just as before, we can also consider inserting into the trace an additional ``small" isometry $U=e^{i\beta(\alpha Q_\xi)}$, where $\xi$ is a Killing vector on $\cM_L$, $Q_\xi$ is its corresponding charge, and $\alpha$ is the corresponding thermodynamic potential. We will be mainly interested in the case where $U$ commutes with the discrete isometry $R$, so we assume this henceforth. The insertion of $U$ can be thought of as a topological defect that wraps $\cM_L$. Consequently, the defect wraps $q$ times around the base of the EFT bundle $\cM_L/\Z_q$, resulting in an effective rotation $U^q$. We conclude that
\be
-\log \Tr\left[g R\right] &\sim -\frac 1 q \log \Tr\left[g^q\right] + \textrm{topological} \qquad (\textrm{if the $R$ action is free}),
\label{eq:generalresultwhenrisfree}
\ee
where $g=e^{-\beta H} U$.

In fact, this argument applies to any global symmetry element $V$ as well, so (\ref{eq:generalresultwhenrisfree}) holds when $g$ is multiplied by a global symmetry group element: $g=e^{-\beta H} U V$. We can think of $V$ as implementing a nontrivial flat connection for a background gauge field coupled to the global symmetry. In this case, the ``topological" terms in (\ref{eq:generalresultwhenrisfree}) could include contributions from nontrivial topology of this connection.

We can also understand the insertion of ``small" isometries geometrically. Again, the idea is to view the mapping torus $M_{\beta,U R}$ as the result of two successive quotients 
\be
\label{eq:mappingtoruswithsmallangles}
M_{\beta,U R} &= ((\cM_L\x \R)/\<h^q\>)/\Z_q = M_{q\beta,U^q}/\Z_q,\nn\\
\textrm{where } h &: (\vec x,\tau) \mto (U R\, \vec x,\tau+\beta),
\ee
where $\Z_q=\<h\>/\<h^q\>$, and we have used $(U R)^q = U^q$. On the right-hand side, we have the mapping torus $M_{q\beta,U^q}$ which is described by the thermal effective action at inverse temperature $q\beta$, with small isometries $U^q$ turned on. The effect of the $\Z_q$ quotient is to multiply the contribution of any integral of a local density by $1/q$. This again leads to (\ref{eq:generalresultwhenrisfree}).\footnote{When $U$ and $R$ don't commute, the same logic works but we have $M_{q\beta,(UR)^q}/\Z_q$ on the right-hand side of (\ref{eq:mappingtoruswithsmallangles}). We can still use thermal EFT, since $(UR)^q$ is $O(\beta)$ close to the identity.}

The work \cite{ArabiArdehali:2021nsx} uses similar ideas to characterize superconformal indices of 4d CFTs near roots of unity. Our novel contribution is to apply these ideas in not-necessarily-supersymmetric, not-necessarily-conformal theories, on general spatial geometries, and also to describe the effects of Kaluza-Klein vortices (see below), which do not appear in superconformal indices.

\subsubsection{Non-free actions and Kaluza-Klein vortex defects}
\label{sec:kkv}

What happens if the action of $R$ is not free? For example, in a 3d QFT on $S^2 \x S^1_\beta$, the action of $(-1)^J$ (where $J$ is the Cartan generator of the rotation group) has fixed points at the north and south poles of $S^2$. In this case, the EFT bundle degenerates at the fixed loci of nontrivial elements of $\Z_q$, namely $R,\dots,R^{q-1}\in \Z_q$. After dimensional reduction, these degeneration loci becomes defects $\mathfrak{D}_i$ (with $i$ labelling the set of defects) in the $d-1$ dimensional thermal effective theory. We call them ``Kaluza-Klein vortex defects" because the KK gauge field $A$ has nontrivial holonomy around them, as we explain in section~\ref{sec:eftgauge}.

Each defect $\mathfrak{D}_i$ contributes to the partition function a coordinate-invariant effective action $S_{\mathfrak{D}_i}$ of the background fields $g, A, \f$ {\it in the infinitesimal neighborhood of $\mathfrak{D}_i$}. We then have the more general result
\be
\label{eq:moregeneralresult}
-\log \Tr\left[g R\right] &\sim -\frac 1 q \log \Tr\left[g^q\right] + \textrm{topological} + \sum_{\mathfrak{D}_i} S_{\mathfrak{D}_i},
\ee
We conjecture that for generic interacting QFTs, the KK vortex defects will be gapped. (In fact, in this work, we will study several examples of free theories where the appropriate defects are still gapped.) In this case, each $S_{\mathfrak{D}_i}$ will be a local functional of $g,A,\f$.

In CFTs, the defect actions $S_{\mathfrak{D}_i}$ are additionally constrained by Weyl-invariance, just like the bulk terms in the thermal effective action.
We will determine the explicit form of $S_{\mathfrak{D}}:=\sum_{\mathfrak{D}_i} S_{\mathfrak{D}_i}$ in CFTs later in section~\ref{sec:defects}. For now, we simply note that the leading term in the derivative expansion of $S_{\mathfrak{D}_i}$ in a CFT is a cosmological constant localized on $\mathfrak{D}_i$:\footnote{$S_{\mathfrak{D}_i}$ itself can also have topological terms; if the topological term has no derivatives (i.e. the Wilson line of the KK photon), it will contribute at the same order in $\beta$ as the defect cosmological constant.}
\be
S_{\mathfrak{D}_i} &= a_{\mathfrak{D}_i}\int_{\mathfrak{D}_i} \frac{d^{n_i}y}{(q\beta)^{n_i}} \sqrt{\hat g|_{\mathfrak{D}_i}} + \textrm{higher derivatives}.
\ee
Here, we assume that $\mathfrak{D}_i$ is $n_i$-dimensional, $y$ are coordinates on the defect, and $\hat g|_{\mathfrak{D}_i}$ denotes the pullback of $\hat g=e^{-2\f} g$ to $\mathfrak{D}_i$. This term behaves like $\beta^{-n_i}$ as $\beta\to 0$. In the case $n_i=0$, i.e.\ when $\mathfrak{D}_i$ is point-like (for example the north/south poles of $S^2$), the ``cosmological constant" becomes simply a constant.

\subsubsection{Example: CFT in general $d$}

As an example application, consider a $d$-dimensional CFT on $S^{d-1} \x S^1_\beta$.
Although our discussion so far has been somewhat abstract, and we have  used only basic geometry and principles of EFT, our conclusion (\ref{eq:moregeneralresult}) makes powerful predictions about CFT spectra. For example, to leading order as $\beta\to 0$, the defect term $S_\mathfrak{D}$ does not contribute, so very generally we obtain a higher dimensional generalization of (\ref{eq:twodresultrationalangle}),
\be
\label{eq:resultforleadingterm}
\log \Tr[e^{-\beta(H-i\vec \Omega\.\vec J)} R] &\sim \frac 1 {q^d} \frac{\vol S^{d-1}}{\prod_{i=1}^n(1+\Omega_i^2)} \frac{f}{\beta^{d-1}}  + \dots,
\ee
valid for any element $R$ of the Cartan subgroup of $\SO(d)$ with order $q$. For example, the relative density of even- and odd-spin operators (with respect to any Cartan generator) grows exponentially at a rate precisely $1/2^d$ times the rate for the un-weighted density of states.

 Unlike in $d=2$, the thermal effective action in $d>2$ can have more than just a cosmological constant term. Consequently, the ``$\ldots$" in (\ref{eq:resultforleadingterm}) includes higher-derivative corrections (in addition to possible vortex defect contributions). However, these higher-derivative corrections can be predicted in the same way: they differ from the un-spin-refined case by replacing $\beta\to q\beta$ and multiplying by $1/q$ to account for the smaller spatial manifold.

The results (\ref{eq:resultforleadingterm}) and (\ref{eq:moregeneralresult}) display an important difference between partition functions weighted by spacetime symmetries and partition functions weighted by global symmetries. If we replace $R$ with a global symmetry element, this corresponds to turning on new background gauge fields in the thermal effective action, whose contributions are captured by Wilson coefficients that are not active when the global symmetry generators are turned off. For example, the density of states weighted by a global symmetry generator $U$ is controlled by a $U$-dependent free energy density $f_U$ with no (obvious) relation to $f$ when $U\neq 1$ (see e.g.\ \cite{Harlow:2021trr,Kang:2022orq}). By contrast, the density of states weighted by different discrete {\it spacetime\/} symmetries are all controlled by the same $f$ (and the same higher Wilson coefficients like $c_1,c_2,\dots$), in a predictable way.

Finally, let us describe the possible discrete rotations $R$ for which (\ref{eq:resultforleadingterm}) applies. Let us write $R=e^{i \vec\theta\. \vec J}$. In order for $R$ to have finite order $q$, we must have $\vec\theta=2\pi(\frac{p_1}{q_1},\dots,\frac{p_n}{q_n})$, where the $p_i/q_i$ are rational numbers (which we assume are in reduced form, so that $p_i$ and $q_i$ are relatively prime). The order of $R$ is $q=\textrm{lcm}(q_1,\dots,q_n)$.

When $d$ is even, the action of $\Z_q$ is free if all $q_i=q$. In this case, the quotient $S^{d-1}/\Z_q$ is a lens space $L(q;p_1,\dots,p_n)$, and there are no vortex defects $\mathfrak{D}$. If instead there exists at least one $q_i \neq q$, then the group element $R^{q_i}$ will have a fixed locus $S^{2k-1}$, where $k$ is the number of $q_j$'s such that $q_j | q_i$, and there will be a corresponding defect $\mathfrak{D}$ at this location (or rather its image after quotienting by $\Z_q$). Note that it is possible for fixed loci to intersect, creating higher codimension defects. For example, if $\vec \theta=2\pi(1,\frac{1}{2},\frac{1}{3})$, the element $R^2$ has a fixed $S^3$, the element $R^3$ has its own fixed $S^3$, and the two $S^3$'s intersect along an $S^1$. Quotienting by $\Z_6$, we obtain a defect localized on $S^3/\Z_3\subset S^5/\Z_6$, a defect localized on $S^3/\Z_2\subset S^5/\Z_6$, and they intersect along an $S^1\in S^5/\Z_6$. In this case, the thermal effective action will include terms localized on the defects and their intersection. When $d$ is odd, any element of $\SO(d)$ necessarily has a nontrivial fixed locus, since there is a direction left invariant by the Cartan generators.

If the theory has a reflection symmetry, then we can more generally consider $R\in O(d)$. The above arguments continue to hold, essentially unmodified. When $R$ includes a reflection, the base of the EFT bundle $S^{d-1}/\Z_q$ can be non-orientable. For example, if we take $R$ to be the parity operator $R:\vec n \to -\vec n$, then $S^{d-1}/\Z_2=\mathbb{RP}^{d-1}$, which is non-orientable in odd $d$. Note that the parity operator acts freely, so in this case we can apply (\ref{eq:generalresultwhenrisfree}).

\section{Kaluza-Klein vortex defects}
\label{sec:defects}

In this section, we explore the form of the defect action $S_\mathfrak{D}$ that contributes whenever the group generated by the discrete rotation $R$ does not act freely. For simplicity, we will restrict our attention to CFT's in $d$-dimensions on a spatial sphere $S^{d-1}$.

\subsection{Background fields and EFT gauge}
\label{sec:eftgauge}

As before, we wish to compute the partition function of a CFT on the geometry $M_{q\beta,U^q}/\Z_q$, where the mapping torus in the numerator is $M_{q\beta,U^q}=(S^{d-1} \x \R)/\<h^q\>$, the group in the denominator is $\Z_q=\<h\>/\<h^q\>$, and the action of $h$ is given by
\be
h:(\vec n,\tau) &\mto (UR \vec n,\tau+\beta) = (e^{i(\frac{2\pi p_a}{q_a} + \beta \Omega_a)J^a} \vec n,\tau+\beta).
\ee

First, let us be more precise about the form of the background fields in this geometry. Following \cite{Benjamin:2023qsc}, we use radius-angle coordinates on the sphere $S^{d-1}$. These are given by a pair of radius and angle $\{r_a,\theta_a\}$ for each orthogonal 2-plane $(a=1,\dots,n=\lfloor \tfrac d 2 \rfloor)$. If $d$ is odd, we have an additional radial coordinate $r_{n+1}$. Together, the radii satisfy the constraint $\sum_{a=1}^{n+\e} r_a^2 = 1$, where $\e=0$ in even $d$ and $\e=1$ in odd $d$.

To write the metric on $M_{q\beta,U^q}$ in Kaluza-Klein form, we switch to co-rotating coordinates
\be
\vf_a &\equiv \th_a - \Omega_a \tau,
\label{eq:phidef}
\ee
where $\tau$ is the coordinate on $\R$.
In co-rotating coordinates, the action of $h$ simplifies to
\be
\label{eq:actionofh}
h:(r_a,\vf_a,\tau)\mto (r_a,\vf_a+\tfrac{2\pi p_a}{q_a},\tau+\beta).
\ee
In particular $h^q$, becomes simply a shift $h^q:\tau \mto \tau+q \beta$. Thus, quotienting by $\<h^q\>$ to obtain $M_{q\beta,U^q}$ makes $\tau$ periodic with period $q\beta$.

The metric of $M_{q\b,U^q}$ in co-rotating coordinates takes the Kaluza-Klein form
\be
\label{eq:metriccorotating}
ds^2 &= g + e^{2\f}(d\tau + A)^2,
\ee
where the fields $g,A,\f$ are given by \cite{Benjamin:2023qsc}
\be
\label{eq:thefields}
e^{2\f} &= 1+\sum_{a=1}^n r_a^2 \Omega_a^2, \\
\label{eq:theafield}
A &= \sum_{a=1}^n \frac{r_a^2 \Omega_a}{1+\sum_b r_b^2 \Omega_b^2} d\vf_a, \\
\label{eq:thelastfield}
g &= \sum_{a=1}^{n+\e} dr_a^2 + \sum_{a,b=1}^n \p{r_a r_b \de_{ab} - \frac{r_a^2 r_b^2 \Omega_a \Omega_b}{1+\sum_{c=1}^n r_c^2 \Omega_c^2}} d\vf_a d\vf_b.
\ee

The metric of the EFT bundle $M_{q\b,U^q}/\Z_q$ is locally the same as (\ref{eq:metriccorotating}). Consequently, we can choose a local trivialization of the EFT bundle such that the fields $g,A,\f$ are identical to (\ref{eq:thefields}) in each patch. However, such a local trivialization will have nontrivial transition functions between patches that contribute to holonomies of the Kaluza-Klein connection along various cycles (including around the defect locus).

If we like, we can perform a gauge transformation that makes the transition functions trivial, at the cost of introducing new contributions to $A$. We refer to such a gauge as ``EFT gauge" because it will be convenient for discussing the EFT limit of the CFT on this geometry. In EFT gauge, the curvature $F=dA$ has $\de$-function type singularities at the fixed-loci of $R$ whose coefficients reflect the topology of the EFT bundle.

\subsubsection{Example: 2d CFT}
\label{sec:exampletwodcft}

\begin{figure}[!ht]
\centering
\begin{tikzpicture}
    \draw[thick] (-3.5,0) -- (4,0);
    \draw[line width=2pt, red] (-0.95,0) -- (1.95,0);
    \node [red] at (-0.92,0) {(};
    \node [red] at (1.92,0) {)};
    \draw[line width=2pt, blue ] (-1.3,0) -- (-0.95,0);
    \draw[line width=2pt, blue!35!red ] (-0.95,0) -- (-0.55,0);
    \node [blue] at (-1.27,0) {(};
    \node [blue] at (-0.58,0) {)};
    \node [red,above] at (0.5,0.2) {$I_1$};
    \node [blue,above] at (-0.9,0.2) {$I_2$};
    \node [] at (-1.05,-0.45) {$-\e\ \,0\ \,\e$};
    \node [] at (1.88,-0.45) {$\tfrac{2\pi}{q}$};
\shift{(0,-2)}{
    \draw[line width=2pt, red] (-0.95,0) -- (1.95,0);
    \node [red] at (-0.92,0) {(};
    \node [red] at (1.92,0) {)};
    \draw[line width=2pt, blue!35!red ] (-0.95,0) -- (-0.55,0);
    \draw[line width=2pt, blue!35!red ] (1.58,0) -- (1.95,0);
    \node [blue] at (-0.58,0) {)};
    \node [blue] at (1.58,0) {(};
    \node [red,above] at (0.5,0.2) {$U_1$};
    \node [blue,below] at (0.5,-0.4) {$U_2$};
    \draw [dashed,thick,blue, ->] (0.8,-0.65) to[out=0,in=-105] (1.72,-0.3);
    \draw [dashed,thick,blue, ->] (0.2,-0.65) to[out=180,in=-95] (-0.75,-0.3);
}
\node [] at (-4,0) {$S^1$};
\node [] at (-4,-2) {$S^1/\Z_q$};
\node [] at (-5,-0.05) {$\phantom{[}\vf\phantom{]}\in$};
\node [] at (-5,-2) {$[\vf]\in$};
\draw [line width=0.6pt,->] (-4,-0.35) -- (-4,-1.6);
\draw [line width=0.6pt,->] (-5.2,-0.35) -- (-5.2,-1.6);
\draw [line width=0.6pt] (-5.28,-0.35) -- (-5.12,-0.35);
\draw [line width=0.6pt,->] (0.5,-0.35) -- (0.5,-1.1);
\end{tikzpicture}
\caption{\label{fig:intervals}The open intervals $I_1$ and $I_2$ are subsets of $S^1$. Their images under the quotient map $S^1 \to S^1/\Z_q$ are $U_1$ and $U_2$, respectively, which together cover $S^1/\Z_q$. We can choose a gauge where the KK fields $g,A,\f$ are given by (\ref{eq:twodfields}) in each of $U_1$ and $U_2$. However, in this gauge, there will be a nontrivial transition function between $U_1$ and $U_2$.}
\end{figure}
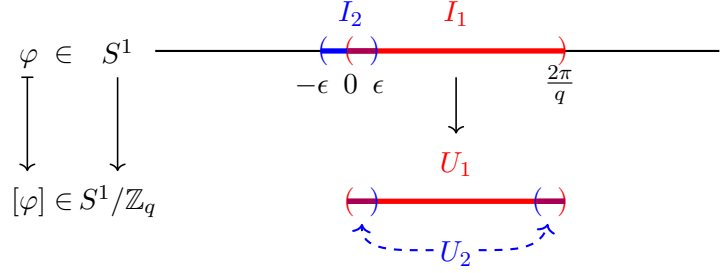

Let us illustrate these ideas with an example. Consider a $2d$ CFT, where the action of $h$ is given by $h:(\vf,\tau)\mto (\vf+\tfrac{2\pi p}{q},\tau+\b)$. The metric on $M_{q\b,U^q}$ is
\be
ds^2 &= d\tau^2 + d\th^2 = d\tau^2 + (d\vf+\Omega d\tau)^2 \nn\\
&=  \underbrace{(1-\tfrac{\Omega^2}{1+\Omega^2})d\vf^2}_{g} + \underbrace{(1+\Omega^2)}_{e^{2\f}} (d\tau + \underbrace{\tfrac{\Omega}{1+\Omega^2} d\vf}_A)^2.
\label{eq:twodfields}
\ee
To choose a local trivialization of the EFT bundle, we first specify two intervals in the $\vf$ coordinate:
\be
I_1=\{\vf : 0<\vf<\tfrac{2\pi}{q}\},\qquad I_2 = \{\vf:-\e<\vf<\e\},
\ee
with $0<\e<\frac{\pi}{q}$.
We denote their images in $S^1/\Z_q$ by $U_1$ and $U_2$, respectively. Together $U_1$ and $U_2$ cover the quotient space $S^1/\Z_q$, see figure~\ref{fig:intervals}.

The bundle projection $\pi:M_{q\beta,U^q}/\<h\> \to S^{1}/\Z_q$ acts by $\pi: [(\vf,\tau)] \mto [\vf]$, where $[(\vf,\tau)]$ denotes an equivalence class modulo the action of $h$, and $[\vf]$ denotes an equivalence class modulo $\frac{2\pi p}{q}$.
Over each open set $U_1,U_2$, we must define trivialization maps
\be
\label{eq:thetrivialization}
\f_{U_i} : \pi^{-1}(U_i) &\ \to\ U_i \x S^1_{q\beta}.
\ee
We choose them as follows. Given $p\in \pi^{-1}(U_i)$, thought of as an equivalence class modulo $\<h\>$, let $(\vf,\tau)$ be a representative of the equivalence class such that $\vf$ is contained in $I_i$. Then we define
\be
\f_{U_i}(p) &= ([\vf],\tau).
\ee
Note that $\tau$ is well-defined modulo $q\beta$ because the only elements of $\<h\>$ that map the $I_i$ to themselves are powers of $h^q$. With this local trivialization, the fields $g,A,\f$ are given by (\ref{eq:twodfields}) in each patch. In particular, we have $A=\tfrac{\Omega}{1+\Omega^2} d\vf$ in both patches.

However, the data of the Kaluza-Klein connection includes both the value of $A$ in each patch, as well as the transition functions between patches. We must also determine these transition functions.

There are two overlap regions to consider. The first is $(0,\e)$. In this region, the transition function is trivial. The second overlap region is the image of $(-\e,0)\subset I_2$ in $S^1/\Z_q$, which coincides with the image of $(\tfrac{2\pi}{q}-\e,\tfrac{2\pi}{q})\subset I_1$ in $S^1/\Z_q$. Note that $(\vf,\tau)$ for $\vf\in (-\e,0)$ is equivalent modulo $\<h\>$ to $(\vf+\frac{2\pi}{q},\tau+(p^{-1})_q \b)$, where $\vf+\frac{2\pi}{q}\in (\tfrac{2\pi}{q}-\e,\tfrac{2\pi}{q})$. Here, $(p^{-1})_q$ denotes the inverse of $p$ mod $q$, i.e.\ it satisfies $p (p^{-1})_q = qn + 1$ for some integer $n$. Thus, the transition function in this second overlap region is
\be
\f_{U_2} \circ \f_{U_1}^{-1} : ([\vf],\tau) &\mto ([\vf],\tau - (p^{-1})_q \beta).
\ee
The holonomy of the connection gets a nontrivial contribution from the transition functions:\footnote{The parallel transport equation is $d\tau+A= 0$, so the holonomy of $\tau$ is computed by $-\oint A$.} 
\be
\label{eq:theholonomy}
-\oint A &= -\frac {2\pi} q \frac{\Omega}{1+\Omega^2} - (p^{-1})_q \beta.
\ee
(The holonomy is an example of the ``topological" terms discussed in section~\ref{sec:2dwithgravitationalanomaly} which can contribute in the thermal effective action, but are not the integral of a local gauge/coordinate invariant density.)

To go to EFT gauge, we perform a gauge transformation (i.e.\ a $\vf$-dependent redefinition of $\tau$) that trivializes the transition functions. One possible choice is 
\be
\tau' &= \tau - (p^{-1})_q\, \beta \left\lfloor \tfrac{ \vf}{2\pi/q}\right\rfloor.
\ee
Note that the function $\left\lfloor \tfrac{ \vf}{2\pi/q}\right\rfloor$ is multi-valued on the entire circle $S^1$, but there is no problem defining it inside the intervals $I_1,I_2$ where we perform the gauge transformation.
In terms of $\tau'$, the transition functions are now trivial in both overlap regions. (A quick way to see why is to note that $\tau'$ is invariant under the $h$-action (\ref{eq:actionofh}).) The gauge field becomes
\be
\label{eq:stackedgaugefield}
A' &= A + d\tau - d \tau' = 
\frac{\Omega}{1+\Omega^2}d\vf + (p^{-1})_q \beta\, \de(\vf) d\vf
 \qquad (\textrm{EFT gauge}).
\ee
The holonomy $-\oint A'$ is still given by (\ref{eq:theholonomy}), but that is now manifest in the local expression for the gauge field (\ref{eq:stackedgaugefield}).

\subsubsection{Example: 3d CFT}

Now consider the same setup in a 3d CFT\@. The metric on $S^2$ is
\be
ds^2_{S^2} &= dr_1^2 + dr_2^2 + r_1^2 \,d\vf^2 \qquad (r_1^2 + r_2^2 = 1).
\ee
Essentially all of the above discussion goes through un-modified, with the radii $r_1,r_2$ coming along for the ride. We can again go to EFT gauge, and the gauge field (\ref{eq:stackedgaugefield}) now gets interpreted as a gauge field on $S^2$. This time, $A$ has $\de$-function-localized curvature at the north and south poles:
\be
dA &= \mp (p^{-1})_q \beta \de(\vec n,\vec n_\pm) d^2 \vec n + \textrm{nonsingular},
\ee
where $\de(\vec n,\vec n')$ represents a $\de$-function on $S^2$, and $\vec n_\pm$ are the north/south poles.

\subsubsection{EFT gauge in general}\label{sec:EFTgaugeInGeneral}

More generally, we go to EFT gauge as follows. First choose a fundamental domain $F$ for the quotient map $S^{d-1} \to S^{d-1}/\Z_q$. Define an integer valued function $k(\vec n)$ by 
\be
k(\vec n) &= 0 \textrm{ if }\vec n \in F,\nn\\
k(R\vec n) &= 1+k(\vec n).
\ee
In words, $k(\vec n)$ counts the power of $R$ needed to move from somewhere in $F$ to $\vec n$. Again, $k(\vec n)$ is multi-valued if we try to define it on the entire sphere, but we only need to define it inside a collection of open sets that cover $S^{d-1}/\Z_q$. For example, in the 2d case considered above, we had $k(\vf)=(p^{-1})_q \lfloor \tfrac{\vf}{2\pi/q}\rfloor$.

Finally, we define
\be
\tau' \equiv \tau - \beta k(\vec n(r_a,\vf_a)).
\ee
In the coordinates $(r_a,\vf_a,\tau')$, $h$ acts simply by shifting angles $\vf_a$:
\be
h:(r_a,\vf_a,\tau') &\mto (r_a,\vf_a+\tfrac{2\pi p_a}{q_a},\tau').
\ee
Consequently, a local trivialization of the EFT bundle defined using the $\tau'$ coordinate has trivial transition functions. The gauge field is given by
\be
A' &= A + d\tau - d\tau' = \sum_{a=1}^n \frac{r_a^2 \Omega_a}{1+\sum_b r_b^2 \Omega_b^2} d\vf_a + \beta d k(\vec n(r_a,\vf_a)).
\ee
The curvature $dA'$ has $\de$-function contributions $\beta d^2 k(\vec n)$ at the fixed loci of powers of $R$.

\subsection{Effective action}

Following the logic of the thermal effective action, let us now equip the EFT bundle with a more general metric $G$ and try to write down a local action of $G$. We will demand that $G$ satisfy the following conditions:
\begin{itemize}
\item It possesses a circle isometry, so that it can be written in Kaluza-Klein form (\ref{eq:metriccorotating}).
\item In EFT gauge, the curvature $dA$ is a sum of $\de$-function singularities of the form $\beta d^2 k(\vec n)$, plus something smooth on $S^{d-1}/\Z_q$. (This ensures that the Kaluza-Klein bundle has the same topology as $M_{q\beta,V^q}/\Z_q$.) \item $g$ and $e^{2\f}$ should be smooth on $S^{d-1}/\Z_q$.
\end{itemize}
Here, a field is ``smooth on $S^{d-1}/\Z_q$" if it lifts to a smooth $\Z_q$-invariant field on $S^{d-1}$.

In the limit $\beta\to 0$, we can separate each of the background fields into a long-wavelength part, with wavelengths much longer than $\beta$, and a short-wavelength part, with wavelengths comparable to (or smaller than) $\beta$. The long-wavelength parts become background fields for the thermal EFT\@. Meanwhile, the short-wavelength parts become operator insertions in that EFT. 

In our case, the $\de$-function curvature singularities $dA\sim \beta d^2 k(\vec n)$ are short-wavelength. They determine the insertion of an operator in the thermal EFT, which is described by the defect action $S_\mathfrak{D}$. This action is a functional of the long-wavelength parts of $g,A,\f$. As mentioned in section~\ref{sec:kkv}, we will assume that the defect is gapped, so that the action functional is local and can be organized in a derivative expansion. To construct it, we should compute curvatures and other invariants of $g,A,\f$, and {\it throw away $\de$-function singularities}. Since $g$ and $\f$ are smooth, this effectively amounts to the replacement
\be
dA &\to d\tl A \equiv dA - \beta d^2 k(\vec n).
\ee
Henceforth, we leave this replacement implicit. In other words, when we write $dA$ in the defect action, we mean its long-wavelength part $d\tl A$, with $\de$-functions thrown away.

The defects live at singularities in the quotient space $S^{d-1}/\Z_q$. How should we write an action for long wavelength fields near these singularities?
Recall that the long-wavelength parts of $g,A,\f$ lift to smooth $\Z_q$-invariant fields on $S^{d-1}$. We will write $S_\mathfrak{D}$ as a functional of these $\Z_q$-invariant lifts, integrated over the preimage of the defect locus modulo $\Z_q$, which we denote by $\tl{\mathfrak{D}}$. We also conventionally divide by $q$, which ensures that Wilson coefficients of defects living at singularities with the same local structure (but possibly different global structure) are the same.

The action should be invariant under gauge/coordinate transformations that preserve the defect locus.  For now, we ignore the possibility of nontrivial Weyl anomalies on the defect $\mathfrak{D}$, and we impose that $S_\mathfrak{D}$ be Weyl-invariant as well. Consequently, it will be a functional of $A$ and the Weyl-invariant combination $\hat g = e^{-2\f} g$.

Consider an $n$-dimensional defect $\mathfrak{D}$ whose preimage $\tl{\mathfrak{D}}$ is the fixed locus of an element $R^l\in \<R\>$ with order $m$. Given a point $p$ on $\tl{\mathfrak{D}}$, we can choose a vielbein $\hat e^a_i$ at $p$ satisfying $\de_{ab}\hat e^a_i \hat e^b_j=\hat g_{ij}$, where $a,b$ are indices for the local rotation group $\SO(d-1)$. The group $\<R^l\>\cong \Z_m$ acts as a subgroup of the local rotation group $\SO(d-1)$, so the $\hat e_i^a$ can be classified into representations of this $\Z_m$. Singlets under $\Z_m$ represent directions parallel to the defect. They are acted upon by an $\SO(n)\subset \SO(d-1)$ that commutes with $\Z_m$. Hence, altogether the $\hat e_i^a$ can be classified into representations of $\Z_m \x \SO(n)$.

To build the defect action, we enumerate curvature tensors built from $\hat g$ and $A$, in a derivative expansion, and contract them with $\hat e_i^a$ to build $\Z_m \x \SO(n)$ invariants $I_i$ with $d_i$ derivatives. The defect action is then
\be
S_\mathfrak{D} &= \frac{1}{q}\int_{\tl{\mathfrak D}} d^n y \sqrt{\hat g|_{\tl{\mathfrak{D}}}} \p{\sum_i a_i (q\beta)^{d_i-n} I_i},
\ee
where $\hat g|_{\tl{\mathfrak{D}}}$ denotes the pullback of $\hat g$ to $\tl{\mathfrak{D}}$, and $y$ are coordinates on the defect. The factors of $q\beta$ are supplied using dimensional analysis.

Finally, to evaluate the defect action on $M_{q\b,V^q}/\Z_q$, we simply plug in the expressions (\ref{eq:thefields}), (\ref{eq:theafield}), (\ref{eq:thelastfield}), which are precisely the $\Z_q$-lifts to $S^{d-1}$ of the long-wavelength parts of $g,A,\f$.

 In what follows, we will sometimes use the notation $\mathfrak{D}$ to refer to {\it both\/} a defect on $S^{d-1}/\Z_q$ and the lift $\tl {\mathfrak{D}}$ of the defect locus to $S^{d-1}$. We hope this will not cause confusion.

\subsection{Example: point-like vortex defects in 3d CFTs}

As an example, consider a 3d CFT, where $R$ acts by the discrete rotation $\vf\to \vf+\tfrac{2\pi p}{q}$. The action of $R$ fixes the north and south poles of $S^2$. Consequently, there are two point-like vortex defects: $\mathfrak{D}_{p/q}$ located at the north pole, and its orientation reversal $\mathfrak{D}_{-p/q}$ located at the south pole. Let us focus on $\mathfrak{D}_{p/q}$.

Classifying the vielbein at the north pole into representations of $\Z_q$, we have basis elements $\hat e_+^i,\hat e_-^i$ with charges $+p$ and $-p$, respectively. We normalize them so that $\hat e_\pm\. (\hat e_\pm)^*=1$. To build basic $\Z_q$-invariant curvatures, we begin with tensors $\hat \nabla_i \cdots \hat \nabla_j \hat R$ and $\hat \nabla_i \cdots \hat\nabla_j F_{kl}$, where $F=dA$, $\hat R$ is the curvature scalar built from $\hat g$, and $\hat \nabla$ denotes a covariant derivative with respect to $\hat g$. We then contract their indices with $\hat e_\pm^i$ in such a way that the total $\Z_q$ charge vanishes. The action at each order in a derivative expansion is a polynomial in these basic $\Z_q$-invariants.

Note that we cannot build valid terms in the action by multiplying two $\Z_q$-charged objects to obtain a $\Z_q$-singlet. For example, $(\hat e^i_+\hat \nabla_i \hat R)(\hat e^i_-\hat \nabla_i \hat R)$ is not admissible. The reason is that $\hat e^i_+\hat \nabla_i \hat R$ individually vanishes, due to $\Z_q$-invariance.

Proceeding in this way, the leading invariants in a derivative expansion are
\be
S_{\mathfrak{D}_{p/q}} &\ni 1, \  \hat \star F, \  \hat R, \  (\hat \star F)^2,\ \dots,
\ee
where $\hat \star F= i \hat e^k_+ \hat e^l_- F_{kl}$ is the Hodge star of $F$ in the metric $\hat g$. Concretely, the action is
\be
S_{\mathfrak{D}_{p/q}} &=  \frac{1}{q}\p{\left.a_{0,p/q} + (q\beta) a_{1,p/q} \hat \star F + (q\beta)^2\p{a_{2,p/q} \hat R + a_{3,p/q}(\hat \star F)^2} + \dots\right|_{\mathfrak{D}_{p/q}}},
\ee
where $(\cdots)|_{\mathfrak{D}_{p/q}}$ denotes evaluation at the preimage of the defect on $S^2$ --- in this case the north pole.  We have written the Wilson coefficients as $a_{i,p/q}$ to emphasize that they depend on the rotation fraction $p/q$. In bosonic theories, the $a_{i,x}$ are periodic in $x$ with period $1$, while in fermionic theories, they are periodic in $x$ with period $2$.

At higher orders in derivatives, we can also include laplacians $\hat \nabla^2$, as well as $q$-th powers of charged derivatives $(\hat e_\pm \. \nabla)^q$. However, note that the background fields on $M_{q\b,U^q}$ given in  (\ref{eq:thefields}), (\ref{eq:theafield}), and (\ref{eq:thelastfield}) are invariant not only under $\Z_q$, but under the full maximal torus $\SO(2)$. Consequently, terms involving charged derivatives $(\hat e_\pm \. \nabla)^q$ will actually vanish on $M_{q\b,U^q}/\Z_q$, and in practice we only need to keep polynomials in $\hat\nabla^{2k} \hat\star F$ and $\hat \nabla^{2k}\hat R$.

Plugging in the fields on $M_{q\b,U^q}$, we find
\be
\hat \star F|_\pm &= \pm 2\Omega, \nn\\
\hat R |_\pm &= 2+10\Omega^2,
\ee
where $(\cdots)|_\pm$ denotes the north/south poles of $S^2$.
Thus, summing up the contributions from the north and south poles, the total defect contribution to $\Tr[e^{-\b H} UR]$ is
\be
\label{eq:predictionforthreeddefectaction}
S_\mathfrak{D} &= \frac{a_{0,p/q}+a_{0,-p/q}}{q} + 2\beta\Omega(a_{1,p/q} - a_{1,-p/q}) \nn\\
&\quad + \frac {(q\beta)^2} q ((2+10\Omega^2)\p{a_{2,p/q}+a_{2,-p/q}) + 4\Omega^2 (a_{3,p/q}+a_{3,-p/q})} + \dots.
\ee
In general, the point-like defect action at each order $\beta^k$ is a polynomial in $\Omega$ that is even if $k$ is even and odd if $k$ is odd. We will verify this structure in several examples below.

There is an important distinction between the terms (\ref{eq:predictionforthreeddefectaction}) arising in the defect action $S_\mathfrak{D}$ and the ``bulk" terms (\ref{eq:evaluatedaction}). Note that the bulk terms contain poles at $\Omega_a = \pm i$. Physically, such poles arise because a great circle $r_a=1$ of the spinning $S^{d-1}$ approaches the speed of light as $\Omega_a\to \pm i$. The measure $\sqrt{\hat g}$ becomes singular at the great circle, and the integral over $r_a$ cannot be deformed away from the singularity because it is at an endpoint of the integration contour. By contrast, the defects $\mathfrak{D}_{\pm p/q}$ are located at the north and south poles of $S^2$, where this phenomenon does not occur, and thus their contributions do not have poles at $\Omega=\pm i$. In general, the action of a defect $\mathfrak{D}$ on $M_{q\b,V^q}/\Z_q$ can have poles at $\Omega_a=\pm i$ if and only if the support of $\mathfrak{D}$ intersects the great circle $r_a=1$. We will see an example in the next subsection.

\subsection{Example: vortex defects in 4d CFTs}

Consider now a 4d CFT, where $R$ acts by discrete rotations on each of the angles $\vf_1\rightarrow \vf_1 + \frac{2\pi p_1}{q_1}$ and $\vf_2\rightarrow \vf_2 + \frac{2\pi p_2}{q_2}$. If $q_1\neq q_2$, we have two 1-dimensional vortex defects $\mathfrak{D}^{(1)}$ and $\mathfrak{D}^{(2)}$. The first defect $\mathfrak{D}^{(1)}$ is located at the fixed locus of $R^{q_1}$, which is given by $(r_1,r_2)=(1,0)$ with $\vf_1\in [0,\frac{2\pi}{q})$, where $q\coloneqq \text{lcm}(q_1,q_2)$. The second defect $\mathfrak{D}^{(2)}$ is located at the fixed locus of $R^{q_2}$, which is given by $(r_1,r_2)=(0,1)$ with $\vf_2\in [0,\frac{2\pi}{q})$.  

Let us focus on $\mathfrak{D}^{(1)}$ for now.  
On $\mathfrak{D}^{(1)}$, the leading term in the effective action is a cosmological constant $\int d\f_1 \sqrt{\hat g|_{\mathfrak{D}^{(1)}}}$, as usual. At the first subleading order in a derivative expansion, we have the term $\int d\vf_1\sqrt{\hat g|_{\mathfrak{D}^{(1)}}}  i \hat e^k_+ \hat e^l_- F_{kl}$, which can be written more simply as $\int \hat \star F$. 

The Wilson coefficients of a defect depend only on the geometry of the singularity where the defect lives. To describe this geometry, it is helpful to introduce the co-prime integers
\be
\begin{aligned}
&P_1:= \frac{p_{1} q_2}{(q_1,q_2)}\,,\quad Q_1:=\frac{q_1}{(q_1,q_2)}\,,\\
&P_2:= \frac{p_{2} q_1}{(q_1,q_2)}\,,\quad Q_2:=\frac{q_2}{(q_1,q_2)}\,.
\end{aligned}
\ee
where $(q_1, q_2)$ is the greatest common divisor of $q_1, q_2$. The structure of the singularity at $\mathfrak{D}^{(1)}$ is determined by the action of $R^{q_1}$, which is
\be
R^{q_1} : \vf_2 \mto \vf_2 + \frac{2\pi P_2}{Q_2}.
\ee
Thus, the Wilson coefficients of $\mathfrak{D}^{(1)}$ should depend only on $P_2/Q_2$. However, there is a subtlety in fermionic theories: Note that $R^{q_1}$ implements a rotation by $2\pi p_1$, which is $(-1)^{p_1 F}$ in fermionic theories. Thus, the Wilson coefficients of $\mathfrak{D}^{(1)}$ can additionally depend on $(-1)^{p_1}$ in that case. Consequently, we will write the Wilson coefficients of $\mathfrak{D}^{(1)}$ as $a_{i,P_2/Q_2,(-1)^{p_1}}$ to emphasize the data they depend on. (We will see subtleties of a similar flavor in section~\ref{sec:fermions}.)

Putting everything together, the action $S_{\mathfrak{D}^{(1)}}$ takes the form
\be
	S_{\mathfrak{D}^{(1)}} &= \frac{1}{q}\left( \frac{a_{0,P_2/Q_2,(-1)^{p_1}}}{q\beta} \int d\vf_1 \sqrt{\hat g|_{\mathfrak{D}}} + a_{1,P_2/Q_2,(-1)^{p_1}}\int \hat \star F 
	 + \dots \right)\nn\\
	&= \frac{2\pi}{q(1+\Omega_1^2)}\left(\frac{a_{0,P_2/Q_2,(-1)^{p_1}}}{q\beta} + 2a_{1,P_2/Q_2,(-1)^{p_1}}\Omega_2
	 + \dots\right),
\ee
where in the second line, we evaluated the action in the background $M_{q\beta,U^q}/\Z_q$.
Note that because $\mathfrak{D}^{(1)}$ lives on the great circle $r_1=1$, its action has poles at $\Omega_1=\pm i $.  

Adding similar terms for $\mathfrak{D}^{(2)}$, the total defect contribution to $\Tr[e^{-\b H} UR]$ is
\be
\begin{aligned}
\label{eq:predictionforthreeddefectaction2}
S_\mathfrak{D} &= \frac{2\pi}{q^2\beta}\left(\frac{a_{0,P_2/Q_2,(-1)^{p_1}}}{1+\Omega_1^2}+\frac{a_{0,P_1/Q_1,(-1)^{p_2}}}{1+\Omega_2^2}\right)\\
			   &   + \frac{4\pi}{q}\left(a_{1,P_2/Q_2,(-1)^{p_1}}\frac{\Omega_2}{1+\Omega_1^2}
			   +a_{1,P_1/Q_1,(-1)^{p_2}}\frac{\Omega_1}{1+\Omega_2^2}\right)+ \dots,
\end{aligned}
\ee
where ``$\dots$" represents higher-order terms in $\beta$ coming from higher dimension operators in the defect action. 

\section{Fermionic theories}
\label{sec:fermions}

In this section, we describe some subtleties associated with partition functions of fermionic theories. Again, for simplicity we mostly restrict our discussion to CFT$_d$ on a spatial sphere $S^{d-1}$, though the final conclusion (\ref{eq:moregeneralresultFERMIONS}) holds in a general QFT\@. In short, the results (\ref{eq:generalresultwhenrisfree}) and (\ref{eq:moregeneralresult}) work in fermionic CFTs as well, but we must take care to keep track of the spin structure of the manifold (in particular whether we have periodic or antiperiodic boundary conditions for fermions around $S^1_\beta$ and $S^1_{q\beta}$), and we must consider the rotation $R$ as an element of $\mathrm{Spin}(d)$.

\subsection{Review of 2d}

Let us first review fermionic CFTs in $2d$. In $2d$, we need to specify the boundary conditions of the fermions around both the space and time circles. This defines four different fermion partition functions:
\begin{align}
Z_{\text{R},+}(\tau, \bar\tau) &\coloneqq \Tr_{\text{R}}\left(e^{2\pi i \tau (L_0 - \frac{c}{24})} e^{-2\pi i \bar\tau (\bar{L}_0 - \frac{c}{24})} \right) \nonumber \\ 
Z_{\text{R},-}(\tau, \bar\tau) &\coloneqq \Tr_{\text{R}}\left((-1)^F e^{2\pi i \tau (L_0 - \frac{c}{24})} e^{-2\pi i \bar\tau (\bar{L}_0 - \frac{c}{24})} \right) \nonumber \\ 
Z_{\text{NS},+}(\tau, \bar\tau) &\coloneqq \Tr_{\text{NS}}\left(e^{2\pi i \tau (L_0 - \frac{c}{24})} e^{-2\pi i \bar\tau (\bar{L}_0 - \frac{c}{24})} \right) \nonumber \\ 
Z_{\text{NS},-}(\tau, \bar\tau) &\coloneqq  \Tr_{\text{NS}}\left((-1)^F e^{2\pi i \tau (L_0 - \frac{c}{24})} e^{-2\pi i \bar\tau (\bar{L}_0 - \frac{c}{24})} \right).
\label{eq:2dspins}
\end{align}
The partition functions in (\ref{eq:2dspins}) are not independent. The partition functions $Z_{\text{R},+}, Z_{\text{NS},+}$, and $Z_{\text{NS},-}$ are invariant under different subgroups of $SL(2,\mathbb Z)$ and can transform into each other. More precisely, $Z_{\text{R},-}$ is invariant under all of $SL(2, \mathbb Z)$; and $Z_{\text{R},+}, Z_{\text{NS},+}$, and $Z_{\text{NS},-}$ are invariant under the congruence subgroups $\Gamma_0(2)$, $\Gamma_{\theta}$, and $\Gamma^0(2)$ respectively, which are defined as:
\begin{align}
\Gamma_0(2) &= \left\{ \begin{pmatrix} ~a~  ~b~ \\ ~c~  ~d~ \end{pmatrix} \in SL(2,\mathbb Z), ~~ c ~\text{even}\right\} \nonumber \\ 
\Gamma_\theta &= \left\{ \begin{pmatrix} ~a~  ~b~ \\ ~c~  ~d~ \end{pmatrix} \in SL(2,\mathbb Z), ~~ a+b ~\text{odd}~, c+d ~\text{odd}\right\} \nonumber \\ 
\Gamma^0(2) &= \left\{ \begin{pmatrix} ~a~  ~b~ \\ ~c~  ~d~ \end{pmatrix} \in SL(2,\mathbb Z), ~~ b ~\text{even}\right\}. 
\end{align}
Finally they transformation into each other as:
\begin{align}
Z_{\text{R},+}(-1/\tau,-1/\bar\tau) &= Z_{\text{NS},-}(\tau,\bar\tau), \nonumber \\
Z_{\text{NS},+}(\tau+1,\bar\tau+1) &= Z_{\text{NS},-}(\tau,\bar\tau). 
\label{eq:FermionSSS}
\end{align}
The NS sector partition function (with or without a $(-1)^F$ insertion) at low temperature is well-approximated by the vacuum state (which is a bosonic state), with Casimir energy $-\frac{c}{12}$:
\begin{equation}
\Tr_{\text{NS}}(e^{-\beta(\Delta - \frac{c}{12})}) \sim e^{\frac{\beta c}{12}}, ~~~~~~\beta \gg 1.
\end{equation}
The Ramond sector ground state, in contrast, has a Casimir energy of $E_{gs} - \frac{c}{12}$ where $E_{gs}$ is the Ramond ground-state energy, a non-negative number that is theory-dependent.\footnote{For supersymmetric theories, $E_{gs} = \frac{c}{12}$, but for generic fermionic theories, $E_{gs}$ can be above or below or equal to $\frac{c}{12}$.} Finally, the Ramond ground-state may not necessarily be unique, so we call the degeneracy $N_{gs} \in \mathbb{N}$. 
\begin{equation}
\Tr_{\text{R}}(e^{-\beta(\Delta - \frac{c}{12})}) \sim N_{gs} e^{-\beta(E_{gs} - \frac{c}{12})}, ~~~~~~\beta \gg 1.
\end{equation}

To study the high temperature behavior of the NS-sector partition function with an arbitrary phase $e^{2\pi i p/q J}$ inserted (with $0 \leq p/q < 2$ and $p, q$ coprime)
\begin{equation}
\Tr_{\text{NS}} (e^{-\beta\left[(\Delta - \frac{c}{12}) - i \Omega J\right]} e^{2\pi i J p/q} ), ~~~~~~~ \beta \ll 1
\label{eq:NSTracewewant}
\end{equation} we can use an $SL(2,\mathbb Z)$ transform. In particular we would like to apply a modular transformation of the form
\begin{equation}
\begin{pmatrix} a & b \\ q & -p \end{pmatrix}
\label{eq:pmatrxiwewant}
\end{equation}
to (\ref{eq:NSTracewewant}). The result crucially depends on the parity of $p, q$. If $p+q$ is odd, then we can choose (\ref{eq:pmatrxiwewant}) to be in $\Gamma_{\theta}$ and map the partition function to the NS sector at low temperature. However, if $p+q$ is even, we map the partition function to the R sector at low temperature instead. We therefore get, for $\beta \ll 1$ and $0 \leq p/q < 2$:
\begin{align}
\Tr_{\text{NS}} (e^{-\beta\left[(\Delta - \frac{c}{12}) - i \Omega J\right]} e^{2\pi i J p/q} ) &\sim e^{\frac{4\pi^2}{q^2\beta (1+\Omega^2)}\frac{c}{12}} ~~~~~~~~~~~~~~~~~ p+q ~\text{odd}, ~\beta \ll 1, \nonumber \\
\Tr_{\text{NS}} (e^{-\beta\left[(\Delta - \frac{c}{12}) - i \Omega J\right]} e^{2\pi i J p/q} ) &\sim N_{gs} e^{\frac{4\pi^2}{q^2\beta (1+\Omega^2)}\left(\frac{c}{12}-E_{gs}\right)}~~~~ p+q ~\text{even}, ~\beta \ll 1.
\label{eq:nssectorbehavior}
\end{align}
Equivalently we can always take $0 \leq \frac pq < 1$ with the insertion of a $(-1)^F$:
\begin{align}
\Tr_{\text{NS}} (e^{-\beta\left[(\Delta - \frac{c}{12}) - i \Omega J\right]} e^{2\pi i J p/q} ) &\sim e^{\frac{4\pi^2}{q^2\beta (1+\Omega^2)}\frac{c}{12}} ~~~~~~~~~~~~~~~~~ p+q ~\text{odd}, ~\beta \ll 1, \nonumber \\
\Tr_{\text{NS}} (e^{-\beta\left[(\Delta - \frac{c}{12}) - i \Omega J\right]} e^{2\pi i J p/q} ) &\sim N_{gs} e^{\frac{4\pi^2}{q^2\beta (1+\Omega^2)}\left(\frac{c}{12}-E_{gs}\right)}~~~~ p+q ~\text{even}, ~\beta \ll 1, \nonumber \\
\Tr_{\text{NS}} ((-1)^F e^{-\beta\left[(\Delta - \frac{c}{12}) - i \Omega J\right]} e^{2\pi i J p/q} ) &\sim e^{\frac{4\pi^2}{q^2\beta (1+\Omega^2)}\frac{c}{12}} ~~~~~~~~~~~~~~~~~ p  ~\text{odd}, ~\beta \ll 1, \nonumber \\
\Tr_{\text{NS}} ((-1)^F e^{-\beta\left[(\Delta - \frac{c}{12}) - i \Omega J\right]} e^{2\pi i J p/q} ) &\sim N_{gs} e^{\frac{4\pi^2}{q^2\beta (1+\Omega^2)}\left(\frac{c}{12}-E_{gs}\right)}~~~~ p ~ \text{even}, ~\beta \ll 1.
\label{eq:withminusf}
\end{align} 
We see that in (\ref{eq:withminusf}), there are two real numbers that can determine the behavior of fermionic partition functions: the central charge $c$ and the Ramond ground state energy $E_{gs}$. Moreover, which of the two high temperature behaviors we get ($\frac{c}{12}$ or $\frac{c}{12} - E_{gs}$ multiplying temperature in the free energy) depends on the parity of $p$ and $q$. We will see this exact same behavior repeat itself for fermionic theories in higher dimensions in section~\ref{sec:higherdfermion}.

In 2d, we can also analyze the behavior of the partition function in the Ramond sector. This does not have a direct analog as far as we are aware in higher dimension, but we include it for completeness. By using the same modular transformation properties discussed earlier, we find:
\begin{align}
\Tr_{\text{R}} (e^{-\beta\left[(\Delta - \frac{c}{12}) - i \Omega J\right]} e^{2\pi i J p/q} ) &\sim e^{\frac{4\pi^2}{q^2\beta (1+\Omega^2)}\frac{c}{12}} ~~~~~~~~~~~~~~~~~ q ~\text{even}, ~\beta \ll 1, \nonumber \\
\Tr_{\text{R}} (e^{-\beta\left[(\Delta - \frac{c}{12}) - i \Omega J\right]} e^{2\pi i J p/q} ) &\sim N_{gs} e^{\frac{4\pi^2}{q^2\beta (1+\Omega^2)}\left(\frac{c}{12}-E_{gs}\right)}~~~~ q ~\text{odd}, ~\beta \ll 1, \nonumber \\
\Tr_{\text{R}} ((-1)^F e^{-\beta\left[(\Delta - \frac{c}{12}) - i \Omega J\right]} e^{2\pi i J p/q} ) &\lesssim N_{gs} e^{\frac{4\pi^2}{q^2\beta (1+\Omega^2)}\left(\frac{c}{12}-E_{gs}\right)}~~~~ \beta \ll 1.
\label{eq:ramondstuff}
\end{align} 
Because the final spin structure (where the fermion is periodic in both space and time directions) is invariant under the full modular group, it has the same universal behavior at high temperature regardless of the phase. In general this spin structure cannot be directly derived from the other three (although there are potentially powerful constraints coming from unitarity, and knowledge of $Z_{\text{R},+}$ \cite{Benjamin:2020zbs}). We write $\lesssim$ rather than $\sim$ in the last line of (\ref{eq:ramondstuff}) due to possible cancellations between fermionic and bosonic Ramond ground states, which would effectively reduce $N_{gs}$ (by an even number) for the final spin structure.

\subsection{Higher $d$}
\label{sec:higherdfermion}

Let us now consider fermionic CFTs in $d>2$. For simplicity let us first consider turning on only a single spin and consider:
\begin{equation}
\Tr[e^{-\beta(H - i \Omega J)} e^{2\pi i \frac{p}{q} J}],
\label{eq:fermiononechem}
\end{equation}
with $\beta \ll 1,~ \Omega \sim O(1), ~ 0 \leq p/q < 2$. As discussed before, this can be computed from a path integral of $S^{d-1} \times S^1_{\beta}$, where we insert a defect that rotates the sphere by $2\pi p/q$. Moreover, due to the fermions in the theory, we need to specify a spin structure on this geometry. In (\ref{eq:fermiononechem}) we compute the path integral with anti-periodic boundary conditions for the fermion around the $S^{1}_{\beta}$. We now imagine the setup as in figure~\ref{fig:basicmanipulation} to reduce the setup into a geometry we can obtain a thermal EFT\@. In particular we need to stack $q$ copies of the CFT, and perform a $2\pi p$ rotation in the spatial direction. The final geometry we get -- in addition to the spatial sphere being modded by $\mathbb{Z}_q$ and the thermal circle increasing by a factor of $q$ -- has different periodicity for the fermions about the time circle depending on the parity of $p+q$. If $p+q$ is odd, the fermions remain antiperiodic, and we can use the original EFT description for fermionic CFTs on a long, thin cylinder. If $p+q$ is even, however, we need a \emph{new} EFT, for fermionic CFTs with periodic boundary conditions on a long, thin cylinder.

We now see there are \emph{two} different thermal EFTs we consider, depending on the periodicity of the fermions around $S^1_{\beta}$. Each EFT comes with its own set of Wilson coefficients. We write them as:
\begin{align}
Z_\textrm{CFT}[S^{d-1}\x S^1_{\b, ~\text{anti-periodic}}] &= Z_\textrm{gapped}[S^{d-1}] \sim e^{-S_\textrm{th}[g_{ij},A_i,\f]},  \nonumber \\
Z_\textrm{CFT}[S^{d-1}\x S^1_{\b, ~\text{periodic}}] &= \tilde{Z}_\textrm{gapped}[S^{d-1}] \sim e^{-\tilde{S}_\textrm{th}[g_{ij},A_i,\f]}.
\end{align}
The two expressions $e^{-S_{\textrm{th}}}$ and $e^{-\tilde{S}_{\textrm{th}}}$ respectively compute the partition function with and without the insertion of $(-1)^F$: 
\begin{align}
\Tr\left[e^{-\beta(H - i\Omega J)} \right] &\sim e^{-S_{\textrm{th}}}, \nonumber \\
\Tr\left[(-1)^F e^{-\beta(H - i\Omega J)}\right] &\sim e^{-\tilde{S}_{\textrm{th}}}.
\end{align}
Let us write the two thermal actions as:
\begin{align}
S_\textrm{th} &= \int \frac{d^{d-1} \vec x}{\beta^{d-1}} \sqrt{\hat g} \p{-f + c_1 \beta^2 \hat R + c_2\beta^2 F^2 + \dots} + S_\textrm{anom}, \nonumber \\
{\tilde S}_\textrm{th} &= \int \frac{d^{d-1} \vec x}{\beta^{d-1}} \sqrt{\hat g} \p{-\tilde f + \tilde{c_1} \beta^2 \hat R + \tilde{c_2}\beta^2 F^2 + \dots} + S_\textrm{anom}. 
\label{eq:twowilsons}
\end{align} 
The most general leading term behavior of (\ref{eq:fermiononechem}) then goes as
\begin{equation}
\log\left[ \Tr(e^{-\beta(H - i \Omega J)} e^{2\pi i \frac{p}{q} J})\right] \sim \begin{cases}
\frac{1}{q^d} \frac{f~\text{vol}~S^{d-1}}{\beta^{d-1}(1+\Omega^2)} + \ldots, ~~~~~~p+q~\text{odd},\\
\frac{1}{q^d} \frac{\tilde{f}~\text{vol}~S^{d-1}}{\beta^{d-1}(1+\Omega^2)}+\ldots, ~~~~~~p+q~\text{even}.
\end{cases}
\label{eq:fermiononechemanswer}
\end{equation}
Note that if we specialize (\ref{eq:fermiononechemanswer}) to $d=2$, this is precisely the behavior we get for (\ref{eq:nssectorbehavior}), if we identify:
\begin{align}
f &= \frac{2\pi c}{12}, \nonumber \\
\tilde f &= 2\pi \left(\frac{c}{12} - E_{gs}\right).
\end{align}
We can think of the difference between the two sets of Wilson coefficients in (\ref{eq:twowilsons}) as the higher-dimensional analog of the ``Ramond ground state energy".

We can generalize (\ref{eq:fermiononechem}) to turning on $n$ spins and consider
\begin{equation}
\Tr\left[e^{-\beta(H - i(\Omega_1 J_1 + \dots + \Omega_n J_n))} e^{2\pi i \frac{p_1}{q_1} J_1} \cdots e^{2\pi i \frac{p_n}{q_n} J_n} \right]
\end{equation}
for $\beta \ll 1$, $\Omega_i \sim O(1)$, $0 \leq \frac{p_i}{q_i} < 2$. If we define 
\begin{equation}
q \coloneqq \text{lcm}(q_1, \ldots, q_n),
\end{equation}
then in our EFT, we will go around the time circle $q$ times and go around the spatial sphere $\sum_{i} \frac{p_i q}{q_i}$ times. Thus depending on if $q + \sum_i \frac{p_i q}{q_i}$ is odd or even, we get the thermal EFT described by $S_{\text{th}}$ or $\tilde{S}_{\text{th}}$.

Finally, all of these results are consistent with (\ref{eq:moregeneralresult}) if we simply interpret $e^{i\beta\Omega\.J}R$ as an element of $\textrm{Spin}(d)$ and interpret $\Tr$ as imposing periodic boundary conditions for both bosonic and fermionic variable. With this understanding, (\ref{eq:moregeneralresult}) is the general recipe.  If we instead interpret $\Tr$ as imposing periodic boundary conditions for bosons and antiperiodic boundary conditions for fermions,  (\ref{eq:moregeneralresult}) should be modified to 
\be
\label{eq:moregeneralresultFERMIONS}
-\log \Tr\left[g R\right] &\sim -\frac 1 q \log \Tr\left[(-1)^{(q-1)F}g^q\right] + \textrm{topological} + \sum_{\mathfrak{D}_i} S_{\mathfrak{D}_i}.
\ee
We will see explicit examples of this in section~\ref{sec:freetheories}.

\section{Free theories}
\label{sec:freetheories}

We now present several examples involving free fields to show explicitly that \eqref{eq:generalresultwhenrisfree} and \eqref{eq:moregeneralresult} hold (along with their appropriate generalizations to fermionic theories). We begin in section~\ref{sec:massive} by checking a \emph{massive} quantum field theory, namely a massive free boson in 2d. We then consider free CFTs in 3d and 4d. Our main tool for computing partition functions in free CFTs is the \textit{plethystic exponential}, which we review in appendix~\ref{appendix:highTexpansion} (see e.g.\ \cite{Feng:2007ur, Melia:2020pzd,Henning:2017fpj}). 
				
In section~\ref{sec:threedexamples},  we consider various examples involving free scalar fields and free fermions in $3$ dimensions.   In section~\ref{sec:4dwithdefectaction},  we explore few more examples in $4$ dimension.  We present additional 4d and 6d examples in appendix~\ref{sec:4d6dwithoutdefectaction}.  Lastly, in section~\ref{sec:2dwithgravitationalanomaly}, we consider 2d CFTs with a local gravitational anomaly. Such theories can include Chern-Simons term in their thermal effective action, which are not integrals of local gauge/coordinate invariant densities. These furnish examples of the ``topological" terms in \eqref{eq:generalresultwhenrisfree} and \eqref{eq:moregeneralresult}. 

\subsection{Massive free boson in 2d}
\label{sec:massive}

As a first check on our formalism (and to illustrate that the basic ideas do not require conformal symmetry), let us study the partition function of a free scalar with mass $m$ in $2d$. We begin by computing the partition function on a rectangular torus $S_L^1 \x S^1_\beta$:
\be
\label{eq:partitionfunctionsum}
\log Z &= -\frac 1 2 \Tr \log(m^2 + \Delta) - S_\textrm{ct} \nn\\
&= -\frac 1 2 \sum_{(r,s)\in \Z^2} \log\p{m^2 + \p{\frac{2\pi r}{L}}^2 + \p{\frac{2\pi s}{\beta}}^2} - S_\textrm{ct} \nn\\
&= -\sum_{r\in \Z} \log 2 \sinh \p{\frac{\beta\sqrt{m^2 + (\frac{2\pi r}{L})^2}}{2}} - S_\textrm{ct},
\ee
where $S_\textrm{ct}$ is a cosmological constant counterterm, and in the third line we performed the sum over $r$ and threw away a constant using the fact that $\sum_{s\in \Z} 1 = 0$ in $\zeta$-function regularization.

Because the summand is slowly-varying in $r$, we can take the thermodynamic limit $L\to \oo$ by replacing the sum over $r$ with an integral over momentum $k=\frac{2\pi r}{L}$:
\be
\log Z &\sim -\frac{L}{\pi} \int_0^\oo dk \p{\log \p{1-e^{-\beta\sqrt{m^2 +k^2}}}
-\frac{\beta\sqrt{m^2+k^2}}{2}
} - S_\textrm{ct}\qquad (L\to \oo).
\ee
The second term in the integrand is UV-divergent, but proportional to $L\beta$. Hence, it takes the form of a cosmological constant and can be removed with an appropriate choice of $S_\textrm{ct}$. We will choose $S_\textrm{ct}$ to simply subtract this contribution, which is equivalent to setting the free energy density to zero in flat $\R^2$. We find
\be
\log Z &\sim \frac{L}{\beta} f(\beta m) \qquad (L\to \oo),
\ee
where minus the effective free energy density is
\be
\label{eq:freenergydensityintegralform}
f(y) &= -\frac{y}{\pi} \int_0^\oo dx \log\p{1-e^{-y\sqrt{1+x^2}}} \sim 
\begin{cases}
\frac{\pi}{6} & y\ll 1,\\
\sqrt{\frac{y}{2\pi}}e^{-y} & y \gg 1.
\end{cases}
\ee
Note that the limit of $f(y)$ as $y\to 0$ is consistent with $f=\frac{2\pi c}{12}$ with $c=1$ for the massless free boson in 2d.

Before continuing to the twisted partition function, let us make two observations. Firstly, the partition function $Z$ obeys a form of modular invariance. To formulate it, let us write $\mathbb{T}^2=\R^2/\Lambda$, where the lattice $\Lambda$ is spanned by basis vectors $\vec e_1,\vec e_2$. We can arrange the basis vectors into a matrix $E\in \R^{2\x2}$ whose columns are $\vec e_1$ and $\vec e_2$, and consider the partition function as a function of $E$. (Above, we studied the case where $\vec e_1=(L,0)$ and $\vec e_2=(0,\beta)$, and thus $E=\begin{psmallmatrix}L & 0\\ 0 & \beta \end{psmallmatrix}$.) Rotational invariance implies that $Z(E)$ is invariant under left-multiplication by an orthogonal group element $g\in \SO(2)$. Modular invariance is the statement that $Z(E)$ is also invariant under an integer change of basis of the lattice $\Lambda$, which is equivalent to right-multiplication by $\g\in \SL(2,\Z)$:
\be
\label{eq:modularinvarianceofZ}
Z(E) &= Z(g E \g).
\ee
In other words, $Z$ is a function on the moduli space $\SO(2)\backslash \R^{2,2}/\SL(2,\Z)$.

The second key observation is that in the thermodynamic limit, $Z$ is unchanged under a small shift of the basis vector that grows as $L\to\oo$: $\vec e_1\to \vec e_1 + \alpha \vec e_2$. To see why, note that the corresponding dual basis shifts by $\hat e_2\to \hat e_2 - \a \hat e_1$ (with $\hat e_1$ unchanged). Thus, the sum (\ref{eq:partitionfunctionsum}) changes by shifting $r\to r-s\a$:
\be
\label{eq:twistthatdoesntmatter}
\log Z\p{\begin{psmallmatrix}
L & 0 \\
\a\b & \b
\end{psmallmatrix}} &= -\frac 1 2 \sum_{(r,s)\in \Z^2} \log\p{m^2 + \p{\frac{2\pi (r-s\a)}{L}}^2 + \p{\frac{2\pi s}{\beta}}^2} - S_\textrm{ct} \nn\\
&\sim \frac{L}{\beta} f(\beta m) \qquad (L\to \oo).
\ee
In the continuum limit, we can identify the momentum $k=\frac{2\pi(r-s\a)}{L}$ and rewrite the sum as an integral over $k$. The shift $r\to r-s\a$ becomes immaterial in this approximation.

Now let us finally consider the partition function with the insertion of a discrete rotation of the spatial circle $S^1_L$ by an angle $2\pi p/q$. This corresponds to the matrix
\be
E' &= \begin{pmatrix}
L & \tfrac {pL} q \\
0 & \beta
\end{pmatrix}
= \begin{pmatrix}
\frac{L}{q} & 0 \\
(p^{-1})_q \beta & q \beta
\end{pmatrix}
\g',\quad\textrm{where}\quad
\g'=\begin{pmatrix}
q & p \\
-(p^{-1})_q & -b
\end{pmatrix}\in \SL(2,\Z).
\ee
As before $(p^{-1})_q$ denotes the inverse of $p$ modulo $q$, and $b$ is chosen so that $\g'$ has determinant $1$. Applying modular invariance (\ref{eq:modularinvarianceofZ}) and the result (\ref{eq:twistthatdoesntmatter}), we find
\be
\label{eq:thingwewant}
\log Z(E') &= \log Z\begin{psmallmatrix}
\frac{L}{q} & 0 \\
(p^{-1})_q \beta & q \beta
\end{psmallmatrix} \sim  \frac{L}{q^2 \beta} f(q\beta m) \qquad (L\to \oo), 
\ee
consistent with (\ref{eq:moregeneralresult}).

Using slightly fancier technology (see e.g.\ \cite{Dondi:2021buw}), we can be more precise about what information is thrown away in the ``$\sim$" in equation (\ref{eq:thingwewant}). We rewrite the partition function in terms of a spectral zeta function
\be
\label{eq:fancypartition}
\log Z(E) &= \frac 1 2 \zeta'_E(0) - S_\textrm{ct},
\ee
where
\be\label{eqn:resumExample}
\zeta_E(s) &= \Tr\left[(m^2+\De)^{-s}\right] = \frac{1}{\G(s)}\int_0^\oo \frac{dt}{t} t^s \Tr\left[e^{-t(m^2 + \Delta)}\right] \nn\\
&= \frac{1}{\G(s)}\int_0^\oo \frac{dt}{t} t^s e^{-t m^2}\sum_{\vec r\in \Z^2} e^{-t(2\pi E^{-T} \vec r)^2} \nn\\
&= \frac{1}{\G(s)}\int_0^\oo \frac{dt}{t} t^s e^{-t m^2} \frac{\det E}{4\pi t}\sum_{\lambda\in \Lambda} e^{-\frac{\lambda^2}{4t}}.
\ee
In the last line, we performed Poisson resummation to obtain a sum over lattice points $\l\in\Lambda$. (Recall that $\L$ is the lattice spanned by the columns of $E$.) We can now perform the integral over $t$ and plug the result into (\ref{eq:fancypartition}). The term $ \l=(0,0)$ precisely cancels against $S_\textrm{ct}$, and we find
\be
\label{eq:reallygeneralformulaforZ}
\log Z(E) &= \frac{m\det E}{2\pi} \sum_{\lambda \in \Lambda -(0,0)} \frac{K_1(m|\lambda|)}{| \lambda|},
\ee
where $K_1(x)$ is a modified Bessel function.

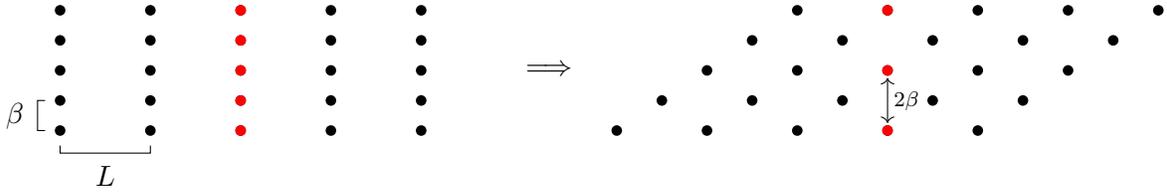
\begin{figure}
\centering
\begin{tikzpicture}
\begin{scope}[shift={(-3.7,0)}]
\foreach \x in {0,0.4,...,1.6} {
  \foreach \y in {0,0.4,...,1.6} {
    \fill (3*\x,\y) circle (2pt);
   }
 }
\foreach \y in {0,0.4,...,1.6} {
  \fill[red] (3*0.8,\y) circle (2pt);
}
\draw (0,-0.2) -- (0,-0.3) -- (1.2,-0.3) -- (1.2,-0.2);
\node at (0.6,-0.6) {$L$};
\draw (-0.2,0) -- (-0.3,0) -- (-0.3,0.4) -- (-0.2,0.4);
\node at (-0.6,0.2) {$\beta$};
\end{scope}
\node at (2.8,0.8) {$\implies$};
\begin{scope}[shift={(3.7,0)}]
\foreach \x in {0,0.4,...,1.6} {
  \foreach \y in {0,0.4,...,1.6} {
    \fill (3*\x+1.5*\y,\y) circle (2pt);
   }
 }
\foreach \y in {0,0.8,...,1.6} {
  \fill[red] (3.6,\y) circle (2pt);
}
\draw[<->] (3.6,0.1) -- (3.6,0.7);
\node[] at (3.85,0.4) {\scriptsize $2\beta$};
\end{scope}
\end{tikzpicture}
\caption{\label{fig:freetheorylattice}{\bf Left:} the lattice $\Lambda$ for a rectangular torus $S^1_L\x S^1_\beta$. In the thermodynamic limit $L\to \oo$, the sum (\ref{eq:reallygeneralformulaforZ}) is dominated by $\l\in \textcolor{red}{\Lambda_\textrm{short}}$ (depicted in red), which are spaced apart by $\beta$. {\bf Right:} the lattice after twisting by a spatial rotation by $\pi$. In the limit $L\to \oo$, the partition function is dominated by $\l\in \textcolor{red}{\Lambda_\textrm{short}'}$, which are now spaced apart by $2\beta$.}
\end{figure}

 Now we can see clearly that in the thermodynamic limit, the lattice vectors $\l\in \L$ that become longer give an exponentially suppressed contribution to $\log Z$ (since $K_1(r)\sim e^{-r}$ for large $r$). Indeed, we have
\be
\label{eq:reallyniceformulaforZ}
\log Z(E) &= \frac{m\det E}{2\pi} \sum_{\l\in \L_\textrm{short}-(0,0)} \frac{K_1(m|\l|)}{|\l|} + O(e^{-L m}) \qquad (L\to \oo),
\ee
where $\L_\textrm{short}$ are the lattice vectors that do not get longer in the thermodynamic limit $L\to \oo$, see figure~\ref{fig:freetheorylattice}. Applying the result (\ref{eq:reallyniceformulaforZ}) to the rectangular torus $S^1_L\x S^1_\beta$, we find another expression for the effective free energy density:
\be
f(y) &= \frac{y}{\pi} \sum_{\ell=1}^\oo \frac 1 \ell K_1(y\ell),
\ee
which agrees with (\ref{eq:freenergydensityintegralform}), as we can see by expanding $\log(1-e^{-y\sqrt{1+x^2}})=-\sum_\ell \frac 1 \ell e^{-\ell y \sqrt{1+x^2}}$ and integrating term-by term.
The result (\ref{eq:reallyniceformulaforZ}) also immediately implies (\ref{eq:twistthatdoesntmatter}). Combining this with modular invariance, we find that (\ref{eq:thingwewant}) holds up to exponential corrections of the form $e^{-m L}$.  Note that this conclusion relies on finiteness of $m$ in the thermodynamic limit. If instead we take $m\to 0$, then the thermal theory has a gapless sector, and we do not expect (\ref{eq:moregeneralresult}) to hold.

We can also understand (\ref{eq:thingwewant}) more directly from (\ref{eq:reallyniceformulaforZ}) as follows. When the partition function has a twist by a rational angle $\frac{2\pi p}{q}$, then a new $\Lambda_\textrm{short}'$ emerges, as depicted in figure~\ref{fig:freetheorylattice}. The emergent $\Lambda_\textrm{short}'$ looks like $\Lambda_\textrm{short}$ in the un-twisted case, but with the replacement $\beta\to q\beta$.

It is also straightforward to generalize this analysis to a massive free scalar in $d$ dimensions. The torus partition function is\footnote{This is an example of an Epstein $\zeta$-function, see e.g.\ \cite{Dunne:2007rt}.}
\be
\log Z(E) &= \det E \p{\frac{m}{2\pi}}^{d/2} \sum_{\l\in \L} \frac{K_{d/2}(m|\l|)}{|\l|^{d/2}}.
\ee
This is again consistent with (\ref{eq:moregeneralresult}) (with vanishing topological and defect contributions) through the mechanism depicted in figure~\ref{fig:freetheorylattice}.

\subsection{3d CFTs}
\label{sec:threedexamples}

\subsubsection{Free scalar}

We now turn our attention to partition functions of higher dimensional CFTs on a spatial $S^{d-1}$.
Let us begin by studying the partition function of a free scalar in 3d, with various discrete rotations inserted. The usual KK reduction of a free scalar on a circle possesses a zero mode. However, in order to apply thermal EFT, the KK reduced theory must be gapped. Thus, in order to avoid zero modes, we will study a {\it complex\/} scalar charged under a $\Z_k$ flavor symmetry. We will turn on the $\Z_k$ fugacity as appropriate to eliminate zero modes, and study the thermal EFT description of the resulting twisted partition functions. Note that in a generic interacting CFT (as opposed to a free theory) we would not expect to have zero modes, and it would not be necessary to consider flavor symmetry.

Let $R(\theta) \in \SO(3)$ denote a rotation around the $z$-axis, and let $\Psi$ denote a reflection in the $z$ direction. We will consider insertions of $R(\theta,s)=R(\theta)\circ \Psi^s \in O(3)$, where $\theta=\frac{2\pi p}{q}$. Note that the reflection potential $s$ takes values in $\{0,1\}$, since $\Psi^2=1$. The partition function is given by
\begin{equation}
\begin{aligned}
&\log\ \Tr \left[e^{-\beta (H-i\Omega J)} e^{\frac{2\pi i \ell Q}{k}} R(\theta,s) \right]\\
&=\sum_{n} \frac{1}{n} \left(e^{\frac{2\pi i \ell n}{k}}+e^{-\frac{2\pi i\ell n}{k}}\right)\frac{e^{-n\beta/2} (1-e^{-2n\beta})}{(1-(-1)^{ns}e^{-n\beta})(1-e^{-n\beta} x^n)(1-e^{-n\beta} x^{-n})},
\end{aligned}
\end{equation}
where $x=e^{2\pi i p/q+i\beta\Omega}$, and $Q$ is the charge of the scalar under $U(1)$ (of which $\Z_k$ is a subgroup).

When we turn on a big rotation $R(\theta,s)$ with order $q$ and include the global symmetry operator $V=e^{\frac{2\pi i \ell Q}{k}}$, a zero mode will be present on the EFT bundle whenever $V^q=1$, since $V$ wraps $q$ times around the base of the EFT bundle --- see the discussion around (\ref{eq:generalresultwhenrisfree}).

Let us consider the case $p/q=1/2$, corresponding to a rotation by an angle $\pi$,  which fixes the north and south poles of $S^2$.  If the flavor group were $\mathbb{Z}_2$, then we would necessarily have a zero mode on the EFT bundle.  Hence we instead consider $\mathbb{Z}_3$ flavor symmetry.

The thermal partition function of a free scalar field with non-zero $\mathbb{Z}_3$ flavor and small rotation-fugacity has the following high temperature expansion
\be\label{eq:usual}
&\log\ \Tr \left[e^{-\beta (H-i\Omega J)} e^{\frac{4\pi i  Q}{3}}\right]
\nn\\
&= 
-\frac{16\zeta(3)}{9(1+\Omega^2)\beta^2}
-\frac{(1+2\Omega^2)\log 3}{12(1+\Omega^2)}
+\frac{(21+4\Omega^2-24\Omega^4)\beta^2}{4320(1+\Omega^2)}+O(\beta^4).
\ee
				
Now we would like to turn on the fugacity for the big rotation with or without a reflection,  leading to absence or presence of non trivial defect action, respectively. 

\subsection*{$\bullet$ Free action: without defect}
Let us first consider an insertion of $R(\pi,1)=R(\pi)\circ \Psi$. This is a parity transformation $\vec n\mto -\vec n$ on $S^2$, and it does not have any fixed points. Thus, the thermal effective action will be free of defects.
Concretely, we find
				\begin{equation}
				\begin{aligned}\label{eq:unusual1}
					&\log\ \Tr \left[e^{-\beta (H-i\Omega J)} e^{\frac{2\pi i  Q}{3}} R(\pi,1) \right]\\
					&=
					-\frac{2\zeta(3)}{9(1+\Omega^2)\beta^2}
					-\frac{(1+2\Omega^2)\log 3}{24(1+\Omega^2)}
					+\frac{(21+4\Omega^2-24\Omega^4)\beta^2}{2160(1+\Omega^2)}+O(\beta^4).
				\end{aligned}
				\end{equation}
Comparing \eqref{eq:usual} with \eqref{eq:unusual1} we find 
\begin{equation}
\log \Tr \left[e^{-\beta (H-i\Omega J)} e^{\frac{2\pi i  Q}{3}} R(\pi,1) \right]\sim \frac{1}{2}\log\Tr \left[e^{-2\beta (H-i\Omega J)} e^{\frac{4\pi i  Q}{3}} \right]\,,
\end{equation}
so \eqref{eq:generalresultwhenrisfree} holds.

\subsection*{$\bullet$ Non-free action: defect}

Without the reflection fugacity, the action of $R$ has fixed points and thermal EFT predicts a non-trivial $S_{\mathfrak{D}}$ as in \eqref{eq:moregeneralresult}.  We find
			\begin{equation}
			\begin{aligned}
					&\log\ \Tr \left[e^{-\beta (H-i\Omega J)} e^{\frac{4\pi i  Q}{3}} R(\pi,0) \right]\\
					&=
					-\frac{2\zeta(3)}{9(1+\Omega^2)\beta^2}
					-\frac{(7+8\Omega^2)\log 3}{24(1+\Omega^2)}
					+\frac{(-69+94\Omega^2+156\Omega^4)\beta^2}{2160(1+\Omega^2)}+O(\beta^4)\,.
				\end{aligned}
				\end{equation}
Comparing with \eqref{eq:usual}, we obtain
			\begin{equation}
			\begin{aligned}
				-\log\ \Tr \left[e^{-\beta (H-i\Omega J)} e^{\frac{2\pi i  Q}{3}} R(\pi,0) \right] \sim -\frac{1}{2} \log\ \Tr \left[e^{-2\beta (H-i\Omega J)} e^{\frac{4\pi i  Q}{3}}  \right] +S_{\mathfrak{D}}\,,
				\end{aligned}
			\end{equation}	
where the total defect action $S_{\mathfrak{D}}(\beta,\Omega)$ is given by
			\begin{equation}
			S_{\mathfrak{D}}(\beta,\Omega)=-\frac{\log 3}{4}
				+\frac{2\Omega^2-1}{24}\beta^2
				+O(\beta^4)\,.
			\end{equation}
Note that $S_\mathfrak{D}$ has precisely the form predicted in (\ref{eq:predictionforthreeddefectaction}), with
\be
a_{0,1/2} = -\frac{\log 3}{4},\quad a_{2,1/2} = -\frac{1}{192},\quad a_{3,1/2} = \frac 7 {384}.
\ee
Furthermore, the linear term in $\beta$ vanishes because $a_{1,1/2} = a_{1,-1/2}$ in bosonic theories.

More generally, we can consider a rotation $R(\frac{2\pi p}{q},0)$, where $p$ and $q$ are coprime and $q$ is not divisible by 3 (so that there is no zero mode upon dimensional reduction). The leading defect Wilson coefficients in this case satisfy: 
 \begin{align}
 	&a_{0,p/q} + a_{0,-p/q}
	=  - \frac 1 3\sum_{k=1}^{q-1}
	\frac{1}{\sin(\tfrac{\pi k p}{3})^2}
	\left[
		\cos\left(\tfrac{2\pi(k+q)}{3}\right)
		\left(	\psi \left(\tfrac{k}{3q}+\tfrac1 3\right)  - \psi \left(\tfrac{k}{3q}\right)  \right)
	\right.\nn\\
	&\left.\hspace{6cm}
	+ \cos\left(\tfrac{2\pi(k+2q)}{3}\right)
	\left(\psi\left(\tfrac{k}{3q}+\tfrac2 3\right) - \psi\left(\tfrac{k}{3q}\right)\right)
	\right],
	\\
	&a_{1,p/q} - a_{1,-p/q} 
	= \frac{1}{6}\sum_{k=1}^{q-1}
	\frac{\cos(\tfrac{\pi k p}{3})}{\sin(\tfrac{\pi k p}{3})^3}
		\left[\cos\left(\tfrac{2\pi(k+q)}{3}\right)+2\cos\left(\tfrac{2\pi(k+2q)}{3}\right)\right],
 \end{align}
where $\psi(z)$ is the digamma function.
			
\subsubsection{Free fermion}		
We can perform a similar exercise with free Dirac fermions.  The partition function is given by 
\begin{equation}
\begin{aligned}
&\log\ \Tr \left[e^{-\beta H} e^{\frac{2\pi i\ell Q}{k}}R(\theta,0) (-1)^{Fw}\right]\\
&=\sum_{n} (-1)^{nw}\frac{(-1)^{n+1}}{n} \left(e^{\frac{2\pi i\ell n}{k}}+e^{-\frac{2\pi i\ell n}{k}}\right)\frac{e^{-n\beta} \left[\left((\sqrt{x})^n+\frac{1}{(\sqrt{x})^n}\right)-e^{-n\beta}\left((\sqrt{x})^n+\frac{1}{(\sqrt{x})^n}\right)\right]}{(1-e^{-n\beta})(1-e^{-n\beta} (\sqrt{x})^{2n})(1-e^{-n\beta} (\sqrt{x})^{-2n})}.
\end{aligned}
\end{equation}

Fermions are antiperiodic in the time direction at finite temperature.  Hence, when no fugacities corresponding to a big rotation or fermion number are turned on, there is no zero mode contribution to the thermal EFT description of the partition function.  In this case, we have the following high temperature behavior:
\begin{equation}\label{eq:usualF}
			\begin{aligned}
				&\log\Tr \left[e^{-\beta (H-i\Omega J)}\right]=
					\frac{3\zeta(3)}{(1+\Omega^2)\beta^2}
					-\frac{(2+\Omega^2)\log 2}{6(1+\Omega^2)}
					+\frac{(24-4\Omega^2-21\Omega^4)\beta^2}{5760(1+\Omega^2)}
					+O(\beta^4).\\
					\end{aligned}
					\end{equation}

However, when $(-1)^F$ and big rotations are turned on,  we must be careful about zero modes.  If the big rotation is given by $e^{2\pi i p/q}$,  then we have various cases depending on the parity of $p,q$ and presence or absence of $(-1)^F$.  We consider all possible cases in the Table~\!\ref{table:fermion}.

				\renewcommand{\arraystretch}{1.5}
\begin{table}[h!]
	\centering
	\begin{tabular}{@{}ll@{}ll@{}ll@{}ll@{}}\toprule
		$(-1)^{F}?$  &  Rotation  & \qquad Boundary Condition &  Flavor?&\quad Equation\\
		  &    $p/q$ & \qquad  $(-1)^{q(w+1)}(-1)^{p}$ &  &\quad \\
		\midrule
		No\ $(w=0)$ &  $1/2$ & \qquad $(-1)^{2} (-1)^1=-1$ &No & \quad  \eqref{eq:qeven}, \eqref{eq:qevenResult}\\
		Yes $(w=1)$ & $1/2$ & \qquad  $(-1)^{4} (-1)^1=-1$ &No & \quad  \eqref{eq:qeven}, \eqref{eq:qevenResult}\\
		\midrule
		No\ $(w=0)$&$1/3$  & \qquad  $(-1)^{3} (-1)^1=+1$  &$ \mathbb{Z}_2$&\quad \eqref{eq:q3p1},  \eqref{eq:q3p1Result}\\
		Yes $(w=1)$& $1/3$ &  \qquad  $(-1)^{6} (-1)^1=-1$ & No &\quad \eqref{eq:q3p1},  \eqref{eq:q3p1Result}\\
		\midrule
		No\ $(w=0)$ & $2/3$& \qquad  $(-1)^{3} (-1)^2=-1$ & No &\quad \eqref{eq:q3p2},  \eqref{eq:q3p2Result}\\
	    Yes $(w=1)$& $2/3$& \qquad  $(-1)^{6} (-1)^2=+1$  & $\mathbb{Z}_2$ &\quad 
\eqref{eq:q3p2},  \eqref{eq:q3p2Result}\\
		\bottomrule
	\end{tabular}
	\caption{\label{table:fermion} Turning on fugacities corresponding to flavor, rotation and fermionic number.  $w=0,1$ refers to turning off and on the fugacity corresponding to $(-1)^F$ respectively.  The fugacity corresponding to big rotation is $e^{2\pi i p/q}$.  The third column lists the effective spin structure on the EFT bundle,  $\pm 1$ in this column refers to periodic and antiperiodic boundary condition respectively.  The fourth column lists whether we need a flavor twist to make sure that there is no zero mode.  Note that antiperiodic boundary condition rules out the presence of zero modes.  The final column refers to relevant equations in the main text.}
\end{table}

Let us consider the case when $q$ is even.  We take $q=2$ for simplicity.  According to table~\!\ref{table:fermion},  we do not need a flavor twist to remove zero modes.  We compute
\begin{equation}\label{eq:qeven}
			\begin{aligned}
					&\log \Tr \left[e^{-\beta  (H-i\Omega J)} R(\pm\pi,0)\right]\\
					&	=\frac{3\zeta(3)}{8(1+\Omega^2)\beta^2}
					-\frac{(2+\Omega^2)\log 2}{12(1+\Omega^2)}
					\mp \frac{\Omega\beta}{4}
					+\frac{(24-4\Omega^2-21\Omega^4)\beta^2}{2880(1+\Omega^2)}
					+O(\beta^3).
					\end{aligned}
				\end{equation}
Note that $\log\Tr \left[e^{-\beta (H-i\Omega J)} (-1)^F R(\pi,0)\right]=\log\Tr \left[e^{-\beta (H-i\Omega J)} R(-\pi,0)\right]$.  
				
Comparing \eqref{eq:usualF} and \eqref{eq:qeven},  we verify  the appropriate generalizations of \eqref{eq:moregeneralresult} to fermionic CFT\@. In particular, see \eqref{eq:moregeneralresultFERMIONS} and note that as an element of spin group, we have $R^2(\pm\pi,0)=(-1)^F$.  We find
				\begin{equation}\label{eq:qevenResult}
			\begin{aligned}
			&-\log\Tr \left[e^{-\beta (H-i\Omega J)} R(\pm\pi,0)\right]\sim -\frac{1}{2}\log\Tr \left[e^{-2\beta (H-i\Omega J)}  \right] + S_{\mathfrak{D}_{\pm}}\,,
				\end{aligned}
				\end{equation}
				where $S_{\mathfrak{D}_{\pm}}$ is given by
				\begin{equation}\label{eq:pirotationSdefect}
			\begin{aligned}
				&S_{\mathfrak{D}_{\pm}}=
			\mp\frac{\Omega\beta}{4} 
			+O(\beta^3)\,.
			\end{aligned}
			\end{equation}
This has precisely the form predicted in (\ref{eq:predictionforthreeddefectaction}), with
\be
a_{1,1/2}-a_{1,-1/2}= \frac{1}{8}\,,\quad a_{0,1/2}+a_{0,-1/2}=0\,.
\ee
				
Next we consider the case when $q$ is odd. Now $p$ can be either even or odd.  For example, let us consider $q=3$ and $p=1$ or $p=2$. We introduce a flavor twist according to Table.~\!\ref{table:fermion}.  Note that,  operationally the insertion of $(-1)^F$ is same as introducing a $\Z_2$ flavor twist.  The insertion of both amounts to inserting nothing.  In what follows,  we choose to insert $(-1)^F$ to eliminate the zero mode.

For $p/q=1/3$,  we find that 
				\begin{equation}
			\begin{aligned}\label{eq:q3p1}
				&\log\Tr \left[e^{-\beta (H-i\Omega J)}  (-1)^F R(2\pi/3,0)\right]\\
				&\sim
				\frac{\zeta(3)}{9(1+\Omega^2)\beta^2}
					-\frac{(10+9\Omega^2)\log 2}{18(1+\Omega^2)}
					+\frac{5\Omega\beta}{3\sqrt{3}}
					+\frac{(-568+1028\Omega^2+1617\Omega^4)\beta^2}{5760(1+\Omega^2)}											+O(\beta^3).
			\end{aligned}
			\end{equation}
Comparing \eqref{eq:usualF} and \eqref{eq:q3p1} we find that 
			\begin{equation}\label{eq:q3p1Result}
			\begin{aligned}
				&\log\Tr \left[e^{-\beta (H-i\Omega J)}  (-1)^F R(2\pi/3,0)\right]\sim 
				\frac{1}{3}\log\Tr \left[e^{-3\beta (H-i\Omega J)}  \right]
				+ S_{\mathfrak{D}}(\beta,\Omega)\,,
			\end{aligned}
			\end{equation}
where the total defect action $S_{\mathfrak{D}}(\beta,\Omega)$ takes the form predicted in \eqref{eq:predictionforthreeddefectaction}:
			\begin{equation}\label{eq:q3p1Sdefect}
			\begin{aligned}
				&S_{\mathfrak{D}}(\beta,\Omega)\sim
				-\frac{4\log2}{9}
				+\frac{5\Omega\beta}{3\sqrt3} 
				-\frac{(8-21\Omega^2)\beta^2}{72}
				+O(\beta^3)\,,
			\end{aligned}
			\end{equation}

The $p/q=2/3$ case can be easily be done using the following identity
\begin{equation}\label{eq:q3p2}
\begin{aligned}
&\log\Tr \left[e^{-\beta (H-i\Omega J)} R(4\pi/3,0)\right]=\log\Tr \left[e^{-\beta (H+i\Omega J)}  (-1)^F R(2\pi/3,0)\right]\,,
\end{aligned}
\end{equation}
from which it follows that 
			\begin{equation}\label{eq:q3p2Result}
			\begin{aligned}
				&\log\Tr \left[e^{-\beta (H-i\Omega J)} R(4\pi/3,0)\right]\sim 
				\frac{1}{3}\log\Tr \left[e^{-3\beta (H-i\Omega J)}  \right]
				+ S_{\mathfrak{D}}(\beta,-\Omega)\,,\\
			\end{aligned}
			\end{equation}
where $S_{\mathfrak{D}}(\beta,\Omega)$ is given by \eqref{eq:q3p1Sdefect}.

\subsection{More examples with a defect action: $4$d CFTs}
\label{sec:4dwithdefectaction}

In 4d, the rotation group has two Cartan generators, which we call $J_1$ and $J_2$. Let us consider inserting $R=\exp(2\pi i  \frac{p_1}{q_1}J_1+2\pi i  \frac{p_2}{q_2}J_2)$ into the trace. When $q_1=q_2$, the action of $R$ is free and there will be no vortex defects. However, when $q_1\neq q_2$, we will have two 1-dimensional vortex defects $\mathfrak{D}^{(1)}$ and $\mathfrak{D}^{(2)}$, whose combined action is given by \eqref{eq:predictionforthreeddefectaction2}.

As a concrete example, consider the free Dirac fermion in 4d. Using plethystic exponentials, we find 
\begin{equation}
\begin{aligned}
-\log \Tr\left[e^{-\beta (H-i\vec{\Omega}\cdot \vec{J})} e^{2\pi i  \frac{p_1}{q_1}J_1+2\pi i  \frac{p_2}{q_2}J_2}\right]
\sim -\frac{1}{q} \log \Tr\left[e^{-q \beta (H-i\vec{\Omega}\cdot \vec{J}) } \right]+S_{\mathfrak{D}}\,.
\end{aligned}
\end{equation}
Here we impose that $q \left(1+\frac{p_1}{q_1}+\frac{p_2}{q_2}\right)$ is odd.  This ensures absence of the $(-1)^F$ insertion in the right hand side above.  Thus the zero mode is absent and the thermal EFT applies.

Furthermore, $S_{\mathfrak{D}}$ has precisely the form predicted by \eqref{eq:predictionforthreeddefectaction2}
with the leading Wilson coefficients given by the following function of $(-1)^{p}$ and coprime integers $P,Q$:
\begin{equation}
\begin{aligned}
a_{0,P/Q,(-1)^{p}}&=\sum_{k=1}^{Q-1}
    \frac{(-1)^{pk}\cos(\pi k \frac{P}{Q})}
    {8\pi\sin^2(\pi k \frac{P }{Q})}
    \left(\psi'\left(\frac{1}{2}+\frac{k}{2Q}\right)
    -\psi'\left(\frac{k}{2Q}\right)\right),
\end{aligned}
\end{equation}
where $\psi'(z)$ is the derivative of the digamma function.  This is a highly nontrivial check of (\ref{eq:predictionforthreeddefectaction2}). Matching (\ref{eq:predictionforthreeddefectaction2}) requires not just reproducing the correct function of $\beta, \Omega_1,\Omega_2$, but also the fact that the Wilson coefficient of $\mathfrak{D}^{(1)}$ depends only on $P_2/Q_2$ and $(-1)^{p_1}$ (and analogously for $\mathfrak{D}^{(2)}$).

Note that when $q_1=q_2$,  we have $Q_1=Q_2=1$,  hence the sum is non-existent, leading to $a_{0,P_1/Q_1,(-1)^{p_2}}=a_{0,P_2/Q_2,(-1)^{p_1}}=0$,  which is consistent with the fact that the action of $R$ becomes free and $S_{\mathfrak{D}}$ should vanish.

\section{Topological terms: example in 2d CFT}
\label{sec:2dwithgravitationalanomaly}

So far, we have focused on cases where the thermal effective action can be written as the integral of a local gauge- and coordinate-invariant density. As discussed in section~\ref{sec:eftgauge}, we can choose a gauge (``EFT gauge") where the background fields $g,A,\phi$ on the EFT bundle look locally the same as on $S^{d-1} \x S^1_\beta$ (possibly with ``small" angular twists turned on). Curvature invariants built out of these fields are then also the same as on $S^{d-1} \x S^1_\beta$, and their integral is not sensitive to global properties of the EFT bundle.

By contrast, when $d$ is even, the thermal effective action can also include Chern-Simons-type terms that are not integrals of local curvature invariants, and hence can be sensitive to global properties of the EFT bundle. The contributions of such terms were called ``topological" in (\ref{eq:generalresultwhenrisfree}) and (\ref{eq:moregeneralresult}). The coefficients of such Chern-Simons terms can be determined systematically from the anomaly polynomial of the CFT \cite{Loganayagam:2012pz,Jensen:2012kj,Loganayagam:2012zg,Jensen:2013kka,Jensen:2013rga,Ng:2014sqa}. The simplest case is when $d=2$, where the thermal effective action contains a 1d Chern Simons term whose coefficient is proportional to the local gravitational anomaly $c_L-c_R$.

In more detail, consider such a 2d CFT with a local gravitational anomaly, $c_L \neq c_R$. We assume that $c_L - c_R = 24k$ with $k \in \mathbb{Z}$. From modular invariance we have a high-temperature expansion of the partition function as:\footnote{To derive (\ref{eq:anomalymod}), we apply the modular transform (\ref{eq:morecomplicatedmod}) to the vacuum state $e^{-\frac{2\pi i \tau c_L - 2\pi i \bar\tau c_R}{24}}$ with $\tau = \frac{i \beta}{2\pi} + \frac{\beta\Omega}{2\pi} + \frac{p}{q}$ and expand in small $\beta$.}
\begin{equation}
\log\left(\Tr\left[e^{-\beta(H-i\Omega J)} e^{2\pi i \frac pq J}\right]\right) \sim \frac{4\pi^2(c_L + c_R - 24 i k \Omega)}{24q^2(1+\Omega^2)\beta} - \frac{2\pi i k (p^{-1})_q}{q},
\label{eq:anomalymod}
\end{equation}
where (\ref{eq:anomalymod}) is accurate to all orders in perturbation theory in $\beta$. This generalizes (\ref{eq:mostgeneral2dphase}) to theories with $c_L \neq c_R$.

In the thermal effective action, we can reproduce the terms in (\ref{eq:anomalymod}) with a Chern-Simons term from the KK gauge field in the action. In particular, we add a term of the form 
\begin{equation}
-\frac{2\pi i k}{q\beta} \oint A
\label{eq:addtoEFT}
\end{equation}
to the thermal effective action. From (\ref{eq:theholonomy}), we see that (\ref{eq:addtoEFT}) precisely reproduces the additional terms in (\ref{eq:anomalymod}). Note that (\ref{eq:addtoEFT}) is properly quantized precisely when $k$ is an integer, since $A$ is a connection on a circle bundle where the fiber has circumference $q\beta$.

\section{Holographic theories}
\label{sec:holography}

In this section we consider CFTs dual to semiclassical Einstein gravity in AdS.  The high temperature behavior of the thermal partition function, $\Tr[e^{-\beta H+i\vec{\theta}\cdot\vec{J}}]$ around $\theta_i=0$,  of such holographic CFTs in $d$ dimensions are captured by the thermodynamics of black hole solutions in AdS$_{d+1}$.  We would like to understand the bulk solution that captures the high temperature behavior of $\mathrm{Tr}[e^{-\beta H+i\vec{\theta}\cdot\vec{J}}]$ near $\theta_i=2\pi p_i/q_i$,  where at least one of the $p_i\neq 0$ and $q_i\geqslant 2$.

First, let us consider a $d=2$ dimensional CFT in the context of  AdS$_{3}$/CFT$_{2}$ duality.  The relevant bulk solution is given by the rotating BTZ black hole with the metric
\begin{equation}
ds^2=-f(r) dt^2+\frac{dr^2}{f(r)}+r^2\left(d\phi-\frac{r_+r_-}{r^2}dt\right)^2\,,\quad f(r):=\frac{(r^2-r_+^2)(r^2-r_-^2)}{r^2}\,,
\end{equation}
where $r_\pm$ are radius of outer and inner horizon respectively.  The BH temperature $\beta^{-1}$ and angular potential $\Omega=\theta/\beta$ are given by 
\begin{equation}
 \Omega=\frac{r_-}{r_+}\,,\quad \beta^{-1}=\frac{r_+^2-r_-^2}{2\pi r_+}\,.
\end{equation}
Asymptotically, the metric is Weyl equivalent to $-dt^2+d\phi^2$.  In Euclidean signature,  $t_E=it$ and we have $(\phi,t_{E}) \sim (\phi-\beta\Omega,t_E+\beta)$. 
The Euclidean action evaluated on this bulk saddle reproduces the high temperature behavior of $\mathrm{Tr}[e^{-\beta(H-i\Omega J)}]$.

Now let us compute $\mathrm{Tr}[e^{-\beta (H-i\Omega J)+2\pi i p/q J}]$ for a holographic $2$D CFT\@.   As explained in section~\ref{sec:EFTbundle},  this can be computed by doing a path integral over $M_{q\beta,\Omega}/\mathbb{Z}_q$,\footnote{In this section, we use the compressed notation $M_{\beta,e^{iJ\Omega}}\to M_{\beta,\Omega}$.} where the action of $\mathbb{Z}_q=\langle h\rangle$ is given by 
\begin{equation}
h: (\phi,t_E) \mapsto (\phi+2\pi p/q-\beta\Omega,t_E+\beta)\,.
\end{equation}
and $M_{q\beta,\Omega}$ is obtained from $S^1\times \R$ by quotienting by $\langle h^q\rangle$. 

Note that the action of $\Z_q$ has a natural extension into the AdS bulk, where the radial direction goes along for the ride
\begin{equation}\label{eq:bulk}
h: (r, \phi,t_E) \mapsto (r, \phi+2\pi p/q-\beta\Omega, t_E+\beta)\,.
\end{equation}
We can use this natural extension to build a bulk dual solution. We start with a BTZ black hole $\Sigma_{q\beta,\Omega}$ with parameters $\tilde{r}_{\pm}=r_{\pm}/q$,  such that the black hole is at inverse temperature $\tilde{\beta}=q\beta$.   The Euclidean metric is given by
\begin{equation}\label{eq:metricqbeta}
ds^2=\tilde{f}(r) dt_E^2+\frac{dr^2}{\tilde{f}(r)}+r^2\left(d\phi+i\frac{\tilde{r}_+\tilde{r}_-}{r^2}dt_E\right)^2\,,\quad \tilde{f}(r):=\frac{(r^2-\tilde{r}_+^2)(r^2-\tilde{r}_-^2)}{r^2}\,,
\end{equation}
Note that $(\phi,t_E)\sim (\phi-q\beta\Omega,t_E+q\beta)$. We can then quotient $\Sigma_{q\beta,\Omega}$ by the action of $\mathbb{Z}_q=\langle h\rangle$, given by  \eqref{eq:bulk}.  The manifold $\Sigma_{q\beta,\Omega}/\mathbb{Z}_q$ is smooth because the action of $h$ in the bulk (\ref{eq:bulk}) is free.

Recall that on the boundary,  the quotient $M_{q\beta,\Omega}/\Z_q$ is a nontrivial bundle over $S^1/ \Z_q$.  The discussion in~\ref{sec:exampletwodcft} applies here, with the radial direction of AdS going along for the ride.  In short,  we consider open sets in the total space and consider the bulk region that asymptotes to this set.  Locally in this bulk region,  the metric is precisely given by \eqref{eq:metricqbeta}.  However,  there are non-trivial transition functions when we go from one open patch to a different one.

By construction,  the quotient $\Sigma_{q\beta,\Omega}/\Z_q$ solves Einstein's equations with the appropriate asymptotic geometry.  Compared to a rotating BTZ black hole at temperature $q\beta$,  the above quotient geometry has its angular variable restricted to $0< \phi<2\pi/ q$. (Note that this almost covers the full manifold except the locus $\phi=0$. This locus is covered by another open patch and we have nontrivial transition function between these two patches.)  Thus the evaluation of the bulk action amounts to an integral over the angular variable in the range $0< \phi<2\pi/ q$ producing a factor of $1/q$, and a replacement $\beta\to q\beta$ which produces another factor of $1/q$ compared to the evaluation of Euclidean action on BTZ black hole at temperature $\beta$.  Thus, the Euclidean action evaluated on the saddle $\Sigma_{q\beta,\Omega}/\mathbb{Z}_q$ reproduces 
\begin{equation}
\log \mathrm{Tr}[e^{-\beta (H-i\Omega J)+2\pi i p/q J}]\sim \log Z_{\texttt{grav}}\left[\Sigma_{q\beta,\Omega}/\mathbb{Z}_q\right]= \frac{1}{q} \log Z_{\texttt{grav}}\left[\Sigma_{q\beta,\Omega}\right]\sim \frac{1}{q}\log \mathrm{Tr}[e^{-q\beta (H-i\Omega J)}],
\end{equation}
where $Z_{\texttt{grav}}\left[\Sigma\right]$ is the gravitational path integral evaluated on the saddle $\Sigma$.

Overall, we have a quotient of a BTZ black hole  whose boundary has a temporal cycle of length $\beta_L=q\beta(1-i\Omega)$ and a spatial cycle of length $L=\frac{2\pi}{q}$.   This naively leads to the modular parameter
\be
\tl\tau\ \text{``=''} \ i L/\beta_L =\frac{2\pi i}{q^2\beta(1-i\Omega)}.
\ee
The above is almost correct, but we must amend it by recalling the presence of nontrivial transition functions.  This leads to the following identification
\be
\tl\tau= \frac{2\pi i}{q^2\beta(1-i\Omega)}-\frac{(p^{-1})_q}{q}\,,\quad \tl{\bar\tau}=-\frac{2\pi i}{q^2\beta(1+i\Omega)}-\frac{(p^{-1})_q}{q}\,,
\ee
which precisely matches the modular parameter $\tl\tau$ obtained after applying a modular transformation, as explained around \eqref{eq:resultfrommodularinvariance}.
Note that we have
\be
\frac{1}{2}\left(\tl\tau + \tl{\bar\tau}\right)= -\frac{(p^{-1})_q}{q} - \frac{2\pi \Omega}{q^2\beta(1+\Omega^2)}\,,
\ee
which is consistent with the holonomy of $A$ derived in \eqref{eq:theholonomy}. We conclude that $\Sigma_{q\beta,\Omega}/\Z_q$ is simply a modular transformation of the BTZ black hole geometry.

In dimensions greater than two, we follow the same prescription.  Let $\Sigma_{\beta,\vec\Omega}$ be a black hole solution that captures the high temperature behavior of the thermal partition function with small angular fugacity --- i.e.\ the AdS-Kerr black hole with inverse temperature $\beta$ and angular velocities $i\vec\Omega$. We claim that the insertion of a big angular  fugacity $R$ with $R^q=1$ is captured by the bulk manifold $\Sigma_{q\beta,\vec\Omega}/\mathbb{Z}_q$,  where $\mathbb{Z}_q$ is the natural extension of the boundary $\Z_q$ into the bulk. To be precise, the AdS-Kerr solution possesses a time-translation isometry, which we parametrize by $t_E$. It furthermore possesses isometries under the Cartan subgroup of the rotation group, which we parametrize by $\f_a$. Then the $\Z_q=\<h\>/\<h^q\>$ group is generated by
\be
\label{eq:bulkzq}
h:(r,r_a,\f_a,t_E) &\mto (r,r_a,\f_a+\tfrac{2\pi p_a}{q_a},t_E+\beta),
\ee
where $r_a$ and $r$ are the remaining bulk coordinates. 

When the bulk action of $h$ is free, $\Sigma_{q\beta,\vec\Omega}/\mathbb{Z}_q$ is a smooth solution to Einstein's equations with the correct asymptotic geometry to describe the partition function with an insertion of $R$. At very high temperatures, the gravitational path integral on this geometry matches the field theory prediction (\ref{eq:generalresultwhenrisfree}) because the quotient by $\Z_q$ just divides the semiclassical gravitational action by $q$. Thus we expect  $\Sigma_{q\beta,\vec\Omega}/\mathbb{Z}_q$ is the dominant solution at high temperatures. As we move away from high temperatures, we conjecture that $\Sigma_{q\beta,\vec\Omega}/\mathbb{Z}_q$ remains the dominant solution down to a finite temperature.

Unlike in $2$d, in higher dimensions it is possible for the bulk $\mathbb{Z}_q$ action (\ref{eq:bulkzq}) to have a fixed locus. Such a fixed locus must occur at the horizon of the black hole (where the Euclidean time circle degenerates), and at a location on the sphere $S^{d-1}$ where the boundary action of $h$ has a fixed point as well. For example, consider a 3d CFT, with a 4-dimensional bulk dual, and consider the case $p_1/q_1=1/2$. The bulk action of $h$ rotates the $\f$ circle by $\pi$, and also shifts $t_E\to t_E+\beta$ (which is equivalent to rotating the thermal circle by $\pi$).  Fixed points of $h$ occur at the north and south poles of the horizon $S^2$. For example, near the north pole, we can choose coordinates so that the metric locally takes the form
\be
dr_1^2 + r_1^2 d\f^2 + dy^2 + y^2 d\psi^2,
\ee
where $\psi=\frac{2\pi}{2\beta} t_E$, and $y\propto \sqrt{r-r_H}$ (with $r_H$ the location of the horizon). In these coordinates, the $\Z_2$ action rotates both angles $\f,\psi$ by $\pi$. Quotienting by this $\Z_2$, we obtain an orbifold singularity of the form $\C^2/\Z_2$.

When $\Sigma_{q\beta,\vec\Omega}/\mathbb{Z}_q$ contains an orbifold singularity, it no longer furnishes a smooth solution to Einstein's equations. To understand the correct bulk dual of the $R$-twisted partition function, we must understand the fate of the orbifold singularity in the bulk theory. The physics of the singularity (or its resolution) essentially determines the defect action $S_\mathfrak{D}$ in holographic CFTs.  Note that in a spacetime with vanishing cosmological constant, a $\C^2/\Z_2$ singularity can be resolved by the ``gravitational instanton" described by the Eguchi-Hansen metric. Perhaps orbifold singularities occurring in the $\Sigma_{q\beta,\vec\Omega}/\mathbb{Z}_q$ can be resolved similarly. Or perhaps the orbifold singularity is resolved by stringy effects. We leave these questions to future work.

For fermionic theories dual to Einstein gravity (which all known examples are), there is another set of Wilson coefficients $\tilde f$, $\tilde c_1$, \ldots that can be defined (see (\ref{eq:twowilsons})) coming from periodic boundary conditions for the fermion. The behavior of the partition function with $(-1)^F$ inserted was explored in \cite{Chen:2023mbc}, which found a black hole solution that lead to an exponentially subleading (in temperature) contribution in the large $N$ limit. Since generically we expect $\tilde f$ to be nonvanishing, this means the black hole solution found in \cite{Chen:2023mbc} should not be the dominant contribution in the $T \gg N \gg 1$ limit. It would be interesting to explore further if there are universal results or constraints on the Wilson coefficients $\tilde f, \tilde c_1, \ldots$ in (fermionic) holographic CFTs (for instance by looking at the theories studied in \cite{Bobev:2023ggk}). 

Finally, let us note that our construction works also in the case when the boundary theory possesses a reflection symmetry and the group element $R$ includes a reflection. In this case, the bulk dual solution $\Sigma_{q\beta,\vec \Omega}/\Z_q$ is non-orientable. This is allowed because the boundary reflection symmetry must be gauged in the bulk, which means we must include contributions from non-orientable geometries. It so happens that a non-orientable geometry dominates in this case.\footnote{See \cite{Harlow:2023hjb} for a recent discussion of gauging spacetime symmetries in the bulk.}

\section{Journey to $\beta=0$}
\label{sec:journey}

So far we have described the structure of CFT partition functions $Z(\beta, \vec \theta)$ at high temperature near rational angles $\frac{\vec\theta}{2\pi}\in \Q^n$. In this section we will use these results to sketch the behavior of the partition function at high temperature around any (possibly irrational) angle. For simplicity let us consider turning on only one angle $\theta$, and study the partition function
	\begin{equation}
Z(\beta,\theta) = \text{Tr}\left[e^{-\beta H + i J \theta} \right]
	\end{equation}
at small $\beta$ with $\theta$ {\it fixed}.
Suppose the angle has the following continued fraction expansion: 
	\begin{equation}
	\frac{\theta}{2\pi} = a_0 + \frac{1}{a_1 + \frac{1}{a_2 + \frac{1}{a_3 + \ldots}}},
	\end{equation}
	where $a_i \in \mathbb{N}$. We also define the fractions $\frac{p_i}{q_i}$ as truncations of the continued fraction, i.e.
	\begin{equation}
\frac{p_i}{q_i} \coloneqq a_0 + \frac{1}{a_1 + \ldots \frac{1}{a_i}}.
\end{equation}
For a generic angle, the probability of $a_i = n$ scales as $\frac{1}{n^2}$ for large $n$, which means that the distribution of $a_i$'s for a generic angle has infinite mean.\footnote{One way to see this scaling is, for a random real number between $0$ and $1$, the probability that $a_1 = n$ is roughly $\frac{1}{n} - \frac{1}{n+1} \sim \frac{1}{n^2}$.}
	
Suppose we have an angle $\theta$ where $a_i \gg 1$ for some $i$. How does the partition function behave at high temperature? 
Since we assume $a_i \gg 1$, we have $q_i \approx a_i q_{i-1} \gg q_{i-1}$. So for
\begin{equation}
q_{i-1} \ll T \ll q_i,
\label{eq:regime}
\end{equation}
we expect
\begin{equation}
\log Z(T) \sim \vol S^{d-1} f T^{d-1} q_{i-1}^{-d}.
\label{eq:scalingsofar}
\end{equation}
However, when
\begin{equation}
T \sim q_i
\end{equation}
the constant or proportionality in (\ref{eq:scalingsofar}) suddenly shrinks. 

\begin{figure}
\centering
\begin{subfigure}{.5\textwidth}
  \centering
  \includegraphics[width=\linewidth]{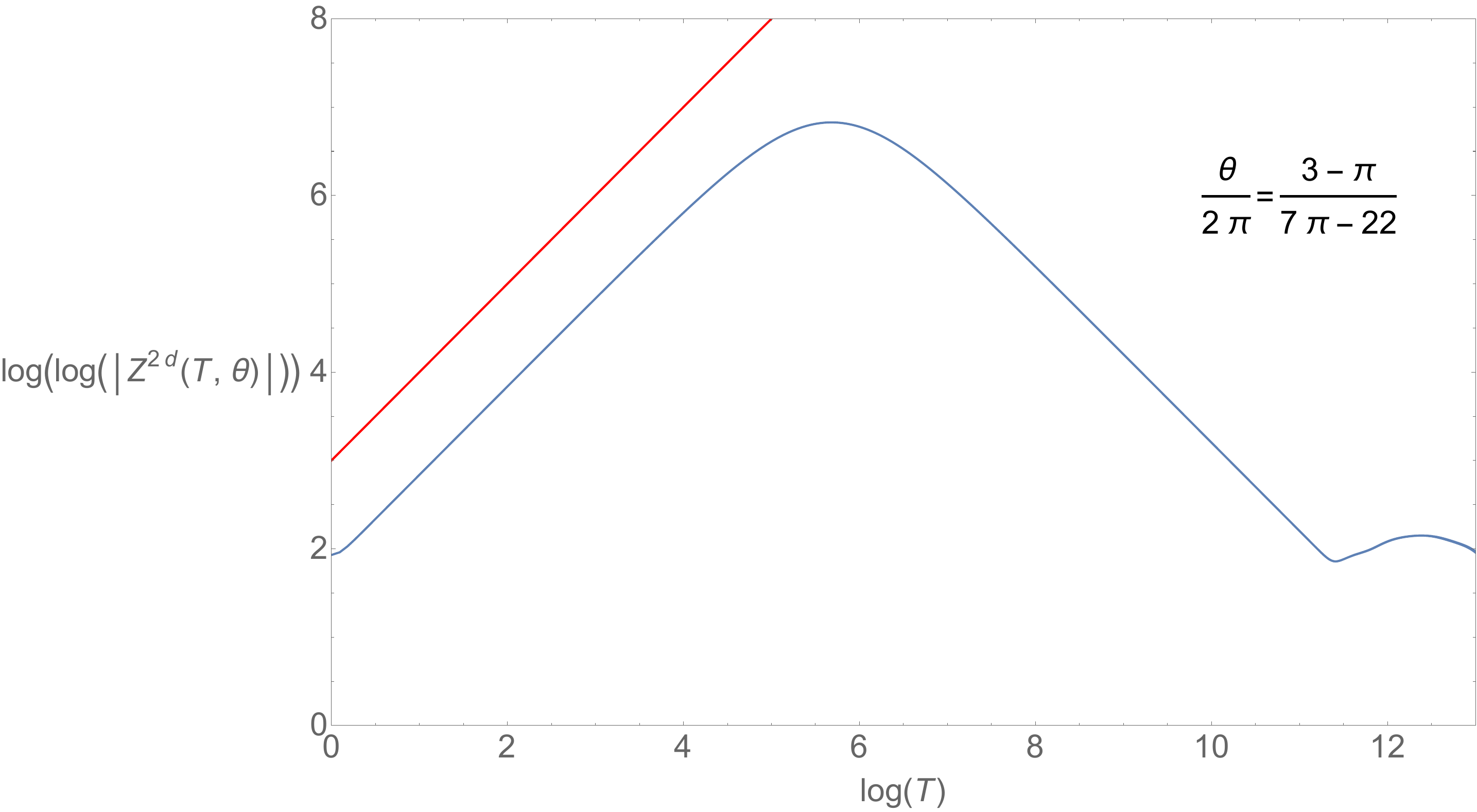}
  \caption{24 free bosons in 2d.}
  \label{fig:sub1}
\end{subfigure}%
\begin{subfigure}{.5\textwidth}
  \centering
  \includegraphics[width=\linewidth]{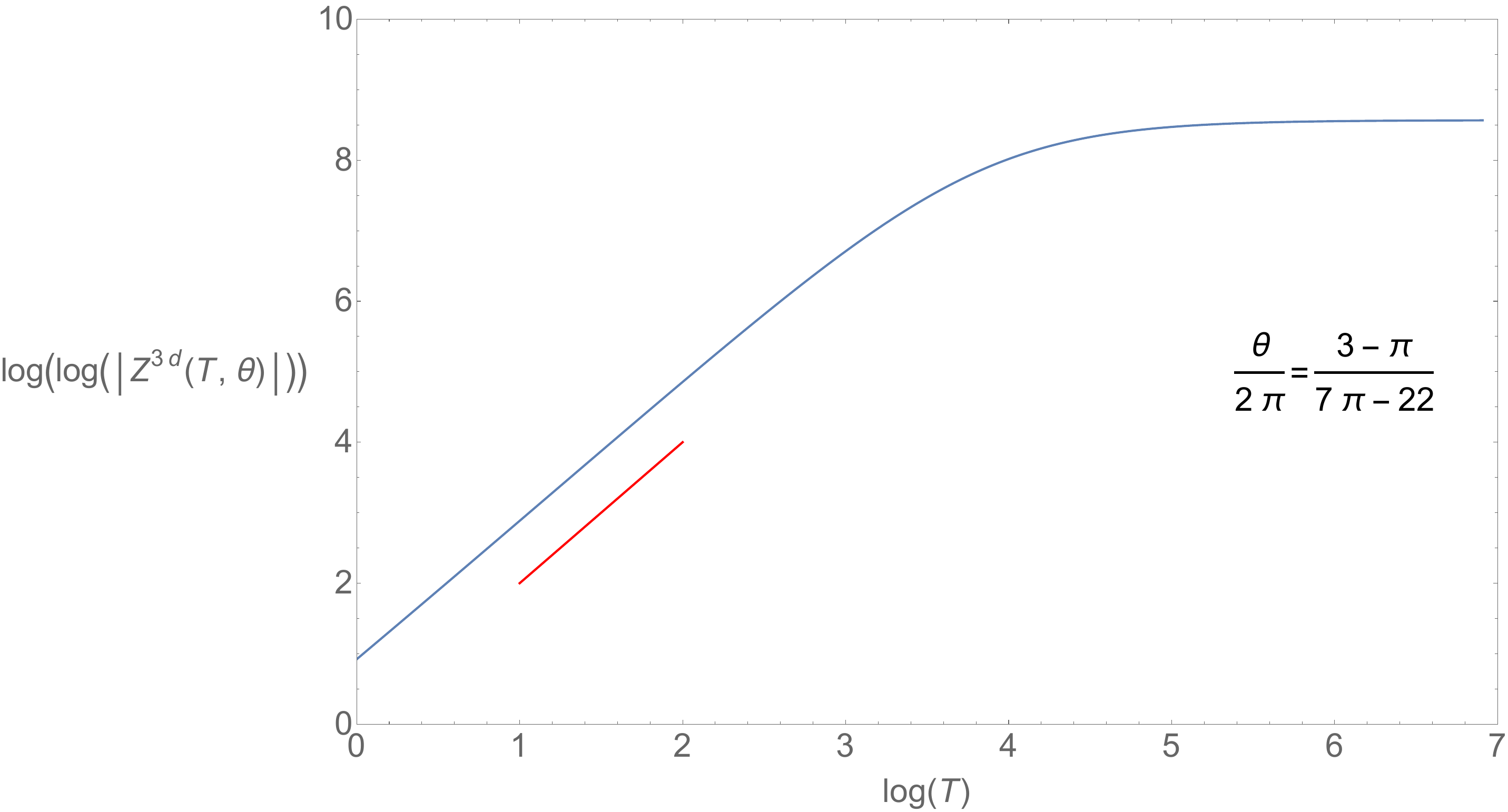}
  \caption{A free boson in 3d.}
  \label{fig:sub2}
\end{subfigure}
\caption{In blue, the log of the free energy of two CFTs evaluated at chemical potential $\frac{\theta}{2\pi} = \frac{3-\pi}{7\pi-22}$ vs. $\log T$. In red, a line with slope $d-1$. (a): $d=2$. (b): $d=3$. We see that at some temperature, there is a region where the slope matches $d-1$, meaning the effective field theory is a good description. If we continue this plot to higher and higher temperatures, there will be infinitely many times the slope matches $d-1$.}
\label{fig:test}
\end{figure}

In figure~\ref{fig:test}, we illustrate this explicitly for two example CFTs. We plot both the Klein invariant $j$ function (which is the partition function of some 2d CFTs at central charge $24$) and the 3d free boson partition function as a function of temperature, with chemical potential $\frac{\theta}{2\pi} = \frac{3-\pi}{7\pi - 22}$. This has the continued fraction expansion:
\begin{equation}
\frac{\theta}{2\pi} = \frac{3-\pi}{7\pi - 22} = 15 + \frac{1}{292 + \frac{1}{1 + \ldots}}.
\label{eq:funnycontfrac}
\end{equation}
(Although any generic real number would serve to illustrate the partition function's behavior, we choose (\ref{eq:funnycontfrac}) for convenience because of the large $292$ showing up immediately.) We see that there indeed is a region of $T$ where the free energies scale as we predict from the effective field theory; but when we increase $T$ further, the scaling breaks down. At large $T$ for generic chemical potential, there will be an infinite number of times the plot in figure~\ref{fig:test} has slope $d-1$, which is when the continued fraction approximation is well-approximated by a rational number (i.e.\ whenever $a_i \gg a_{i-1}$ in the continued fraction). 

\begin{figure}
\centering
\begin{subfigure}{.5\textwidth}
  \centering
  \includegraphics[width=\linewidth]{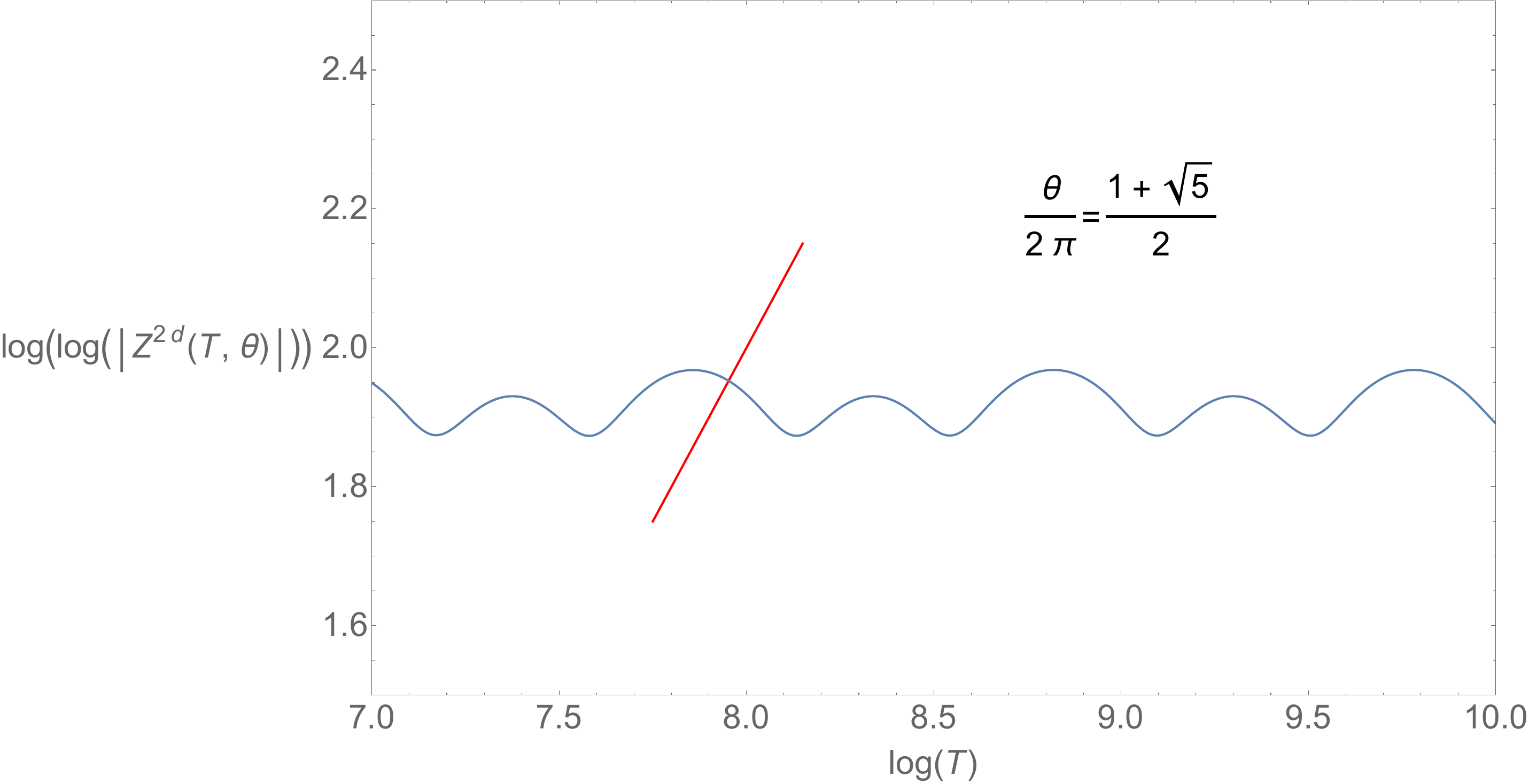}
  \caption{24 free bosons in 2d.}
  \label{fig:sub1phi}
\end{subfigure}%
\begin{subfigure}{.5\textwidth}
  \centering
  \includegraphics[width=\linewidth]{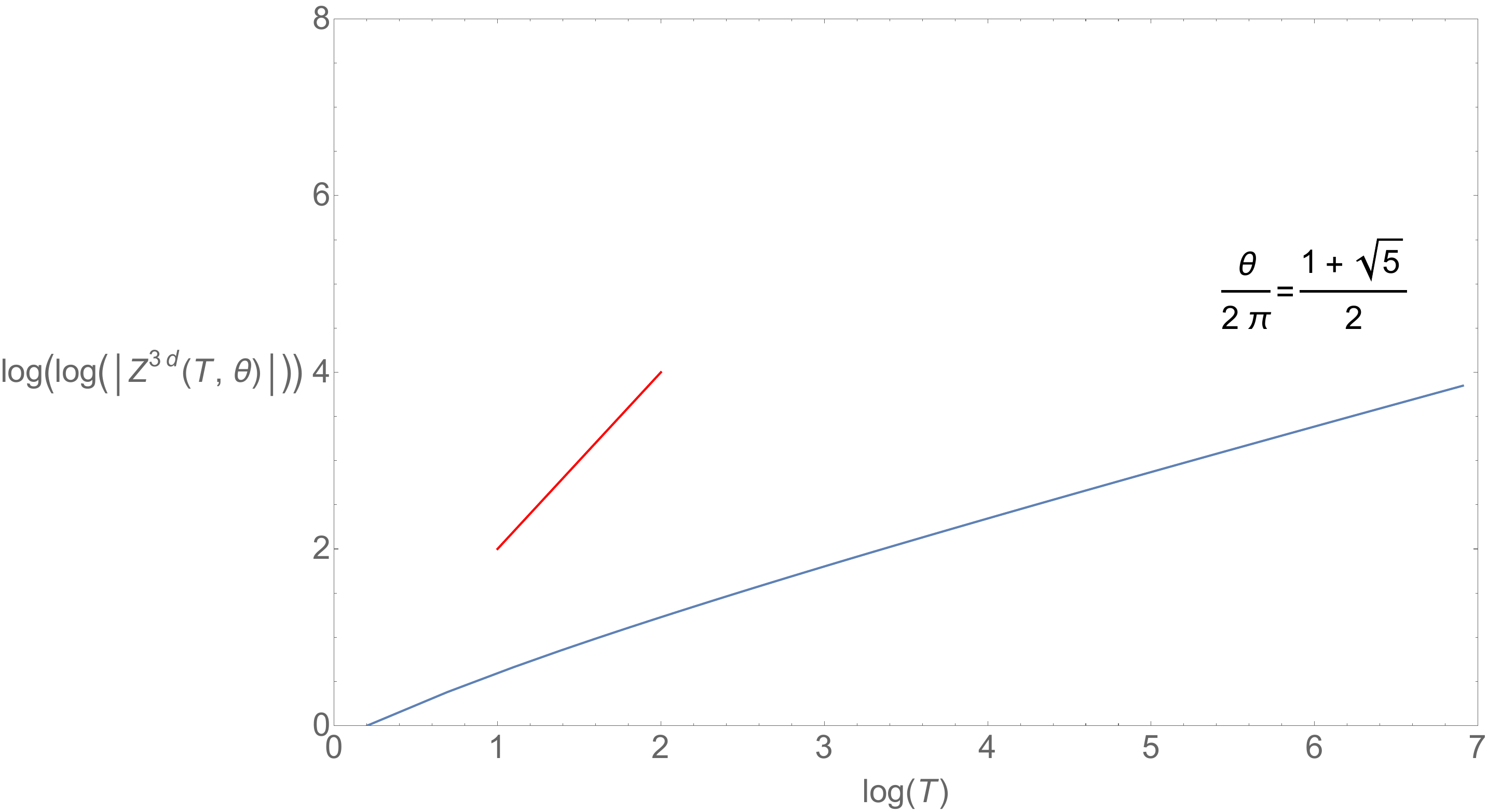}
  \caption{A free boson in 3d.}
  \label{fig:sub2phi}
\end{subfigure}%
\caption{In blue, the log of the free energy of two CFTs evaluated at chemical potential $\frac{\theta}{2\pi} = \frac{1+\sqrt 5}{2}$ vs. $\log T$. In red, a line with slope $d-1$. (a): $d=2$. (b): $d=3$. We see that at no temperature is there a region where the slope matches $d-1$, meaning the effective field theory is never a good description. Since we have fine-tuned the chemical potential to be a real number whose continued fraction has no large numbers in it, we do not guarantee any region in temperature where we have a good EFT description of our theory.}
\label{fig:test2}
\end{figure} 

Finally we note that if we fine-tune the angle $\theta$ so that none of the $a_i$'s ever become large, we can make there be no regime where the partition function obviously grows as $T^{d-1}$. For example, choosing the angle to be the golden ratio 
\begin{equation}
\frac{\theta}{2\pi} = \frac{1+\sqrt{5}}2 = 1 + \frac{1}{1+ \frac{1}{1+ \ldots}}
\end{equation}
gives an angle where there's never a large enough regime in $T$ trust our effective field theory. In figure~\ref{fig:test2}, we again plot the Klein invariant $j$ function and the free boson partition functions as a function of temperature, but this time with the chemical potential set to $\frac{\theta}{2\pi} = \frac{1+\sqrt{5}}2$. We see that the slope of $\log \log |Z|$ against $\log T$ never matches $d-1$ in a large region, so there is no good EFT description of our system.

\section{Nonperturbative corrections}
\label{sec:nonpert}

In this section, we consider nonperturbative corrections to the thermodynamic limit $L\to \oo$ at finite $\beta$. For concreteness, we focus on a CFT$_d$ and its dimensional reduction on $S^1_\beta$ to a $d{-}1$-dimensional gapped theory. By conformal symmetry, the thermodynamic limit is equivalent to $\beta\to 0$ (with a fixed-size spatial manifold). For simplicity, we will not turn on ``small" angular twists $\beta\vec\Omega$, though it would be straightforward to incorporate them.

The thermal effective action essentially captures the dynamics of the ground state of the  $d{-}1$ dimensional gapped theory, while nonperturbative corrections come from particle excitations. On the geometry $\R^{d-1}\x S^1_\beta$, the excitations can be classified into irreps of the $d{-}1$-dimensional Poincar\'{e} group and the Kaluza-Klein $U(1)$ that rotates the $S^1_\beta$. Irreps of the Poincar\'{e} group are labeled by a mass and a little group representation --- for simplicity, we will focus on scalars. Thus, each excitation of interest is labeled by a mass $m_i$ and a KK charge $\mathfrak{q}_i\in \Z$.
The lightest mass $m_{\mathfrak{q}}$ for each KK charge $\mathfrak{q}$ is sometimes called the ``$\mathfrak{q}$-th screening mass", while the lightest nonzero mass overall is the ``thermal mass" of the theory. Note that when $d=2$, the spectrum of masses $(m_i,\mathfrak{q}_i)$ are simply $(\frac{2\pi}{\beta}\De_i,\ell_i)$, where $(\De_i,\ell_i)$ are scaling dimensions and spins of local operators. However, in higher dimensions, the masses $m_i$ are not related in an obvious way to the local operator spectrum.

In the partition function $Z_\textrm{CFT}[\cM_{d-1} \x S^1_\beta]$, the leading nonperturbative effects at small $\beta$ are expected to come from ``worldline instantons" associated with particles of mass $m_i$ propagating along geodesics of $\cM_{d-1}$, see e.g.\ \cite{Dondi:2021buw,Grassi:2019txd,Hellerman:2021yqz,Hellerman:2021duh,Caetano:2023zwe}. Such contributions can be computed from the worldline path integral
\be
\label{eq:worldlineinstantontgeneral}
	\log Z\Big|_{\text{single-particle}}&=\sum_{(m_i,\mathfrak{q}_i)}\int Dx^\mu(\tau)\ \exp\left(-m_i \int d\tau \sqrt{\dot{x}_\mu\dot{x}^\mu}+\frac{2\pi i \mathfrak{q}_i}{\beta}\oint A\right).
\ee
Here, for each particle, we have included a length term $-m_i\int ds$ proportional to the mass, along with a coupling $\frac{2\pi i \mathfrak{q}_i}{\beta}\oint A$ to the background KK gauge field. (Note that we include a factor of $\frac{2\pi}{\beta}$ because in our conventions, $A$ is a connection on a circle bundle where the fiber has circumference $\beta$, instead of the usual $2\pi$.) 

In appendix~\ref{sec:worldlineinstantons}, we compute the wordline path integral (\ref{eq:worldlineinstantontgeneral}) on some geometries of interest. For example, by computing (\ref{eq:worldlineinstantontgeneral}) on $S^{d-1}$, we find that the leading nonperturbative terms in the partition function $Z_\textrm{CFT}[S^{d-1} \x S^1_\beta]$ have the form
\begin{equation}\label{eqn:nonpertMain1}
\log(\Tr_{\mathcal{H}_{S^{d-1}}}[e^{-\beta H}])\supset \sum_{m_i} e^{-2\pi m_i}\frac{(\pm i)^{d-2}m_i^{d-2}}{\Gamma (d-1)}\left(1+ O\left(\frac{1}{m_i}\right)\right).
\end{equation}
Note that the effect of each particle is exponential in the mass $e^{-2\pi m_i}$, where $2\pi$ is the length of a great circle on $S^{d-1}$. By dimensional analysis, the masses are proportional to $1/\beta$, and hence these are indeed nonperturbative corrections in $\beta$. In addition to the exponential dependence, the worldline path integral makes an unambiguous prediction for the leading coefficient of the exponential, coming from a gaussian determinant. We immediately see that the interpretation of (\ref{eqn:nonpertMain1}) is subtle because the coefficient becomes {\it imaginary\/} when $d$ is odd (even when the partition function must be real). We discuss this phenomenon and its interpretation in appendix~\ref{section:oddDResurgence}.

Specializing to 4d, we can similarly compute leading nonperturbative corrections to a partition function on a lens space $L(q;1)$, coming from a ``short" geodesic of length $2\pi/q$:
\begin{equation}\label{eqn:nonpertMain2}
	\log(\Tr_{\mathcal{H}_{L(q;1)}}[e^{-\beta H}])\supset \begin{cases}\sum_{m_i}
	e^{-\frac{2\pi m_i}{q}} \frac{m_i}{2\sin(\frac{2\pi }{q})}\left(1+O\left(\frac{1}{m_i}\right)\right)  & (\text{$q\neq 2$}),
	\\
	\sum_{m_i}
	e^{-\pi  m_i}\ \frac{m_i^2}{2}\left(1+O\left(\frac{1}{m_i}\right)\right) & (\text{$q=2$}).
\end{cases}
\end{equation}
By our discussion of the EFT bundle, this result is closely related to the partition function on $S^3$, with a twist by a rational angle of order $q$. To obtain the latter, we must replace $\beta\to q\beta$, and account for the presence of a nontrivial holonomy for the KK gauge field (since $S^1_{q\beta}$ is nontrivially fibered over $L(q;1)$). We find
\begin{equation}\label{eqn:nonpertMain3}
	\log(\Tr_{\mathcal{H}_{S^{3}}}[e^{-\beta H-\frac{2\pi i}{q}J_{12}-\frac{2\pi i}{q}J_{34}}])	 \supset\begin{cases} \sum_{(m_i,\mathfrak{q_i})}
		e^{-\frac{2\pi m_i}{q^2}+\frac{2\pi i  \mathfrak{q}_i }{q}} \frac{m_i}{2q\sin(\frac{2\pi }{q})}\left(1+O\left(\frac{1}{m_i}\right)\right) &  (\text{$q\neq 2$}),\\
		\sum_{(m_i,\mathfrak{q}_i)}
	e^{-\frac{\pi  m_i}{2}+\pi i \mathfrak{q}_i}\ \frac{m_i^2}{8}\left(1+O\left(\frac{1}{m_i}\right)\right) & (\text{$q=2$}).
		\end{cases}
\end{equation}
In particular, nonperturbative corrections to the twisted partition function go like $e^{-2\pi m_i/q^2}$, as opposed to $e^{-2\pi m_i}$ in the un-twisted case.

More generally, in any dimension $d$, when the quotient by $\Z_q$ creates a short geodesic of length $\ell_\textrm{short} = 2\pi/q$, the leading nonperturbative corrections will behave like $e^{-(m_i/q) \ell_\textrm{short}} = e^{-2\pi m_i/q^2}$, where $m_i/q$ comes from the replacement $\beta\to q \beta$. Note that this matches the result from modular invariance in 2d. In 2d, after applying a modular transformation, the twisted partition function becomes $\mathrm{Tr} \left[e^{2\pi i \tl \tau \left(L_0-\frac{c}{24}\right)-2\pi i \tl{\bar\tau} \left(\bar L_0-\frac{c}{24}\right)} \right]$, where $\tl\tau = \textrm{const} + \frac{2\pi i}{q^2 \beta}$. Thus, the leading $\beta$-dependence of the contribution of excited states to the twisted partition function is $e^{-4\pi^2 \De_i/(q^2\beta)}=e^{-2\pi m_i/q^2}$, where $m_i=\frac{2\pi \De_i}{\beta}$.

Note that the action (\ref{eq:worldlineinstantontgeneral}) is only the leading approximation to the effective action of a worldline instanton in the small $\beta$ limit. In particular, there can be power-law corrections in $\beta$ coming from higher curvature terms. Thus, while the tree-level and $1$-loop terms (\ref{eqn:nonpertMain1}), (\ref{eqn:nonpertMain2}), and (\ref{eqn:nonpertMain3}) in the worldline path integral are universal, the subleading corrections in $1/m_i$ are not necessarily universal, since they get contributions both from (computable) loops and from higher curvature terms.\footnote{Furthermore, our choice of $\zeta$-function regularization does not in general respect coordinate-invariance of the worldline path integral. To restore coordinate invariance one must add non-coordinate-invariant counterterms like $g^{\mu\nu} g^{\rho\tau}g_{\alpha\beta}\Gamma^{\alpha}_{\mu\rho}\Gamma^{\beta}_{\nu\tau}$ with the appropriate coefficients. See \cite{Bastianelli_van_Nieuwenhuizen_2006} for discussion.}

In appendix~\ref{sec:nonpertmore}, we derive \eqref{eqn:nonpertMain1}, \eqref{eqn:nonpertMain2} and \eqref{eqn:nonpertMain3} by performing the worldline path integral explicitly. We also verify the universal leading terms in several examples from free field theory.

\section{Discussion and future directions}
\label{sec:discuss}

In this work, we found that the high-temperature partition function, twisted by a finite-order discrete rotation $R$, is captured by the same thermal EFT as the un-twisted partition function. One consequence is that ``spin-refined" densities of states (like the difference between the density of even-spin and odd-spin operators) are determined by the same Wilson coefficients as the usual density of states, up to subleading contributions from Kaluza-Klein vortices. Furthermore, the partition function $Z(\beta,\vec\theta)$ itself has an intricate fractal-like structure as a function of angle at small $\beta$, with the same universal asymptotics controlling the neighborhood of every rational angle.

These results follow entirely from effective field theory, together with the assumption that generic CFTs develop a mass gap when dimensionally-reduced. It would be interesting to revisit this assumption and understand how our results are modified in the presence of potential gapless modes. It would also be interesting to investigate whether Kaluza-Klein vortex defects can support gapless excitations, contribute to Weyl anomalies, and/or have nontrivial topological terms in their effective action\footnote{See e.g. \cite{Benini:2012ui, Closset:2018ghr} for a study of similar vortex defects in the context of supersymmetric quantum field theories.}.

Our central construction is simple: it is essentially the observation that it is useful to construct a mapping torus $M_{\beta,R}=(\cM_L \x \R)/\Z$ from two successive quotients: first quotienting by $q\Z$, and then by $\Z_q=\Z/q\Z$. This idea is applicable on other geometries besides $S^{d-1} \x \R$, and it would be interesting to explore its implications for other types of partition functions. For example, one could explore ``spin-refined" statistics of OPE coefficients by studying the behavior of discrete spacetime symmetries on the ``genus-2" geometry of \cite{Benjamin:2023qsc}, or spin-refined lens-space partition functions \cite{Razamat:2013opa}, or the interaction of discrete spacetime symmetries with other forms of higher-dimensional ``modular invariance" \cite{Shaghoulian:2015kta,Shaghoulian:2016gol,Luo:2022tqy,Allameh:2024qqp}.  Supersymmetric partition functions have been studied on a wide variety of geometries, see e.g.\ \cite{Pestun:2016zxk}. It is an enduring challenge to understand observables of non-supersymmetric (potentially nonperturbative) theories on these geometries.

 One can also consider applying thermal EFT to BCFTs to study the asymptotic spectra of boundary operators.  In $2$ dimensions, this boils down to studying the partition function on a finite cylinder in the $\beta\to 0$ limit and writing down an EFT on a finite interval with two end points.  The end points will become point-like defects in the thermal effective action.  In higher dimensions, by introducing defects one may break the $SO(d)$ invariance down to some subgroup $H$.  We can imagine turning on a rotation belonging to $H$ and applying thermal EFT ideas in this context.

So far, our main tool for computing partition functions has been thermal EFT, which is organized in an expansion in small $\beta$. This expansion is likely asymptotic in general. In fact we can see its asymptotic nature explicitly in odd-dimensional free theories. It is an important question whether one can obtain more precise results about high temperature partition functions, potentially including convergent expansions and/or numerical bounds (as is possible in 2d using the modular bootstrap \cite{Hellerman:2009bu}).

One possible approach is through a better understanding of resurgence in the small-$\beta$ expansion. In particular, it would be nice to better understand the structure of nonperturbative terms beyond the worldline instantons discussed in section~\ref{sec:nonpert}. We expect that in an interacting theory, there should also be contributions from an infinite sum of ``instanton graphs," representing massive particles propagating and interacting. An old example of instanton graphs are L\"uscher corrections in field theories on torii \cite{Luscher:1985dn,Luscher:1986pf}. However, to our knowledge, the rules for general instanton graphs on general geometries in general massive QFTs have not been spelled out.

Relatedly, modular invariance in 2d CFTs constrains some of the nonperturbative behavior of the partition function. Given some input light spectra, modular invariance highly constrains the resulting spectra, which in effect forces the partition functions with any phase inserted to behave in a certain way. Techniques such as Poincare series and Rademacher series have been used to complete the light spectrum of a 2d CFT (see e.g.\ \cite{Dijkgraaf:2000fq, Maloney:2007ud, Keller:2014xba, Alday:2019vdr}), which roughly take the form of a (convergent) sum over rational angles. It would be interesting if there were related techniques in higher dimensions, resumming all rational angle insertions in the partition function trace.

Another potential avenue to making the high-temperature expansion precise is using Tauberian techniques, which have been applied successfully to correlation functions \cite{Pappadopulo:2012jk,Qiao:2017xif,Das:2020uax,Pal:2022vqc} and torus partition functions in 2d \cite{Pal:2019zzr,Mukhametzhanov:2020swe,Pal:2020wwd}.  An essential ingredient in Tauberian methods is positivity,\footnote{See \cite{Marchetto:2023xap} where, even if OPE coefficent can become negative, Tauberian theorems were used along with some boundedness conditions from below,  to predict asymptotics of OPE coefficent averaged over a large microcanonical window. However, it is not clear how to extend the result rigorously for an order one window.  The same theorem is used in \cite{Pal:2019yhz} as well.} which has not yet played an important role in applications of thermal EFT.

\section*{Acknowledgements}	

We thank Arash Ardehali, Luca Delacretaz, Jake McNamara, Hirosi Ooguri, Edgar Shaghoulian, Joaquin Turiaci,  and Yifan Wang for helpful discussions. We thank Yifan Wang for very helpful comments on a draft.
This material is based upon work supported by the U.S. Department of Energy, Office of Science, Office of High Energy Physics, under Award Number DE-SC0011632. In addition, 
NB and SP are supported in part by the Sherman Fairchild Foundation. 
DSD, JL, and YX are supported in part by Simons Foundation grant 488657 (Simons Collaboration on the Nonperturbative Bootstrap).

\appendix

\section{Qualitative picture of the 3d Ising partition function}
\label{sec:pictureappendix}

In this appendix, we explain our procedure for producing figure~\ref{fig:isingcartoon}. Let $S_{p/q}$ denote the leading contribution to the thermal effective action around the angle $\frac{2\pi p}{q}$ in 3d:
\be
-S_{p/q} &= \frac{1}{q^3} \frac{f \vol\, S^2}{\beta^2 + (\theta-\frac{2\pi p}{q})^2}.
\ee
We expect $-S_{p/q}$ to be a good estimate for the action when it is large. Thus, to patch together different EFT descriptions, we roughly want to choose the action $-S_{p/q}$ that is largest for each $(\beta,\theta)$. This would lead to the approximation $-S\approx \max_{p,q} -S_{p/q}$, where $p,q$ runs over co-prime pairs of integers. However, such an estimate would not be smooth, so we instead combine the different actions with a root-mean-square:
\be
\label{eq:sapprox}
-S\approx \p{\sum_{p,q} (-S_{p/q})^2}^{1/2}.
\ee
In figure $1$, we plot $\log \log Z = \log (-S)$, for $S$ given in (\ref{eq:sapprox}), with coprime pairs of integers up to denominator $15$. We use the value of minus the free energy density $f\approx 0.153$ determined from Monte-Carlo simulations \cite{vasilyev2009universal, PhysRevE.53.4414,PhysRevE.56.1642}. 

Note that the approximation (\ref{eq:sapprox}) has some unrealistic features. Firstly, its expansion around each rational angle contains subleading corrections in $\beta$ that do not conform with the expectation from thermal EFT (\ref{eq:moregeneralresult}). Secondly, it does not incorporate the nonperturbative effects discussed in section~\ref{sec:nonpert}. Our goal with this approximation is simply to give a qualitative picture of the partition function. It is interesting to ask whether there is a natural basis of functions for $Z$ that naturally incorporates these constraints.

In principle, the qualitative features of figure~\ref{fig:isingcartoon} can be checked. For instance, one can explicitly build the partition function of the 3d Ising model with a phase (e.g.\ $(-1)^J$) inserted, by simply using the operator scaling dimensions that have been estimated from the conformal bootstrap (or other methods). One can then plot it as a function of $\beta, \Omega$, and check that, for instance, the leading Wilson coefficient is approximately $\frac{0.153}{8}$ (similar to what was done in appendix A of \cite{Iliesiu:2018fao} and appendix D of \cite{Benjamin:2023qsc}). An important technical obstacle we run into when attempting this is: when computing the partition function with the phase $(-1)^J$, the EFT is valid when $2\beta \ll 1$ (rather than $\beta \ll 1$), so in effect, one needs to keep more operators in the partition function to get a trustworthy estimate. It would be interesting if other techniques to estimate the scaling dimensions of the 3d Ising model know about enough high energy operators to see explicitly the qualitative features of figure~\ref{fig:isingcartoon}.

\section{Review of plethystic sums}\label{appendix:highTexpansion} 
In this appendix, we review some facts about the high temperature expansion of plethystic sums. 
 Plethysic sums can be written as derivatives of various spectral zeta functions and to compute 
the high temperature expansion, one ``resums'' the zeta functions following a procedure very similar to (\ref{eqn:resumExample}). The same resummation technique is also used to compute the high temperature expansion of 
massive free field free energy, as we discussed in the section~\ref{sec:closerlookatfreetheory}.
\paragraph*{Plethystic sums and spectral zeta functions}

We start with some general discussion. Let $f(\beta)=\sum_{k} d_k \ e^{-\beta \lambda_k}$ be a generating function.  We define a few related quantities:
\begin{itemize}
\item Plethystic sum: \be\log(\text{PE}[f]):=\sum_{n=1}\frac{1}{n}f(n \beta). \ee
\item Spectral zeta function: \be\zeta(s;f):=\sum_{n\in \mathbb{Z}}\sum_{k} d_k\left[\left(\frac{2\pi n}{\beta}\right)^2+\lambda_k^2\right]^{-s}.\ee
\item Heat trace: \be H(t;f):=\sum_{k} d_k \ e^{-t \lambda_k^2}.\ee
\end{itemize}
The plethystic sum is related to the spectral zeta function in a simple way:
\be\label{eqn:claim1}\log(\text{PE}[f])-\frac{\beta}{2}\sum_{k} d_k \lambda_k=\frac{1}{2}\frac{d}{ds}\Big|_{s=0}\zeta(s;f).\ee
where the second term on the left hand side is the zeta function regularized Casimir energy.\footnote{Note that the Casimir energy 
computed using zeta function regularization is NOT the same as the Casimir energy included in appendix~\ref{sec:4d6dwithoutdefectaction}. See section 3.1 in \cite{Benjamin:2023qsc} for 
a discussion of schemes and \cite{Herzog:2013ed} for the difference with zeta function regularization.}

To derive \eqref{eqn:claim1}, we start by using the Schwinger parametrization to obtain:
\be\begin{split}
\zeta(s;f)&=\sum_{n\in \mathbb{Z}}\sum_k d_k \left[\left(\frac{2\pi n}{\beta}\right)^2+\lambda_k^2\right]^{-s}=\frac{1}{\Gamma(s)}\sum_{n\in\mathbb{Z}}\sum_{k} d_k \int_{0}^\infty \frac{dt}{t} t^s e^{-t(\frac{2\pi n}{\beta})^2-t\lambda_k^2}\\
&=\frac{1}{\Gamma(s)}\frac{\beta}{2\sqrt{\pi}}\sum_{m\in\mathbb{Z}}\sum_{k} d_k \int_{0}^\infty \frac{dt}{t}\ t^{s-\frac{1}{2}}\ e^{\frac{m^2\beta^2}{4 t}-t\lambda_k^2},
\end{split}\ee
where in the last line we've Poisson resummed with respect to $n$.
Now we split the summation over $m$ into an ``excited part" with $m\neq 0$ and a ``vacuum part" where $m=0$:
\be
 \zeta_{\text{excited}}(s;f)&=\frac{1}{\Gamma(s)}\frac{\beta}{\sqrt{\pi}}\sum_{m=1}^\infty \sum_{k} d_k\int_{0}^\infty \frac{dt}{t} \ t^{s-\frac{1}{2}}\ e^{\frac{m^2\beta^2}{4 t}-t\lambda_k^2},\nn\\
\zeta_{\text{vacuum}}(s;f)&=\frac{1}{\Gamma(s)}\frac{\beta}{\sqrt{\pi}}\sum_{k} d_k\int_{0}^\infty \frac{dt}{t} \ t^{s-\frac{1}{2}}\ e^{-t\lambda_k^2}.
\ee
Performing the integral over $t$, we get:
\be\begin{split}\frac{d}{ds}\Big|_{s=0}\zeta_{\text{excited}}(s;f)&=\frac{\beta}{\sqrt{\pi}} \sum_{m=1}^\infty \sum_{k}d_k\int_{0}^\infty\frac{dt}{t}\ t^{-1/2} \ e^{-\frac{m^2\beta^2}{4t}-t\lambda_k^2}\\&=2\sum_{m=1}^\infty\frac{1}{m}\sum_{k}d_k e^{-\beta m \lambda_k}=2 \log(\text{PE}[f]).\end{split}\ee
For $\zeta_{\text{vacuum}}(s;f)$, we invert the Schwinger parametrization: 
\be
\frac{d}{ds}\Big|_{s=0}\zeta_{\text{vacuum}}(s;f)&= \frac{\beta}{2\sqrt{\pi}}\Gamma(-1/2)\sum_{k}d_k \lambda_k=-\beta\sum_{k}d_k \lambda_k.
\ee
Putting everything together we arrive at (\ref{eqn:claim1}).

To compute partition functions with flavor twists and rational angular fugacities we need to consider plethystic sums taking the following form:
\be
\sum_{n=1}^\infty \frac{e^{-in\theta}}{n} F(n\beta,\omega_q^n), \ \ \ \ \omega_q=\exp\left(\frac{2\pi i}{q}\right).
\ee
Some formal manipulation shows that when the ``twisted generating function'' $F(\beta,\omega_q)$ satisfies $F(\beta,\omega_q)=F(\beta,\omega_q^{-1})$, the following relation
holds:
\be\label{eqn:claim2}
\begin{split}
\sum_{n=1}^\infty \frac{e^{-in\theta}}{n} F(n\beta,\omega_q^n)&+\sum_{n=1}^\infty \frac{e^{in\theta}}{n} F(n\beta,\omega_q^n)=\frac{1}{q}\frac{d}{ds}\Big|_{s=0}\zeta\lbrack q\theta,s;F(q\beta,1)\rbrack+\frac{\beta}{q}\sum_{k}d_k \lambda_k\\
&+\frac{1}{2q}\sum_{\ell=1}^{q-1}\frac{d}{ds}\Big|_{s=0}\left[ L\left(q \theta,\frac{\ell}{q},s;F(q\beta,\omega_q^{\ell})\right)+L\left(-q \theta,\frac{\ell}{q},s;F(q\beta,\omega_q^{\ell})\right)\right].
\end{split}
\ee
where:
\begin{itemize}
	\item In the second sum on the right hand side, $d_k$ and $\lambda_k$ are supposed to be read from $F(q\beta, 1)$, that is, $F(q\beta,1)=\sum_{k=0}d_k e^{-\beta \lambda_k}$
	\item \be\label{eqn:zetaFlavor}  \zeta(\theta,s;f)=\sum_{n\in \mathbb{Z}}\sum_k d_k \left[\left(\frac{2\pi n}{\beta}+\frac{\theta}{\beta}\right)^2+\lambda_k^2\right]^{-s}, \ \ \ \ \text{with } f(\beta)=\sum_{k=0} d_k e^{-\beta \lambda_k}.\ee
	\item \be  L(\theta,a,s;f)=\sum_{n\in \mathbb{Z}}\sum_{k}d_k \frac{e^{2\pi i n a}}{\left[\left(\frac{2 \pi n}{\beta}+\frac{\theta}{\beta}\right)^2+\lambda_k^2\right]^s} , \ \ \ \ \text{with } f(\beta)=\sum_{k=0} d_k e^{-\beta \lambda_k}.\ee
\end{itemize}
Now that we've related different plethystic sums to various spectral zeta functions,the high temperature expansion can be computed by ``resumming'' all these zeta functions.
For concreteness, we look at some examples:
\paragraph*{Example: free scalar on $S^1\times S^3$}

The relevant generating function is\footnote{This is the spectral generating function of the operator $\Delta+\frac{(d-2)}{4(d-1)}R$ with $\Delta$ being the Laplacian on $S^3$ and $d=4$}
\begin{equation}
	f(\beta)=\frac{e^{-\beta}(1-e^{-2\beta})}{(1-e^{-\beta})^4}=\sum_{k=1} k^2 e^{-\beta k}.
\end{equation}
The spectral zeta function is:
\be\label{eqn:spectralZetaSum}\zeta(s;f)=\frac{1}{\Gamma(s)}\sum_{n\in\mathbb{Z}} \int_{0}^\infty \frac{dt}{t} t^s e^{-t(\frac{2\pi n}{\beta})^2} H(t;f).\ee
with 
$$ H(t;f)=\sum_{k=1}^{\infty}k^2 e^{-tk^2}=\frac{\sqrt{\pi}}{4}t^{-3/2}+\sqrt{\pi}\sum_{l=1}^\infty e^{-\ell^2\pi^2/t}\left(\frac{1}{2}t^{-3/2}-\ell^2\pi^2 t^{-5/2}\right).$$
Here we performed a Poisson resummation with respect to $k$ and separated the piece with $\ell=0$ from $\ell\neq 0$.
\begin{itemize}
	\item The term with $n=0$ in the sum (\ref{eqn:spectralZetaSum}) is the contribution of zero mode:
	\begin{equation}
		\begin{split}
	&\zeta(s;f)\Big|_{n=0}=\sum_{k=1}^\infty d_k \lambda_k^{-s}=\sum_{k=1}^\infty k^{2-2s}=\zeta_R(2s-2).
		\end{split}
	\end{equation}
	\item Terms with $n\neq 0$ and $\ell=0$ generate the perturbative series:
	\be
     \zeta(s;f)\Big|_{n\neq0,\ell=0}=\frac{\sqrt{\pi}}{4\Gamma(s)}\sum_{n\in \mathbb{Z},n\neq0}\int_{0}^{\infty}\frac{dt}{t} \ t^s\ e^{-t(\frac{2\pi n}{\beta})^2} t^{-3/2}=\sqrt{\pi} \frac{\Gamma(s-\frac{3}{2})}{2 \Gamma(s)} \left( \frac{2\pi}{\beta}\right)^{3-2 s}\zeta_R(2s-3). 
	\ee
	\item Terms with $n\neq 0$ and $\ell\neq 0$ generate the non-perturbative corrections:
	{\footnotesize\be
	\begin{split}
	\zeta(s;f)\Big|_{n\neq0,\ell\neq0}=\frac{\sqrt{\pi}}{\Gamma(s)}\sum_{n=1}^\infty \sum_{\ell=1}^\infty\left[ 2^{\frac{5}{2}-s}\left(\frac{n}{\ell\beta}\right)^{\frac{3}{2}-s}K_{-\frac{3}{2}+s}\left(\frac{4 \pi^2 \ell n}{\beta}\right)-2^{\frac{9}{2}-s}\pi^2 \ell^2 \left(\frac{n}{\ell\beta}\right)^{\frac{5}{2}-s} K_{-\frac{5}{2}+s}\left(\frac{4\pi^2 \ell n }{\beta}\right)\right].
    \end{split}
	\ee}
\end{itemize}
Following (\ref{eqn:claim1}), we find:
\be
\log(\text{PE}[f])-\frac{\beta}{240}=\frac{\pi^4}{45\beta^3}-\frac{\zeta(3)}{4 \pi^2}-\frac{1}{2\pi^2}\sum_{n=1}^\infty\sum_{\ell=1}^\infty e^{-\frac{4 \pi^2 n \ell}{\beta}}(\frac{1}{\ell^3}+\frac{4 n \pi^2}{\ell^2 \beta}+\frac{8 n^2 \pi^4}{\ell\beta^2}).
\ee
With the same $d_k$ and $\lambda_k$, we can construct a slightly different zeta function:
\be\label{eqn:massiveZeta}
\zeta_m(s)=\sum_{k=0}^{\infty}\frac{d_k}{\left(\lambda_k^2+m^2\right)^s}.
\ee
this spectral zeta function satisfies
\be \frac{1}{2}\frac{d}{ds}\Big|_{s=0}\zeta_m(s)=\log(Z_{\text{massive}}[S^3])\ee where $Z_{\text{massive}}[S^3])$ is given in eqn(\ref{eqn:massivePartitionfunction}).
Applying the same resummation procedure, we get (\ref{eqn:freeMassive4d}).
The lens space partition functions $ Z_{\text{massive}}[L(q;1)]$ can be computed in an almost identical way. One simply replace $d_k=k^2$ by
 \cite{Asorey:2012vp}:
\be
d_k&=
\begin{cases}
	k\left[\frac{k}{q}\right],& k-q\left[\frac{k}{q}\right]\in 2\mathbb{Z},\\
    k\left(\left[\frac{k}{q}\right]+1\right),& k-q\left[\frac{k}{q}\right]\in 2\mathbb{Z}+1,
    \end{cases}
    &&(q\in 2\mathbb{Z}+1),
\nn\\
d_k &=   
\begin{cases} 
	0,&  k\in 2\mathbb{Z},\\
	k\left(2\left[\frac{k}{q}\right]+1\right),& k\in 2\mathbb{Z}+1,
\end{cases}
&&(q\in 2\mathbb{Z}).
\ee
which are the degeneracy for $\Delta+\frac{(d-2)}{4(d-1)}R$ with $\Delta$ being the Laplacian in $L(q;1)$ and $d=4$. The same
calculations will yield (\ref{eqn:oddLensSapceFreePartitionFunctionComplete}) and (\ref{eqn:evenLensSapceFreePartitionFunctionComplete}). 
 \paragraph*{Example: free scalar on $S^1\times S^5$}\ 
 \\
 The relevant generating function is:
 $$f(\beta)=\frac{q^2(1-q^2)}{(1-q)^6}=\sum_{k=1}^\infty \frac{k^2(k^2-1)}{12}e^{-\beta k}.$$
 and the heat trace is:
 \begin{equation}
\begin{split}
H(t;f)&=\frac{1}{24}\sum_{k\in \mathbb{Z}}k^2(k^2-1) e^{-t k^2}\\
&=\sqrt{\pi}\left(\frac{1}{32 t^{5/2}}-\frac{1}{48 t^{3/2}}\right)+\sqrt{\pi}\sum_{l=1}^\infty e^{-\frac{l^2 \pi^2}{t}}\left[\frac{\pi ^4 l^4}{12 t^{9/2}}-\frac{\pi ^2 l^2}{4 t^{7/2}}+\frac{1}{t^{5/2}}\left(\frac{\pi ^2
   l^2}{12}+\frac{1}{16}\right)-\frac{1}{24 t^{3/2}}\right].
\end{split}
\end{equation}
Following the same procedure as in the example above, we obtain:
 \begin{align}
&\log(\text{PE}[f]-\frac{31}{60480}\beta=\frac{2 \pi^6}{945 \beta^5}-\frac{\pi^4}{540 \beta^3}+\frac{\zeta(5)}{16 \pi^4}+\frac{\zeta(3)}{48 \pi^2}\\&+\sum_{n=1}^\infty \sum_{\ell=1}^\infty e^{-\frac{4 \pi^2 n \ell}{\beta}}
\left[\frac{4 \pi ^4 n^4}{3 \beta ^4 \ell}+\frac{4 \pi ^2 n^3}{3 \beta \ell^2}+\frac{1}{\beta ^2}\left(\frac{n^2}{\ell^3}+\frac{\pi ^2 n^2}{3\ell}\right)+\frac{1}{\beta}\left(\frac{n}{2 \pi ^2 \ell^4}+\frac{n}{6 \ell^2}\right)+\frac{1}{8 \pi ^4 \ell^5}+\frac{1}{24 \pi ^2 \ell^3}\right].
\end{align}
if we replace $\zeta(s;f)$ by $\zeta_m(s)$, we reproduce (\ref{eqn:freeMassive6d}).
\paragraph*{Example: free scalar on $S^1\times S^3$ with $\mathbb{Z}_2$ flavor twist}\ 
\\
The plethystic sum in interest is:
\be
\sum_{n=1}^\infty \frac{(-1)^n}{n} f(n\beta),\quad f(\beta)=\frac{e^{-\beta}(1-e^{-2\beta})}{(1-e^{-\beta})^4}=\sum_{k=1}k^2 e^{-\beta k}.
\ee
Setting $q=1$ and $\theta=\pi$ in (\ref{eqn:claim2}), we find that:
\be
\sum_{n=1}^\infty \frac{(-1)^n}{n} f(n\beta)=\frac{1}{2}\frac{d}{ds}\Big|_{s=0}\zeta(\pi,s;f)+\frac{\beta}{2}\sum_{k}d_k \lambda_k.
\ee
where $\zeta(\pi,s;f)$ is defined in (\ref{eqn:zetaFlavor}). After resumming the zeta function, we find:

\begin{equation} \sum_{n=1}^\infty \frac{(-1)^n}{n} f(n\beta)-\frac{\beta}{240}=-\frac{7 \pi^4}{360 \beta^3}-\sum_{n=1}^\infty\sum_{\ell=1}^\infty e^{-\frac{(4n-2)\ell\pi^2}{\beta}}\left(\frac{(2n-1)^2 \pi^2}{\ell \beta^2}+\frac{2n-1}{\ell^2 \beta}+\frac{1}{\ell^3 \pi^2}\right).\end{equation}
Finally, we discuss twisted partition function with rational angular fugacity. These are the main objects of interest in this paper. Even though the techniques outlined in this appendix can be very easily generalized to 
higher dimensions and to more complicated combinations of angles, we will focus on 4d scalar partition functions twisted by $e^{-\frac{2\pi i}{q}J_{12}-\frac{2\pi i}{q}J_{34}}$ for simplicity. As explained in section~\ref{sec:closerlookatfreetheory}, the 
subtle difference between the lens space partition function and the twisted partition function is related to the non-trivial topology of the EFT bundle. 

\paragraph{Example: free scalar in $S^1\times S^3$ with $\mathbb{Z}_q$ rotation, $q$ odd} \ 
\\
We focus on the twisted partition function $\log\left(\Tr[e^{-\beta H-\frac{2\pi i }{q}J_{12}-\frac{2\pi i }{q}J_{34}}]\right)$ and restrict 
to the case where $q$ is odd. The plethystic sum in interest is:
\be
\sum_{n=1}^\infty\frac{1}{n} F(n\beta,\omega_q^n),\quad F(\beta,\omega_q)=\frac{e^{-\beta}(1-e^{-2\beta})}{(1-\omega_q e^{-\beta})^2(1-\omega_q^{-1} e^{-\beta})^2}.
\ee
It is easy to verify:
\be 
F(q \beta,\omega_q^\ell)=\begin{cases}\sum_{k=1}^\infty k^2 e^{-q \beta  k}, \quad \ell=0 \\  \\ \sum_{k=1}^\infty \frac{k \sin(\frac{2\pi k \ell}{q})}{\sin(\frac{2\pi \ell}{q})}e^{-q\beta k}, \quad \ell\neq 0 \end{cases}
\ee
Setting $\theta=0$ in (\ref{eqn:claim2}),we find:
\be \begin{split}
\sum_{n=1}^\infty\frac{1}{n} F(n\beta,\omega_q^n)&=\frac{1}{2q}\frac{d}{ds}\Big|_{s=0}\zeta\lbrack 0,s;F(q\beta,1)\rbrack+\frac{\beta}{q}\sum_{k}d_k \lambda_k\\
&+\frac{1}{2q}\sum_{\ell=1}^{q-1}\frac{d}{ds}\Big|_{s=0}L\left(q \theta,\frac{\ell}{q},s;F(q\beta,\omega_q^{\ell})\right).
\end{split}\ee
where $d_k=k^2$, $\lambda_k=qk$ for the second sum on the right hand side. After resumming all the zeta functions, we find:
\begin{equation}\label{eqn:PEexpansionqodd}
\begin{split}
   &\sum_{n=1}^{\infty}\frac{1}{n} \frac{e^{-n\beta}(1-e^{-2n\beta})}{(1-\omega_q^n e^{-n\beta})^2(1-\omega_q^{-n} e^{-n\beta})^2}=\frac{ \pi ^4}{45 \beta ^3 q^4}+\frac{\beta}{240}-\frac{\zeta(3)}{4 \pi^2 q}\\
   &-\frac{1}{q}\sum_{n=1}^{\infty}\sum_{\ell=1}^{\infty} e^{-\frac{4 \pi ^2 \ell n}{\beta  q}}\left(\frac{1}{2 \pi ^2 \ell^3}+\frac{2 n}{\beta  \ell^2 q}+\frac{4 \pi ^2 n^2}{\beta ^2 \ell q^2}\right)\\
    &+\sum_{n=1}^\infty\sum_{\substack{\ell=1\\ \ell\notin q\mathbb{Z}}}^\infty\frac{ \cos(\frac{2\pi n \ell}{q})}{\sin(\frac{2\pi \ell}{q})}e^{-\frac{4 \pi^2 n \ell}{q^2 \beta}}\left(\frac{2\pi n}{\ell q \beta}+\frac{q}{ 2 \ell^2 \pi }\right)\\
    &+\sum_{\substack{\ell=1\\ \ell\notin q \mathbb{Z}}}^\infty\frac{1}{4 \pi \ell^2} \frac{1}{\sin(\frac{2 \pi \ell}{q})}.
    \end{split}
\end{equation}
\paragraph*{Example: free scalar in $S^1\times S^3$ with $\mathbb{Z}_q$ rotation and $\mathbb{Z}_2$ flavor twist, $q$=odd}\ \ \ \ \ \ 
\\
The relevant plethystic sum is:
\be 
\sum_{n=1}^\infty \frac{(-1)^n}{n} F(n\beta,\omega_q^n),\quad  F(\beta,\omega_q)=\frac{e^{-\beta}(1-e^{-2\beta})}{(1-\omega_q e^{-\beta})^2(1-\omega_q^{-1} e^{-\beta})^2}.
\ee
Setting $\theta=\pi$ in (\ref{eqn:claim2}) and resum all the zeta functions, we get:
\begin{equation}\label{eqn:PEexpansionqoddZ2twist}
\begin{split}
&\sum_{n=1}^{\infty}\frac{(-1)^n}{n} \frac{e^{-n\beta}(1-e^{-2n\beta})}{(1-\omega_q^n e^{-n\beta})^2(1-\omega_q^{-n} e^{-n\beta})^2}=-\frac{7 \pi^4}{360 \beta^3 q^4 }+\frac{\beta}{240}\\
&-\sum_{n=1}\sum_{\ell=1}e^{-\frac{2 \ell(2n-1) \pi^2}{q \beta}}\left(\frac{(2n-1)^2 \pi^2}{\ell q^3 \beta^2}+\frac{2n-1}{\beta \ell^2 q^2}+\frac{1}{2 \ell^3 q \pi^2}\right)\\
&+\sum_{\substack{\ell=1\\\ell\notin q\mathbb{Z}}}^\infty \sum_{n=0}^\infty\frac{\cos(\frac{2 \pi \ell}{q}(\frac{q-1}{2}-n))}{\sin(\frac{2\pi \ell}{q})} e^{-\frac{2 \ell(2n+1)\pi^2}{q^2 \beta}}\left(\frac{(2n+1)\pi}{\ell q \beta}+\frac{q}{2 \ell^2 \pi}\right).
\end{split}
\end{equation}

\section{More examples for 4d and 6d CFTs}\label{sec:4d6dwithoutdefectaction}

In this appendix, we compute a few more examples of the high temperature expansion of CFTs with a large rotation inserted. We will focus on insertions $R$ without fixed points (so that $S_{\mathcal{D}}$ vanishes). In odd $d$, $SO(d)$ necessarily has a nontrivial fixed locus, so we focus on even $d$. In even $d$, if we insert $R = \exp\left(2\pi i \left(\frac{p_1}q J_1 + \dots + \frac{p_n}{q} J_n\right)\right)$ (namely all the denominators are equal, with $\text{gcd}(p_i, q)=1$), then $R$ has no fixed points and there is no defect action. Previously in section~\ref{sec:4dwithdefectaction}, we described $4$d examples with a defect coming from $q_1 \neq q_2$; in this example we will consider the (simpler) case with no defect action. 

We note that the formulas in this section are accurate to all orders in perturbation theory in $\beta$; this is because the free energy of free theories in even dimensions truncates at $O(\beta)$. This is an accident of free theories in even dimensions and does not generalize.

Finally, we note that in even dimensions, the Hamiltonian $H=D+\varepsilon_0$ includes a contribution from the Casimir energy, which is not accounted for in the sum over characters using plethystic exponentials (see section 3.1 of \cite{Benjamin:2023qsc} for details on this scheme, \cite{Herzog:2013ed} for a calculation of the Casimir energy, and e.g.\ \cite{Giombi:2014xxa} for values of the $a$ anomaly). We have included this contribution to the final expression (although terms linear in $\beta$ are invariant when acting on with (\ref{eq:moregeneralresult})).

 \paragraph*{Example: 4d $\mathbb{Z}_2$-twisted complex scalar} \ 
\\
Here we consider a 4d free complex scalar. We insert a $\mathbb{Z}_2$ twist (which we denote as $e^{\pi i Q}$) where we identify the field $\phi$ with $-\phi$ as we go around the thermal circle, in order to remove the gapless sector. Without any insertion of a rotation $R$, we get the following free energy: 
\begin{equation}
 \begin{aligned}
				&\log\ \Tr \left[ 
				e^{-\beta (H-i\Omega_1 J_1-i\Omega_2 J_2)} e^{\pi i Q} \right]\\
				&\sim\frac{1}{\prod_{i=1}^2(1+\Omega_i^2)}
				\left(
					-\frac{7\pi^4}{180\beta^3}
					-\frac{\pi^2\sum_{i=1}^2\Omega_i^2}{36\beta}
					+\frac{(
					3+ 
					6\sum_{i=1}^2\Omega_i^2 
					- 10\Omega_1^2\Omega_2^2 + 6\sum_{i=1}^2\Omega_i^4)\beta}{720}
				\right).
\end{aligned}
\end{equation}

Now let us consider a rotation by $2\pi/3$ in each Cartan direction. As expected, when we insert the large rotation, we find:
\begin{equation}
		\log\ \Tr \left[e^{-\beta (H-i\Omega_1 J_1-i\Omega_2 J_2)} e^{\pi i Q} e^{2\pi i \left(\frac{J_1}{3} + \frac{J_2}{3}\right)} \right]
		\sim
		\frac{1}{3}\log\ \Tr \left[ e^{-3\beta (H-i\Omega_1 J_1-i\Omega_2 J_2)} e^{3 \pi i Q} \right].
\end{equation}

\paragraph*{Example: 4d Dirac fermion} \
\\
As discussed in section~\ref{sec:fermions}, for fermionic theories we need to compute the partition function without and with a $(-1)^F$ insertion: 
\begin{equation}\begin{aligned}
			&\log\ \Tr \left[e^{-\beta (H - i\Omega_1 J_1 - i\Omega_2 J_2)}\right]\\
			&\sim\frac{1}{\prod_{i=1}^2(1+\Omega_i^2)}
			\left(
				\frac{7\pi^4}{90\beta^3}
				-\frac{\pi^2(3+\sum_{i=1}^2\Omega_i^2)}{36\beta}
				+\frac{(
				18+ 
				6\sum_{i=1}^2\Omega_i^2 
				+10\Omega_1^2\Omega_2^2 - 21\sum_{i=1}^2\Omega_i^4)\beta}{1440}
			\right)
\end{aligned}\end{equation}
\begin{equation}\begin{aligned}
			&\log\ \Tr \left[ (-1)^F e^{-\beta  (H - i\Omega_1 J_1 - i\Omega_2 J_2)}
			e^{4\pi i Q/3}\right]\\
			&\sim\frac{1}{\prod_{i=1}^2(1+\Omega_i^2)}
			\left(
				\frac{52\pi^4}{1215\beta^3}
				-\frac{\pi^2(3+\sum_{i=1}^2\Omega_i^2)}{54\beta}
				+\frac{(
				18+  
				6\sum_{i=1}^2\Omega_i^2 
				+10\Omega_1^2\Omega_2^2 - 21\sum_{i=1}^2\Omega_i^4)\beta}{1440}
			\right).
\end{aligned}\end{equation} 

Fermions carry $U(1)$ charge.  We use a convention where the particle carries charge $+1$ and anti-particle carries charge $-1$.  To make sure that the zero mode does not contribute in the thermal effective field theory description,  we turned on the fugacity corresponding to $U(1)$ and set it to $e^{4\pi i/3}$,  along  with a $(-1)^F$ insertion.   This amounted to turning on $\mathbb{Z}_3\subset U(1) $ flavor symmetry for the partition function with a $(-1)^F$ insertion.
 After turning on a large rotation $e^{2\pi i J_1 p_1/q}e^{2\pi i J_2 p_2/q}$, we get the following results (exactly as predicted in section~\ref{sec:fermions}): 
\begin{align}
			&\log\ \Tr\ \left[ e^{-\beta(H - i\Omega_1 J_1 - i\Omega_2 J_2)} e^{2\pi i \left(\frac{J_1}{3} + \frac{J_2}{3}\right)}\right]
			\sim
			\frac{1}{3}\log\ \Tr \left[e^{-3\beta (H - i\Omega_1 J_1 - i\Omega_2 J_2)}\right]\\
			&\log\ \Tr \left[e^{-\beta(H-i\Omega_1J_1-i\Omega_2J_2)}e^{2\pi i Q/3}e^{2\pi i \left(\frac{J_1}{2} + \frac{J_2}{2}\right)}\right]
			\sim
			\frac{1}{2}\log\ \Tr \left[ (-1)^F e^{-2\beta  (H - i\Omega_1 J_1 - i\Omega_2 J_2)}e^{4\pi i Q/3}\right].
\end{align}

 \paragraph*{Example: 6d twisted complex scalar}\ 
\\
Here we consider a 6d free scalar with a $\mathbb{Z}_2$ twist:
{\footnotesize
\begin{equation}\begin{aligned}
			&\log\ \Tr \left[e^{-\beta (H-i\sum_{k=1}^3 \Omega_k J_k) }e^{\pi i Q}\right]\\
			&\sim\frac{1}{\prod_{i=1}^3(1+\Omega_i^2)}
			\left(
				-\frac{31\pi^6}{7560\beta^5}
				+\frac{7\pi^4(1-\sum_{i=1}^3\Omega_i^2)}{2160\beta^3}
				+\frac{\pi^2(8\sum_{i=1}^3\Omega_i^2 
				- 5\sum_{cyc}\Omega_1^2\Omega_2^2 - 3\sum_{i=1}^3\Omega_i^4)}{2160\beta}
			\right.\\
			&\left.
				-\frac{(
				37+  
				62\sum_{i=1}^3\Omega_i^2
				-154\sum_{cyc}\Omega_1^2\Omega_2^2 - 104\sum_{i=1}^3\Omega_i^4
				+ 70\Omega_1^2\Omega_2^2\Omega_3^2 
				+ 42\sum_{i=1}^3\Omega_i^2\sum_{i=1}^3\Omega_i^4
				-22\sum_{i=1}^3\Omega_i^6
				)\beta}{60480}
			\right).
\end{aligned}\end{equation}
}
With the insertion of a large rotation, we get:
\begin{equation}
			\log\ \Tr \left[e^{-\beta (H-i\sum_{k=1}^3 \Omega_k J_k) }e^{\pi i Q} e^{2\pi i \left(\frac{J_1}{3} + \frac{J_2}{3} + \frac{J_3}{3}\right)} \right]
			\sim
			\frac{1}{3} \log\ \Tr \left[e^{-3\beta (H-i\sum_{k=1}^3 \Omega_k J_k) }e^{3\pi i Q}\right]
\end{equation}

\paragraph*{Example: 6d Dirac fermion}\
\\
Finally we consider a 6d free Dirac fermion. We get:
		
{\footnotesize
\begin{equation}\begin{aligned}
			&\log\ \Tr \left[e^{-\beta (H-i\sum_{k=1}^3 \Omega_k J_k) }\right]\\
			&\sim\frac{1}{\prod_{i=1}^3(1+\Omega_i^2)}
			\left(
					\frac{31\pi^6}{1890\beta^5}
					-\frac{7\pi^4(5+\sum_{i=1}^3\Omega_i^2)}{1080\beta^3}
					+\frac{\pi^2(135 + 26\sum_{i=1}^3\Omega_i^2 
					+ 10\sum_{cyc}\Omega_1^2\Omega_2^2 
					- 21\sum_{i=1}^3\Omega_i^4)}{4320\beta}
			\right.\\
			&\left.
					+\frac{(
					-880+ 
					55\sum_{i=1}^3\Omega_i^2
					-14\sum_{cyc}\Omega_1^2\Omega_2^2 -743\sum_{i=1}^3\Omega_i^4
					- 70\Omega_1^2\Omega_2^2\Omega_3^2 
					+ 147\sum_{i=1}^3\Omega_i^2\sum_{i=1}^3\Omega_i^4
					- 302\sum_{i=1}^3\Omega_i^6
					)\beta}{120960}
			\right).
\end{aligned}\end{equation}
} 
With an insertion of a rotation $R$, we get the following:
\begin{align}
\log\ \Tr \left[e^{-\beta(H- i\sum_{k=1}^3\Omega_k J_k)} e^{2\pi i \left(\frac{J_1}{2} + \frac{J_2}{2} + \frac{J_3}{2}\right)} \right]
\sim
\frac{1}{2}\log\ \Tr \left[ e^{-2\beta  (H - i\sum_{k=1}^3\Omega_k J_k)}\right]
\end{align}

\section{More on nonperturbative terms}
\label{sec:nonpertmore}

In this appendix, we discuss more details on the nonperturbative parts of the partition function outlined in section~\ref{sec:nonpert}. In appendix~\ref{sec:worldlineinstantons}, we derive equations \eqref{eqn:nonpertMain1}, \eqref{eqn:nonpertMain2} and \eqref{eqn:nonpertMain3} by performing the worldline path integral. In appendices~\ref{sec:closerlookatfreetheory}, \ref{sec:functionaldeterminants}, and~\ref{section:oddDResurgence}, we verify the universal leading terms in several examples from free field theory.

\subsection{Worldline path integral}
\label{sec:worldlineinstantons}

\subsubsection{Worldline path integral on the sphere}

\begin{figure}
\centering
\begin{tikzpicture}[
    decoration={markings, mark=at position 0.5 with with {\arrow[line width=1pt]{latex}}},
    dashdot/.style={dash pattern=on 4pt off 3pt on 1pt off 3pt}
]
\draw (0,0) circle (2cm);
\draw[ postaction=decorate](-2,0) arc (180:360:2 and 0.6);
\draw[dashed] (-2,0) arc (180:0:2 and 0.6);
\node at (0,-1.) {$\theta(\tau)$};
\draw[->] (-0.8,-0.55)--(-0.8,0.1);
\node at (-0.2,0.1) {$y^a(\tau)$};
\end{tikzpicture}
\caption{\label{fig:coordinatesforsphere}Coordinates on $S^{d-1}$. The classical path is $\theta(\tau)=2 \pi \tau$ and  perpendicular fluctuations $y^a(\tau)$ are 
integrated over.}
\end{figure}
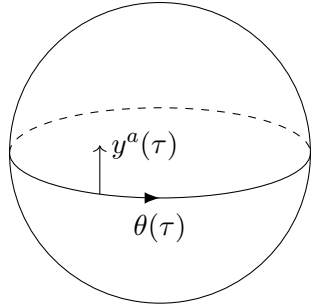

Let us begin by computing the worldline path integral on $S^{d-1}$.
Let $\theta$ denote a coordinate along a great circle and let $y^i$ ($i=1,\dots,d-2$) parametrize the perpendicular directions, see figure~\ref{fig:coordinatesforsphere}. The metric on $S^{d-1}$ takes the form:
\begin{equation}
ds^2=dy^2+\cos^2|y|\, d\theta^2.
\end{equation}

In the absence of a background gauge field, a single worldline instanton contribution is given by
\begin{align}\label{ng-1}
  \int Dx^\mu(\tau)\ e^{-m_i\int ds} 
   &=\int Dy^i(\tau)\  \exp\left[-m_i \int_0^1 d\tau \sqrt{\cos^2|y|\left(\frac{d\theta}{d\tau}\right)^2+\left(\frac{dy}{d\tau}\right)^2}\right]
   \nn\\
    &\sim e^{-2 \pi m_i}\int Dy^i(\tau) \exp\left[- m_i \int_0^1 d\tau \left( \frac{\dot{y}^2}{4 \pi }-\pi  y^2\right)\right],
\end{align}
where in the last line we Taylor-expanded the square root in fluctuations around the great circle.\footnote{This approximation is allowed as long as the thermal wavelength $\sqrt{2\pi/m}$ is much smaller than the size of the sphere.} We also used reparameterization invariance of the Nambu-Goto action to set $\theta(\tau)=2 \pi \tau$.  In the path integral, we sum over periodic paths  $y^i(\tau)$ satisfying $y^i(0)=y^i(1)$, with the following mode expansion:
\begin{equation}
y^i(\tau)=\sum_{n=1}^\infty a_n^i \sin(2 \pi n \tau)+\sum_{n=1}^\infty b_n^i \cos(2 \pi n \tau)+\frac{b_0^i}{\sqrt{2}}.
\end{equation}
In writing this mode expansion, we have defined the mode numbers and coefficients $a^i_n,b^i_n$ so that the measure on the moduli space of geodesics is locally $\prod_i da_1^i db_1^i$. We furthermore imposed that each mode should be unit-normalized under the inner product $\<y,y\>=2\int d\tau |y|^2$. The path integral can then be reduced to a product of ordinary integrals over $a_n^i$, $b_n^i$:\footnote{The path integral measure $Dy(\tau)$ is induced from a choice of inner product. This inner product must be local --- i.e.\ an integral over $\tau$ --- in order for the path integral to satisfy proper cutting and gluing rules. However, the overall coefficient of the inner product does not matter because $\det(\textrm{const})=\prod_{n\in \Z}\textrm{const} = 1$ in $\zeta$-function regularization. As long as the modes are all normalized the same way with respect to the inner product, the measure factorizes into an integral over each mode coefficient $Dy=\prod_i db_0^i \prod_{n,i} da_n^i db_n^i$.}
\begin{align}\label{eqn:PImeasure}
   & \int Dy^i(\tau) \exp\left[- m_i \int_0^1 d\tau \left( \frac{\dot{y}^2}{4 \pi}-\pi y^2\right)\right]
   \nn\\
   &=\prod_{i=1}^{d-2}\left\{\prod_{n=1}^\infty \int da_n^i db_n^i\ \exp\left[{-\frac{m_i \pi }{2}\left(n^2-1\right)\left((a_n^i)^2+(b_n^i)^2\right)}\right]\times \int db_0^i \exp\left[{\frac{m_i\pi }{2}}(b_0^i)^2\right]\right\}.
\end{align}

Modes with $n>1$ are massive and correspond to oscillations around the geodesic. The modes with $n=0$ are tachyonic. After adding these tachyonic fluctuations, the trajectory of the particle will deviate 
from the great circle and shrink towards a point, reducing its length (see e.g.\ figure 4  and the surrounding discussions in \cite{Dondi:2021buw}).
Consequently, the classical trajectory we are expanding around is an {\it unstable\/} saddle point of the Nambu-Goto action.  To define the integral over the tachyonic directions, we need to make sense of integral of the form $\int_{-\infty}^\infty dx \ e^{\alpha^2 x^2}$ where $\Re(\alpha^2)>0$.  We do so by analytically continuing from the region where $\Re(\alpha^2)<0$ and defining:
	\begin{equation}\label{eqn:analyticContinuation}
		\int_{-\infty}^{\infty} dx\ e^{\alpha^2 x^2}:=(\pm i) \frac{\sqrt{\pi}}{|\alpha|},\ \ \ \ \Re{\alpha^2>0}.
	\end{equation}
	
Furthermore,  we make the same choice of sign in  (\ref{eqn:analyticContinuation}) for integrals over all such tachyonic modes.
This prescription, however, does not remove all ambiguities. Since there are $d-2$ perpendicular directions and each 
of them contributes a single tachyonic mode, the overall phase factor is $(\pm i)^{(d-2)}$. When $d$ is even, both choices of sign give the 
same result. But when $d$ is odd, the ambiguity remains. As we will see in appendix~\ref{section:oddDResurgence}, this remaining two-fold 
ambiguity is related to a contour ambiguity in performing a Borel resummation. 

The zero modes with $n=1$ represent rotations of the great circle. The corresponding mode coefficients $a^i_1,b^i_1$ are local coordinates on the moduli space of geodesics. Every geodesic on $S^{d-1}$ is an intersection of $S^{d-1}$ with a 2-plane in $\mathbb{R}^d$ containing the origin, so this moduli space is the same as the Grassmannian $\textrm{Gr}(2,\R^d)$. Hence, the zero mode integral is the volume of this Grassmannian:
\begin{equation}
    \prod_{i=1}^{d-2}\int da_{1}^i db_1^{i}=\text{vol}(\text{Gr}(2,\mathbb{R}^d))=\frac{\text{vol}(O(d))}{\text{vol}(O(2))\times \text{vol}(O(d-2))}=\frac{\text{vol}(S^{d-1})\times \text{vol}(S^{d-2}) }{2 \text{vol}(S^1)}.
\end{equation}
We can finally explicitly calculate the path-integral over $y^i(\tau)$:
\begin{align}
& \int Dy^i(\tau) \exp\left[- m_i \int_0^1 d\tau \left( \frac{\dot{y}^2}{4 \pi}-\pi  y^2\right)\right]
\nn\\
&=\text{vol}(\text{Gr}(2,\mathbb{R}^d))\prod_{i=1}^{d-2}\left[(\pm i)\sqrt{\frac{2}{m_i }}\times {\prod_{n=1}^{\infty}}^{\prime}\frac{2 }{m_i}\ {\prod_{n=1}^{\infty}}^{\prime}\frac{1}{n^2-1}\right]
\nn\\
&=
\frac{(\pm i)^{d-2}m_i^{d-2}}{\Gamma (d-1)},
\end{align}
where ${\prod_{n}}^{\prime}$ means we skip the mode $n=1$. We computed the infinite product above using zeta function regularization \cite{2020EPJWC.24401008A}. In particular, the following formulas are useful for such computations:
\begin{equation}
	\prod_{n=1}^\infty n= \sqrt{2\pi}, \quad \textrm{and}\quad \prod_{n=1}^\infty a=a^{-1/2} \quad (a>0).
\end{equation}
Putting everything together, we obtain (\ref{eqn:nonpertMain1}).

\subsubsection{Worldline path integral on lens space}

Three dimensional homogeneous lens space $L(q;1)$ can be defined as the quotient of the three sphere $S^3=\left\{(z_1,z_2)\in \mathbb{C}^2\big| |z_1|^2+|z_2|^2=1\right\}$ by the equivalence relation $(z_1,z_2)\sim (e^{2\pi i /q} z_1,e^{2\pi i /q} z_2)$.
Geodesics in $L(q;1)$ comes in two types: the contractable ``long geodesics'' with length $2\pi$, and the non-contractable ``short geodesics'' with length
$2\pi/q$. If $q\neq 2$, each short geodesic is an intersection of a complex line 
$\alpha z_1+\beta z_2=0$ with $S^3$, so the corresponding moduli space is $\text{Gr}(1,\mathbb{C}^2)$. When $q=2$, every real two-plane $\alpha z_1+\beta z_2+\gamma \bar{z}_1+\delta \bar{z}_2$ in $\mathbb{R}^4$ intersects $S^3$ at a short geodesic,
so the corresponding moduli space is $\text{Gr}(2,\mathbb{R}^4)$.

To compute worldline instanton contributions to the lens space partition function, we choose the classical trajectory $\theta(\tau)$ to wind around 
a short geodesic and integrate over perpendicular fluctuations obeying a slightly different boundary condition:
\begin{equation}
\begin{pmatrix}y^1(1) \\ y^2(1)\end{pmatrix}=
\begin{pmatrix}
\cos(\frac{2\pi }{q}) & -\sin(\frac{2\pi }{q})\\
 \sin(\frac{2\pi }{q}) & \cos(\frac{2\pi }{q}) 
 \end{pmatrix}
 \begin{pmatrix}y^1(0)\\ y^2(0)\end{pmatrix}.
\end{equation}
Introducing the complex variable $z(\tau)=y^1(\tau)+i y^2(\tau)$, we have the following mode expansion:
\begin{equation}
    z(\tau)=\sum_{n\in \mathbb{Z}}c_n\,e^{2\pi i(n+1/q)\tau}.
\end{equation}
Plugging in $\theta(\tau)=\frac{2\pi \tau}{q}$, we find the worldline path integral
\begin{align}
    &\int Dz(\tau)\  \exp\left[-m_i \int_0^1 d\tau \sqrt{(1-|z|^2)\left(\frac{d\theta}{d\tau}\right)^2+\left|\frac{dz}{d\tau}\right|^2}\right]\nn\\
    &=e^{-\frac{2m_i\pi }{q}}\prod_{n\in \mathbb{Z}}\int dc_n d\bar{c}_n \exp\left[-m_i \pi \left(n^2q+ 2n \right)|c_n|^2\right].
\end{align}

When $q\neq 2$, the only zero mode is $c_0\,e^{(2\pi i /q)\tau}$. In this case, following the treatment of zero modes outlined above, we find
\begin{align}
    \prod_{n\in \mathbb{Z}}\int dc_n d\bar{c}_n \exp\left[-m_i \pi \left(n^2q +2n \right)|c_n|^2\right]
    &=\text{vol}(\text{Gr}(1,\mathbb{C}^2))\prod_{n\in \mathbb{Z}_{\neq 0}}\frac{1}{m_i}\frac{1}{n\left(2+n q\right)}=\frac{m_i}{2\sin(\frac{2\pi}{q})},
\end{align}
where we've used 
\begin{equation}
\text{vol}(\text{Gr}(k,\mathbb{C}^d))=\frac{\text{vol}(U(d))}{\text{vol}(U(k))\times \text{vol}(U(d-k))}=\frac{\pi ^{k (d-k)}G(k+1)  G(d-k+1)}{G(d+1)},
\end{equation}
where $G(z)$ is the Barnes G-function. 

When $q=2$, there are two zero modes $c_0 e^{\pi i \tau}$ and $c_{-1}e^{-\pi i \tau}$. Thus,
\begin{align}
\prod_{n\in \mathbb{Z}}\int dc_n d\bar{c}_n \exp\left[-m_i \pi \left(2n^2+2n \right)|c_n|^2\right]
&=\int dc_0 d\bar{c}_0 dc_{-1} d\bar{c}_{-1}\prod_{n\in \mathbb{Z}_{\neq 0,-1}}\frac{1}{2 m_i}\frac{1}{n\left(1+n\right)}.
\end{align}
The integral over zero mode coefficients $c_0$ and $c_{-1}$ is not exactly the same as the volume of $\text{Gr}(2,\mathbb{R}^4)$: the correct local coordinates on $\text{Gr}(2,\mathbb{R}^4)$ should be the coefficient in front of $\cos(\pi \tau)$, $i\cos(\pi \tau)$
$\sin(\pi \tau)$ and $i\sin(\pi \tau)$, therefore:
\begin{equation}
	d\text{Vol}(\text{Gr}(2,\mathbb{R}^4))=4\ dc_0d\bar{c}_0dc_{-1}d\bar{c}_{-1}.
\end{equation}
Combing zero modes with massive modes, we get:
\begin{equation}
	\prod_{n\in \mathbb{Z}}\int dc_n d\bar{c}_n \exp\left[-m_i \pi \left(2n^2+2n \right)|c_n|^2\right]=\frac{m_i^2}{2}.
\end{equation}
Notice that tachyonic modes never appeared in the computations above. This is consistent with the fact that short geodesics are non-contractable!

\subsubsection{World line contribution to spin-refined partition function in $4d$}

We would like to compute single-particle non-perturbative corrections for a spin-refined sphere partition function in 4d. As 
explained in section~\ref{sec:foldingunfolding}, the corresponding EFT bundle is non-trivial. By winding around the fundamental cycle in $L(q;1)$, one
also advances by $\beta$ along the thermal circle whose total length is $q\beta$. Thus, after Kaluza-Klein reduction, we have a flat connection on an $S^1_{q\beta}$ bundle over $L(q;1)$ whose holonomy along a short geodesic is given by $\oint A=\beta$. The corresponding phase in the worldline action is thus $\exp(\frac{2\pi i \mathfrak{q}_i}{q \beta} \beta) = \exp(\frac{2\pi i \mathfrak{q}_i}{q})$.
Since the connection is flat, the classical equations of motion 
will not be affected, and we can compute the ``Nambu-Goto part" of the worldline path integral in exactly the same way as for the lens space partition function. Putting everything together, we arrive at (\ref{eqn:nonpertMain3}).

\subsection{A closer look at free field theory}
\label{sec:closerlookatfreetheory}

In this subsection, we discuss non-perturbative corrections in free field partition functions. In particular, we will show that they indeed have the structure predicted in section~\ref{sec:nonpert}. Actually, thanks to the Fock space structure of the Hilbert space, we can easily predict the leading terms in all non-perturbative corrections to free theory partition functions  using worldline instantons. The ``free-theory upgraded predictions'' in (\ref{eqn:nonPertFreeMain1}), (\ref{eqn:nonPertFreeMain2}), and (\ref{eqn:nonPertFreeMain3})
can also be checked explicitly, as we will do in section~\ref{sec:functionaldeterminants}.

In the language of section~\ref{sec:worldlineinstantons}, there is a worldline instanton correction for each Poincar\'{e} irrep in the $d{-}1$ dimensional massive theory on $\R^{d-1}$. In a free theory, the Hilbert space furthermore has the structure of a Fock space, so that for any symmetry generator $g$, we have
\begin{equation}
	\text{Tr}_{\mathcal{H}}[g]=\sum_{n=1}^\infty \text{Tr}_{\mathcal{H}_n}[g]=\sum_{n=1}^\infty \text{Tr}_{\text{Sym}^n(\mathcal{H}_1)}[g]=\exp\left(\sum_{\ell=1}^{\infty}\frac{1}{\ell}\text{Tr}_{\mathcal{H}_1}[g^\ell]\right),
\end{equation}
where $\cH_1$ is the single-particle Hilbert space.

We can apply this result to a free massive theory on $S^{d-1}$ as follows. Consider the sphere $S^{d-1}$ in ``angular" quantization, where we choose an angle $\f\in [0,2\pi)$ and slice the path integral on slices of constant $\f$. Each spatial slice is a $d{-}2$-dimensional hemisphere, with some boundary conditions at the equator of the hemisphere. (Our arguments will be heuristic because we will be vague about the nature of these boundary conditions, see \cite{Agia:2022srj} for a recent discussion.) Let the corresponding Hamiltonian be $H$. The sphere partition function is then
\be
\Tr_{\cH}(e^{-2\pi H}) = \exp\p{\sum_{\ell=1}^\oo \frac 1 \ell \Tr_{\cH_1}[e^{-2\pi \ell H}]}.
\ee
Each single-particle worldline instanton discussed in section~\ref{sec:worldlineinstantons} computes the leading contribution to $\Tr_{\cH_1}[e^{-2\pi H}]$. To instead compute $\Tr_{\mathcal{H}_1}[e^{-2 \pi \ell H}]$, we should expand around a classical trajectory $\theta(\tau)=2 \pi \ell \tau$ that winds $\ell$ times around the great circle. Expanding the worldline path integral around this trajectory, we find:
\be
\label{eq:multiplewinding}
\Tr_{\mathcal{H}_1}[e^{-2\pi \ell H}]&=\sum_{m}e^{-2\pi \ell  m}\frac{(\pm i)^{(d-2)(2\ell -1)}m^{d-2}}{\Gamma (d-1)}\left(1+ O\left(\frac{1}{m}\right)\right).
\ee

For a free scalar CFT on $S^{d-1}\times S^1_{\beta}$, the single-particle states are KK modes, so that we expect
\be
\label{eqn:nonPertFreeMain1}
\log(Z_{\text{free}}[S^{d-1}\times S^1_{\beta}])&\sim\sum_{m}\sum_{\ell=1}^\infty \frac{1}{\ell}\text{Tr}_{\mathcal{H}_1} e^{-2 \pi \ell H},
\ee
where each $\text{Tr}_{\mathcal{H}_1} e^{-2 \pi \ell H}$ is given by (\ref{eq:multiplewinding}), and $m$ runs over the spectrum of KK masses. Here, ``$\sim$" means that the right-hand side displays the nonperturbative terms in the left-hand side.

We can derive similar results for a free massive theory on the lens space $L(q;1)$.
If we consider an instanton on the locus $|z_1|=1$, wrapping $\ell$ times around the ``short geodesic," then the lens space worldline path integral evaluates $\Tr_{\mathcal{H}_{1}}[e^{-\frac{2\pi \ell }{q}H-\frac{2\pi \ell i}{q}J}]$, where $J$ generates a rotation in the $z_2$ plane. 
Hence for the free scalar on $L(q;1)\times S^1_{\beta}$, when $q$ is odd:
\begin{equation}\label{eq:oddLensSapceFreePartitionFunction}
	\begin{split}
\log(Z_{\text{free}}[L(q;1)\times S^1_{\beta}])&=\sum_{m}\sum_{\ell=1}^\infty \frac{1}{\ell}\Tr_{\mathcal{H}_{1}}[e^{-\frac{2\pi \ell }{q}H-\frac{2\pi \ell i}{q}J}]\\
&=\sum_{m}\sum_{\ell=1}^{\infty}\frac{1}{q \ell}\Tr_{\mathcal{H}_1}[e^{-2 \pi\ell H}]+
\sum_{m}\sum_{\ell\notin q \mathbb{Z}}\frac{1}{\ell} \Tr_{\mathcal{H}_1}[e^{-\frac{2\pi \ell }{q}H-\frac{2\pi i \ell}{q}J}].
	\end{split}
\end{equation}
In the last line, we singled out terms where $\ell\in q\mathbb{Z}$. Up to a $1/q$ factor, the first sum is exactly the same as the nonperturbative terms in $\log(Z_{\text{free}}[S^{d-1}\times S^1_\beta])$. Indeed, since the short geodesic generates $\pi_1(L(q;1))\simeq\mathbb{Z}_q$,
after winding around $q$ times, the loop becomes contractable. Notice that each contractable long geodesic in $L(q;1)$ will be 
lifted to $q$ different geodesics in $S^3$, so the moduli space of a long geodesic in $L(q;1)$ should have $1/q$-th the volume of $\text{Gr}(2,\mathbb{R}^4)$ and this is the origin of the $1/q$ factor.

Combining everything, the leading order terms are:
\be\label{eqn:nonPertFreeMain2}
\log(Z_{\text{free}}[L(q;1)\times S^1_{\beta}])&\sim\sum_m
\Bigg(
-\frac{1}{q}\sum_{\ell=1}^\infty e^{-2\pi \ell m}\frac{m^2}{2\ell }
		+\sum_{\substack{\ell=1\\ \ell\notin q\mathbb{Z}}}^\infty e^{-\frac{2\pi \ell m}{q}} \frac{m}{2\ell \sin(\frac{2\pi \ell}{q})}
\Bigg)
\quad (\textrm{odd $q$}),
\ee
where ``$\sim$" means we show nonperturbative corrections. Similarly when $q$ is even, we find
\be
&\log(Z_{\text{free}}[L(q;1)\times S^1_\beta])
\nn\\
\label{eqn:nonPertFreeMain3}
&\sim\sum_{m}\Bigg(-\frac{1}{q}\sum_{\ell=1}^\infty e^{-2\pi \ell m}\frac{m^{2}}{2\ell}
			+\sum_{\substack{\ell =1\\ \ell\notin (q/2)\mathbb{Z}}}^\infty e^{-\frac{2\pi \ell m}{q}} \frac{m}{2\ell\sin(\frac{2\pi \ell}{q})}
+\frac{1}{q}\sum_{\ell=1}^\infty e^{-\pi (2\ell-1) m } \frac{m^2}{2\ell-1}
\Bigg)
\quad (\text{even $q$}).
\ee

\subsection{Partition function from functional determinants}
\label{sec:functionaldeterminants}

In this section, we use the methods of \cite{Asorey:2012vp} to compute the partition function of free scalar theories 
by explicitly summing over the Laplacian spectrum. We will find agreement with (\ref{eqn:nonPertFreeMain1}), (\ref{eqn:nonPertFreeMain2}), and (\ref{eqn:nonPertFreeMain3}). In particular, this will verify the general predictions of section~\ref{sec:worldlineinstantons} for free theories.

Consider a massive scalar field on $S^{d-1}$ with the action
\begin{equation}
	S_{\text{free}}=\frac{1}{2}\int d^{d-1}x\sqrt{g} \left[g_{\mu\nu}\partial^\mu \phi\partial^\nu \phi+m^2 \phi^2 +\frac{1}{4}\frac{(d-2)}{(d-1)}R \phi^2\right],
\end{equation}
with $R=(d-1)(d-2)$.  Here, we have chosen the curvature coupling $\frac 1 2 \xi R\f^2$ to have the form appropriate for a conformally-coupled scalar in $d$-dimensions, dimensionally reduced to $d{-}1$ dimensions. However, this coupling will not affect the leading form of nonperturbative corrections that are our main focus. The partition function is the functional determinant
\begin{equation}\label{eqn:massivePartitionfunction}
	Z_{\text{massive}}[S^{d-1}]=\det\left[\Delta+m^2+\frac{(d-2)^2}{4}\right]^{-1/2},
\end{equation}
where $\Delta$ is the Laplace operator on $S^{d-1}$, which has eigenvalues $\lambda_k=k(k+d-2)$ with degneracies $d_k=\frac{(d-3+k)!(d-2+2k)}{k!(d-2)!}$. Let us focus on $d=4$ and $d=6$ i.e the sphere partition function of a massive theory on $S^3$ and $S^5$.

Using zeta function regularization, we find
\be\label{eqn:freeMassive4d}
\log(Z_{\text{massive}}[S^3])&= \frac{\pi m^3}{6}
 -\sum_{\ell}^{\infty}\frac{e^{-2\pi \ell m}}{2 \ell}\left(m^2+\frac{m}{\pi \ell^2}+\frac{1}{2\pi^2 \ell^3}\right),
\ee  
\be\label{eqn:freeMassive6d}
\log(Z_{\text{massive}}[S^5])&=-\frac{\pi}{120}m^5-\frac{\pi}{72}m^3+
\sum_{\ell=1}^{\infty}\frac{e^{-2\pi \ell m}}{2 4\ell}
\Bigg(m^4+\frac{2 m^3}{\pi \ell}+m^2\left(1+\frac{3}{\pi^2\ell^2}\right)
\nn\\&\quad+m\left(\frac{1}{\pi \ell}+\frac{3}{\pi^2\ell^2}+\frac{3}{2\pi^4 \ell^4}+\frac{1}{2\pi^2\ell^2}\right)\Bigg).
\ee

We now need to sum over the KK masses to obtain $Z_{\text{free}}[S^{d-1}\times S^1_\beta]$. Let us consider $\mathbb{Z}_2$-twisted free scalar fields as an example, where the $\Z_2$ twist is introduced to remove a zero mode upon dimensional reduction. The KK mass spectrum is $m=|(2n-1)/\beta|$ for $n\in\mathbb{Z}$. Thus, the free energy in $d=4$ and $d=6$ is
\begin{equation}
	\log(Z_{\text{free}}[S^{3}\times S^1_\beta])\sim 
	-\sum_{n=1}\sum_{\ell=1}^\infty e^{-\frac{2\pi^2(2n-1)\ell}{\beta}}\frac{1}{\ell}\left( \frac{(2n-1)^2\pi^2 }{\beta^2}+\frac{2n-1}{\beta\ell^2}+\frac{1}{2 \pi^2 \ell^3}\right),
\end{equation}
\be
&\log(Z_{\text{free}}[S^{5}\times S^1_\beta])
\nn\\
&\sim 
\sum_{n=1}^\infty \sum_{\ell=1}^\infty \frac{e^{-2\pi^2(2n-1) \ell/\beta }}{12\ell}
	\Bigg(\frac{\pi^4 (2n-1)^4}{\beta^4}+\frac{2 \pi^2 (2n-1)^3}{\ell \beta^3}
\nn\\
	&\quad+\frac{\pi^2 (2n-1)^2}{\beta^2}\left(1+\frac{3}{\pi^2\ell^2}\right)+\frac{\pi (2n-1)}{\beta}\left(\frac{1}{\pi \ell}+\frac{3}{\pi^2\ell^2}+\frac{3}{2\pi^4 \ell^4}+\frac{1}{2\pi^2\ell^2}\right)\Bigg).
\ee
where ``$\sim$" means we only show non-perturbative corrections.
These results are in agreement with the general form (\ref{eqn:nonPertFreeMain1}) predicted from worldline instantons, together with the Fock-space structure of the free theory.

The same approach can be generalized to compute a lens space partition function in 4d. We display 
the result here for a more direct comparison with the twisted partition function:
\be
\label{eqn:oddLensSapceFreePartitionFunctionComplete}
\log(Z_{\text{massive}}[L(q;1)])&= \frac{\pi m^3}{6 q}
-\frac{1}{q}\sum_{\ell}^{\infty}\frac{e^{-2\pi \ell m}}{2 \ell}\left(m^2+\frac{m}{\pi \ell}+\frac{1}{2\pi^2 \ell^2}\right)
\nn\\
&\quad+\sum_{\substack{\ell=1\\\ell \notin q \mathbb{Z}}}^\infty \frac{e^{-\frac{2 \pi \ell m}{q}}}{\sin\left(\frac{2\pi \ell}{q}\right)}\left(\frac{m}{2\ell}+\frac{q}{4\pi \ell^2}\right) \ \ \ \ \ \ (q\in 2\mathbb{Z}+1),
\ee 
\be \label{eqn:evenLensSapceFreePartitionFunctionComplete}
 \log(Z_{\text{massive}}[L(q;1)])&= \frac{\pi m^3}{6 q}
 -\frac{1}{q}\sum_{\ell}^{\infty}\frac{e^{-2\pi \ell m}}{2 \ell}\left(m^2+\frac{m}{\pi \ell}+\frac{1}{2\pi^2 \ell^2}\right)
 \nn\\
 &\quad+\frac{1}{q}\sum_{\ell}^{\infty}\frac{e^{-\pi(2\ell-1) m}}{2 \ell-1}\left(m^2+\frac{2m}{\pi (2\ell-1)}+\frac{2}{\pi^2 (2\ell-1)^2}\right)
 \nn\\
 &\quad+\sum_{\substack{\ell=1\\\ell \notin \frac{q}{2} \mathbb{Z}}}^\infty \frac{e^{-\frac{2 \pi \ell m}{q}}}{\sin\left(\frac{2\pi \ell}{q}\right)}\left(\frac{m}{2\ell}+\frac{q}{4\pi \ell^2}\right) \ \ \ \ \ \ (q\in 2\mathbb{Z}).
\ee

In appendix~\ref{appendix:highTexpansion}, we computed the high temperature expansion of a free scalar twisted by $e^{-\frac{2\pi i }{q}J_{12}-\frac{2\pi i }{q}J_{34}}$. 
Restricting to odd $q$ and inserting a $\Z_2$ twist to remove the zero mode, we find
\be
\label{eqn:spinRefinedZ2twist}
&\log\left(\Tr\left[e^{-\beta H-\frac{2\pi i }{q}J_{12}-\frac{2\pi i }{q}J_{34}}(-1)^N\right]\right)
\nn\\
&\sim 
 -\sum_{n=1}\sum_{\ell=1}e^{-\frac{2 \ell(2n-1) \pi^2}{q \beta}}\left(\frac{(2n-1)^2 \pi^2}{\ell q^3 \beta^2}+\frac{2n-1}{\beta \ell^2 q^2}+\frac{1}{2 \ell^3 q \pi^2}\right)
\nn\\
&\quad+\sum_{\substack{\ell=1\\\ell\notin q\mathbb{Z}}}^\infty \sum_{n=0}^\infty\frac{\cos(\frac{2 \pi \ell}{q}(\frac{q-1}{2}-n))}{\sin(\frac{2\pi \ell}{q})} e^{-\frac{2 \ell(2n+1)\pi^2}{q^2 \beta}}\left(\frac{(2n+1)\pi}{\ell q \beta}+\frac{q}{2 \ell^2 \pi}\right),
\ee
where ``$\sim$" means we only show non-perturbative corrections and $N$ is the $\f$-number operator, so that $(-1)^N$ implements the $\Z_2$ twist.
Note that this is not quite identical to the lens space partition function (\ref{eqn:oddLensSapceFreePartitionFunctionComplete}) summed over the mass spectrum $m=\frac{|2n+1|\pi}{q\beta}$.
To obtain the twisted partition function (\ref{eqn:spinRefinedZ2twist}), we must additionally modify the lens space result to account for the nontrivial background gauge field. To do so, we multiply each term in the summation over short geodesics by a phase $\exp(\frac{2\pi i \ell \mathfrak{q}_i}{q})$. 
For a $\mathbb{Z}_2$ twisted free scalar field, the KK charge spectrum is $\mathfrak{q}=n+\frac{1}{2}+\frac{q}{2}$. After putting in all 
the phases, we recover (\ref{eqn:spinRefinedZ2twist}). Finally, note that this result is consistent with the general prediction (\ref{eqn:nonpertMain3}) from worldline instantons.

\subsection{Free theories in odd dimension}
\label{section:oddDResurgence}

The examples presented in section~\ref{sec:functionaldeterminants} were all in even $d$. In odd $d$, we face the puzzle that the contribution of a worldline instanton (\ref{eqn:nonpertMain1}) is imaginary (even when the partition function should be real), and furthermore its phase depends on how we choose the integration contours for tachyonic modes. The proper interpretation of these kinds of contributions is explained for example in \cite{Dondi:2021buw}. They can be understood as characterizing singularities in the Borel plane when computing the partition function via Borel resummation. 

In this appendix, we provide a quick summary of the discussion from \cite{Dondi:2021buw} in the example of a massive free scalar on $S^2$. (We can think of this theory as the contribution of a single KK mode to the partition function of the 3d free scalar on $S^2\x S^1_\beta$.) We can compute the partition function in terms of the heat trace, following appendix~\ref{appendix:highTexpansion}.
The heat trace on $S^2$ is:
\begin{equation}
	\begin{split}
\Tr\left[e^{-t(\Delta+\left(\frac{d-2}{2}\right)^2)}\right]&=\sum_{k=0}^\infty (2k+1)e^{-t(k+\frac{1}{2})^2}=\sum_{k\in\mathbb{Z}}|k+\frac{1}{2}| e^{-t(k+\frac{1}{2})^2}\\
&=\sum_{\ell\in \mathbb{Z}}\left(\frac{1}{t}-\frac{2\pi \ell}{t^{3/2}}F\left(\frac{\pi\ell}{\sqrt{t}}\right)\right),
	\end{split}
\end{equation}
where $F(z)$ is Dawson's function which admits the following asymptotic expansion near $z=\infty$:
\begin{equation}
	F(z)\sim \sum_{n=0}^\infty \frac{(2n-1)!!}{2^{n+1}}\left(\frac{1}{z}\right)^{2n+1}.
\end{equation}
The heat trace therefore has the following expansion near $t=0$:
\begin{equation}
	\Tr\left[e^{-t(\Delta+\left(\frac{d-2}{2}\right)^2)}\right]\sim\sum_{n=0}^\infty a_n t^{n-1},\ \ \ a_n=\frac{(-1)^{n+1}(1-2^{1-2n})}{n!}B_{2n},
\end{equation}
where $B_{2n}$ are the Bernoulli numbers.

Let us now perform a Borel resummation of the series $\Phi\equiv\sum_{n=0}^\infty a_n t^n$:
\begin{equation}\label{eqn:BorellResum}
	\mathcal{S}[\Phi](t)=\frac{2}{\sqrt{t}}\int_0^\infty d\zeta \ e^{-\zeta^2/t} \mathcal{B}\Phi(\zeta),
\end{equation}
where the Borel transformed series $\mathcal{B}\Phi(\zeta)$ is:
\begin{equation}
\mathcal{B}\Phi(\zeta)=\sum_{n=0}^\infty \frac{a_n}{\Gamma(n+\frac{1}{2})}\zeta^{2n}=\frac{1}{\sqrt{\pi}}\frac{\zeta}{\sin(\zeta)}.
\end{equation}
With this expression for $\mathcal{B}\Phi$, the integral in (\ref{eqn:BorellResum}) is apparently ill-defined since 
there are poles located at $\zeta=\ell\pi$, $\ell\in\mathbb{Z}$ on the positive real axis. A contour prescription is needed to define the integral.
 
One natural option is to deform the integration contour to pass over (we will refer to the corresponding contour as $\mathcal{C}_{+}$) or
under (we will refer to the corresponding contour as $\mathcal{C}_{-}$) the poles. However, these two choices are not equivalent. 
Their difference is the sum of residues at the poles:
\begin{equation}
	\frac{2}{\sqrt{t}}\int_{\mathcal{C}_{+}-\mathcal{C}_{-}}d\zeta \ e^{-\zeta^2/t} \mathcal{B}\Phi(\zeta)=2 i t \left(\frac{\pi}{t}\right)^{3/2}\sum_{\ell\neq 0} (-1)^{\ell+1}|\ell| e^{-\ell^2 \pi^2/t}.
\end{equation}
 The ambiguity in integration contour leads to an ambiguity in the Borel-resummed heat trace:
\begin{equation}
\label{eq:thingwithnonpertterms}
\Tr\left[e^{-t(\Delta+\left(\frac{d-2}{2}\right)^2)}\right]=\frac{2}{\sqrt{\pi}t^{3/2}}\int_{\mathcal{C}_{\pm}}d\zeta \ \frac{\zeta e^{-\zeta^2/t}}{\sin\zeta}+2 i \left(\frac{\pi}{t}\right)^{3/2}\sum_{\ell\neq 0}\sigma_\ell^{\pm} (-1)^{\ell+1}|\ell|e^{-\frac{\ell^2 \pi^2}{t}},
\end{equation}
where $\sigma_\ell^{\pm}$ are arbitrary coefficients which jump as we switch from $\mathcal{C}_{+}$ to $\mathcal{C}_{-}$. If we require the heat trace 
to be real when $t\in\mathbb{R}_{+}$, then the $\sigma^{\pm}_{\ell}$ are fixed to be $\pm\frac{1}{2}$. This is equivalent to a principal value prescription for the Borel integral:
\be
\label{eq:principalvalue}
\Tr\left[e^{-t(\Delta+\left(\frac{d-2}{2}\right)^2)}\right]=\frac{2}{\sqrt{\pi}t^{3/2}}\int_{0}^\oo d\zeta\,\mathrm{P}\left[\frac{\zeta e^{-\zeta^2/t}}{\sin\zeta}\right].
\ee

To compute $\log Z$, we must supply a factor of $e^{-t m^2}$ and integrate $\int dt/t$. If we start with (\ref{eq:principalvalue}), we get a valid integral representation for the partition function. However, 
if we start with (\ref{eq:thingwithnonpertterms}) and perform the $t$-integral term by term, we obtain the series of nonperturbative corrections
\begin{equation}
\log(Z_{\text{free}}[S^2])\Bigg|_{\text{non-perturbative}}=\pm i\sum_{\ell=1}^\infty \frac{(-1)^{\ell}e^{-2\pi m \ell}}{\ell}\left(m+\frac{1}{2\pi\ell}\right),
\end{equation}
whose leading terms in $m$ agree precisely with the worldline instanton predictions (\ref{eq:multiplewinding}) and (\ref{eqn:nonPertFreeMain1}) when $d=3$.

To summarize, worldline instantons encode residues of certain singularities in the Borel plane. We conjecture that this remains true in interacting theories. In general, the thermal effective action gives an asymptotic expansion in $\beta$. When we Borel-resum this expansion, we encounter singularities in the Borel plane coming from worldline instantons (together with other nonperturbative effects like instanton graphs).

\bibliographystyle{JHEP}
\bibliography{refs}
\end{document}